%% file: main.tex
\documentclass[apj]{openjournal}

\usepackage{xcolor}
\usepackage{textgreek}
\usepackage[utf8]{inputenc}
\usepackage[english]{babel}

\usepackage{hyperref}
\hypersetup{
    unicode, 
    colorlinks=true,
    linkcolor=linkcolor,
    citecolor=linkcolor,
    filecolor=linkcolor,
    urlcolor=linkcolor,
}
\usepackage{color,colortbl}
\definecolor{linkcolor}{rgb}{0.0,0.3,0.5}
\usepackage{tensind}
\tensordelimiter{?}
\DeclareGraphicsExtensions{.bmp,.png,.jpg,.pdf}
\usepackage{verbatim}
\usepackage[normalem]{ulem}
\usepackage{orcidlink}
\usepackage{soul}

\usepackage{amssymb}
\usepackage{amsmath}
\usepackage{amsthm}
\usepackage{array}
\usepackage{longtable}
\newcounter{algorithm}
\newenvironment{algorithm}[1][]{%
  \par\medskip\noindent\begin{minipage}{\linewidth}\hrule\medskip
  \renewcommand{\caption}[1]{\refstepcounter{algorithm}\noindent\textbf{Algorithm~\thealgorithm.} ##1\par\smallskip}%
}{\medskip\hrule\end{minipage}\par\medskip}
\usepackage{makecell}

\graphicspath{ {./figures/} }

\newcommand{\new}[1]{{#1}}

\newcommand{\vecx}{\mathbf{x}}
\newcommand{\vecv}{\mathbf{v}}

\newcommand{\vecG}{\mathbf{G}}
\newcommand{\vecU}{\mathbf{U}}
\newcommand{\vecF}{\mathbf{F}}
\newcommand{\tenP}{\mathsf{P}}
\newcommand{\tenA}{\mathsf{A}}
\newcommand{\tenS}{\mathsf{S}}
\newcommand{\tenI}{\mathsf{I}}
\newcommand{\tenG}{\mathsf{G}}
\newcommand{\tenZ}{\mathsf{Z}}
\providecommand{\tenD}{\mathsf{D}}
\providecommand{\tenB}{\mathsf{B}}
\newtheorem{lem}{Lemma}
\newtheorem{prop}{Proposition}
\theoremstyle{definition}

\newenvironment{pf}{\begin{proof}}{\end{proof}}
\newenvironment{pfprop}{\begin{proof}[Proof of Proposition~\ref{prop:energy_block}]}{\end{proof}}
\newenvironment{pfcb}{\begin{proof}[Proof of Proposition~\ref{prop:checkerboard}]}{\end{proof}}
\newenvironment{pfm1}{\begin{proof}[Outline of the proof of Proposition~\ref{prop:m1}]}{\end{proof}}

\providecommand{\tenH}{\mathsf{H}}

\providecommand{\tenQ}{\mathsf{Q}}
\providecommand{\tenW}{\mathsf{W}}
\providecommand{\tenX}{\mathsf{X}}
\providecommand{\tenC}{\mathsf{C}}
\providecommand{\tenV}{\mathsf{V}}
\providecommand{\tenY}{\mathsf{Y}}
\newcommand{\tenJ}{\mathsf{J}}
\newcommand{\tenM}{\mathsf{M}}
\newcommand{\hatt}{\hat{t}}
\newcommand{\hrho}{\hat{\rho}}
\newcommand{\he}{\hat{e}}
\newcommand{\hk}{\hat{k}}
\newcommand{\homega}{\hat{\omega}}
\newcommand{\hE}{\hat{E}}
\newcommand{\hnabla}{\hat{\nabla}}
\newcommand{\hp}{\hat{p}}
\newcommand{\hv}{\hat{v}}

\newcommand{\hF}{\hat{F}}
\newcommand{\hx}{\hat{x}}

\newcommand{\hP}{\hat{P}}
\newcommand{\hT}{\hat{T}}
\newcommand{\hOmega}{\hat{\Omega}}

\newcommand{\hvecv}{\hat{\mathbf{v}}}
\newcommand{\hvecx}{\hat{\mathbf{x}}}
\newcommand{\hvecF}{\hat{\mathbf{F}}}
\newcommand{\hvecG}{\hat{\mathbf{G}}}
\newcommand{\htenP}{\hat{\mathsf{P}}}
\newcommand{\hvecFU}{\hat{\mathbf{F}}_{\hat{\mathbf{U}}}}
\newcommand{\hvecU}{{\hat{\mathbf{U}}}}
\newcommand{\hvecW}{{\hat{\mathbf{W}}}}
\newcommand{\hvecS}{\hat{\mathbf{S}}}
\newcommand{\calC}{\mathcal{C}}
\newcommand{\calP}{\mathcal{P}}
\newcommand{\calM}{\mathcal{M}}
\newcommand{\calL}{\mathcal{L}}

\newcommand{\calD}{\mathcal{D}}

\begin{document}

\title{An Adaptive Reduced Speed of Light Approximation}

\author{Chong-Chong He\,\orcidlink{0000-0002-2332-8178}}
\email{chongchong.he@anu.edu.au}
\affiliation{Research School of Astronomy and Astrophysics, Australian National University, Mt. Stromlo Observatory, 233 Mt. Stromlo Rd., Stromlo, ACT 2612, Australia}

\author{Mark R. Krumholz \orcidlink{0000-0003-3893-854X}}
\affiliation{Research School of Astronomy and Astrophysics, Australian National University, Mt. Stromlo Observatory, 233 Mt. Stromlo Rd., Stromlo, ACT 2612, Australia}

\author{Benjamin D. Wibking \orcidlink{0000-0003-3175-2291}}
\affiliation{The Observatories of the Carnegie Institution for Science, 813 Santa Barbara Street, Pasadena, CA 91101, USA}

\begin{abstract}
The reduced speed of light approximation (RSLA) is a common approach in radiation-magnetohydrodynamic (RHD) simulations of non-relativistic systems where the gas speed $v_{\rm gas} \ll c$. In the RSLA one eases the severe time step constraint imposed by the speed of light $c$ by replacing it with a reduced value $\hat{c}$ chosen so that $v_{\rm gas} \ll \hat{c} \ll c$. This  approach works well in flows with low or moderate optical depth, but fails in optically thick media, where it makes thermal relaxation too slow and gives incorrect RHD wave modes. Here we propose an Adaptive Reduced Speed of Light Approximation (ARSA) that removes these problems by replacing the global speed reduction factor $\phi_0 \equiv \hat{c}/c$ with a \textit{local}, cell-by-cell reduction factor $\phi$ derived from a von~Neumann stability analysis of the implicit-explicit update used by modern two-moment RHD codes. The resulting limiter has no tunable parameters, can be computed using only local operations, and improves on the classic RSLA by allowing $\phi \to 1$ in cells where a reduced speed of light is not required because the exchange with the gas damps unstable modes, allowing recovery of the correct thermal relaxation and wave-damping rates in highly-opaque regions. We compare ARSA to the classic RSLA in a range of standard RHD test problems, and show that in opaque regions where gas holds most of the heat capacity the ARSA is much more accurate than the RSLA at the same time step, or that it can achieve the same accuracy as the RSLA with up to 100 times fewer time steps. However, gains are limited in regions where radiation holds a large share of the heat capacity, as it does in some astrophysical systems.
\end{abstract}


\maketitle

\section{Introduction}
\label{sec:intro}

The exchange of energy and momentum between matter and radiation is a central process in a wide range of astrophysical systems, including star and star cluster formation \citep[e.g.,][]{Krumholz12b, Skinner15a, Rosen16a, Menon22b}, the envelopes and winds of massive stars \citep[e.g.,][]{Jiang15a, Jiang17a}, accretion discs \citep[e.g.,][]{Jiang13a}, core-collapse supernovae \citep[e.g.,][]{Skinner19a}, and the reionization of the universe \citep[e.g.,][]{Gnedin01a, Rosdahl13a, Kimm14a}. Because direct solution of the time-dependent equation of radiative transfer over the full six-dimensional phase space of position, direction, and frequency remains prohibitively expensive when coupled to hydrodynamics, most numerical treatments of radiation-hydrodynamics (RHD) instead evolve angular moments of the radiation intensity, closing the moment hierarchy either by flux-limited diffusion \citep[FLD; e.g.,][]{Turner01a, Howell03a, Krumholz07b, Kuiper10a}, by an assumed local closure such as M1 \citep[e.g.,][]{Gonzalez07a, Rosdahl13a, Skinner13a, Kannan18a}, or by a variable Eddington tensor computed from a separate approximate solution of the transfer equation \citep[e.g.,][]{Gnedin01a, Jiang12a, Davis12a, Menon22a}.

The chief numerical difficulty presented by the moment equations is the disparity between the speed of light $c$ and the characteristic matter signal speeds. This disparity manifests in two ways: the radiation transport terms carry signals at up to speed $c$, while the matter-radiation exchange terms can be stiff, with equilibration times many orders of magnitude smaller than any flow timescale. The traditional response to this challenge is to integrate the moment equations implicitly \citep[e.g.,][]{Turner01a, Howell03a, Krumholz07b, Jiang21a}, which renders the update unconditionally stable but requires the solution of a globally-coupled non-linear system at each time step. Implicit transport solvers are communication-intensive and increasingly ill-matched to modern hardware, where the bulk of the available computing power resides in GPUs that favor local, regular operations. Consequently a number of modern RHD codes instead treat the transport terms with an explicit, shock-capturing update -- which is local, vectorizes well, and achieves high throughput on GPUs -- while retaining an implicit treatment only of the purely local exchange terms \citep[e.g.,][]{Skinner13a, Rosdahl13a, Kimm14a, Kannan18a, Skinner19a, Wibking22a, He24b, He24a}.

The price of this approach is a Courant-Friedrichs-Lewy (CFL) time step constraint $\Delta t \lesssim \Delta x / c$, which for non-relativistic systems is typically several orders of magnitude smaller than the hydrodynamic time step. To mitigate this cost it is common to adopt the reduced speed of light approximation (RSLA), introduced by \citet{Gnedin01a} in the context of cosmological reionization: one simply replaces the speed of light $c$ appearing in the radiation moment equations with a reduced value $\hat{c} \ll c$, chosen to remain well above all matter signal speeds so that the ordering of timescales in the problem is preserved. The CFL constraint is correspondingly relaxed by a factor $c/\hat{c}$, which typically renders explicit transport affordable.

The approximation is, however, uncontrolled, and its limits of validity have been the subject of considerable discussion. \citet{Skinner13a} derive the criterion $\hat{c} \gg v_{\max}\max(1, \tau_{\max})$, where $\tau_{\rm max}$ is the maximum system optical depth; this condition keeps radiation diffusion fast compared with the gas dynamics in the static diffusion regime. \citet{Gnedin16a} shows that in reionization calculations the RSLA must not be applied to the metagalactic background, and \citet{Deparis19a} and \citet{Ocvirk19a} demonstrate that ionization front speeds and post-reionization neutral fractions are corrupted when $\hat{c}$ is chosen too small. \citet{Wibking22a} point out a still more fundamental problem: the RSLA cannot simultaneously conserve energy and momentum, and in the dynamic diffusion regime, where radiation is advected by optically thick flows, it produces order-unity errors in the radiation field no matter how $\hat{c}$ is chosen.

In part to limit such errors, several authors have experimented with letting the reduced speed of light vary, or with making other corrections for its effects. \citet{Katz17a} and \citet{Rosdahl18a} divide $\hat{c}$ by two at each level of refinement in an adaptive mesh refinement (AMR) simulation, which keeps the radiation time step the same on every level. \citet{Cain24a} shows that in reionization simulations the RSLA is equivalent to a time-dependent rescaling of the ionizing emissivity and corrects for it, and \citet{Gnedin25a} shows that the RSLA does not conserve photons in some regimes and proposes a partial remedy. These methods set $\hat{c}$ by the resolution or correct its effect on the photon budget of optically thin gas. The optically thick regime, where \citet{Skinner13a} and \citet{Wibking22a} locate the greatest problems of the RSLA, has received less attention: \citet{Rosdahl15a} split the radiation in unresolved thick cells into a trapped part, advected with the gas, and a streaming part, a sub-grid model that keeps $\hat{c}$ fixed. \citet{2026ApJ...999..257K} adopted a similar approach by introducing an unmodified radiation component that dominates in optically thick regions.

A separate line of work treats the stiff exchange with implicit-explicit (IMEX) schemes \citep{Jin96a} and with asymptotic-preserving (AP) schemes, which stay stable and consistent in the diffusion limit at a time step independent of the stiffness \citep{Jin99a}. An HLL flux is not AP on its own, since its numerical diffusion can exceed the physical diffusion in thick cells \citep{lowrie01, He24a}. These schemes keep the true speed of light, and so, with explicit transport, are limited to the step $\Delta t \lesssim \Delta x/c$.

In this paper we propose a method that improves on the standard RSLA significantly in exactly this regime. The key insight that enables this improvement is that, when radiation and matter are tightly coupled and matter holds most of the heat capacity, no reduction of the signal speed is needed for stability, because exchange with the gas damps unstable radiation modes. We therefore develop a variant of the RSLA that replaces the global reduction factor $\phi_0 = \hat{c}/c$ by a local one, $\phi(\mathbf{x})$, which every cell sets close to the largest value that the numerical scheme supports stably at the time step of the standard RSLA. This removes the failures of the RSLA in the high-opacity regime while retaining the favorable cost and parallelization properties of explicit transport.

Our plan for the remainder of this paper is as follows: \autoref{sec:analysis} analyzes the classic RSLA and motivates our approach. In \autoref{sec:irsla} we derive the stability boundary and the ARSA, and in \autoref{sec:implementation} we provide an algorithmic implementation of the method. We present tests of the new method in \autoref{sec:tests}, and summarize our findings in \autoref{sec:summary}.

\section{Analysis of the Reduced Speed of Light Equations}
\label{sec:analysis}

In this section we show why the standard RSLA fails in optically thick gas, and what a better approximation must do.

\subsection{The Non-Dimensional Equations of Radiation-Hydrodynamics}
\label{ssec:nondim}

We start from the equations of RHD for non-relativistic flows,
\begin{eqnarray}
\frac{\partial}{\partial t}\rho + \nabla \cdot \left(\rho\vecv\right) & = & 0 \\
\frac{\partial}{\partial t}\left(\rho \vecv\right) + \nabla\cdot\left(\rho \vecv\otimes\vecv\right) + \nabla p & = & \vecG, \\
\frac{\partial}{\partial t}\left(\rho e\right) +  \nabla \cdot\left[\left(\rho e + p\right)\vecv\right]  & = & c G^0 \\
\frac{\partial}{\partial t}E + \nabla \cdot \vecF & = & -c G^0 \\
\frac{1}{c^2}\frac{\partial}{\partial t}\vecF + \nabla \cdot \tenP & = & -\vecG ,
\end{eqnarray}
where $\rho$, $\vecv$, $p$, and $e$ are the matter density, velocity, pressure, and energy per unit mass, and $E$, $\vecF$, and $\tenP$ are the radiation energy density, flux, and pressure tensor. The coupling terms on the right-hand sides are the time-like and space-like components of the radiation four-force density $(G^0, \vecG)$. This four-vector can be expressed in multiple reference frames; here we adopt the mixed-frame formulation whereby radiation quantities are expressed in the inertial frame, while the emissivity and opacity of the matter are Taylor-expanded about their comoving frame values in powers of $v/c$ \citep{Mihalas99a, Lowrie99a, Mihalas01a, Krumholz07b}. For matter in local thermodynamic equilibrium (LTE), neglecting scattering, and to first order in $v/c$, this formulation gives
\begin{eqnarray}
cG^0 & = & \rho c\left(\kappa_E E - \kappa_P a_r T^4\right) + \rho\left(\kappa_F - 2\kappa_E\right)\frac{\vecv\cdot\vecF}{c},
\label{eq:G0_int}
\\
\vecG & = & \rho\kappa_F\frac{\vecF}{c} - \frac{\rho}{c}\left[\kappa_P a_r T^4\,\vecv + \left(\kappa_F - \kappa_E\right)E\,\vecv + \kappa_F\,\vecv\cdot\tenP\right],
\label{eq:G_int}
\end{eqnarray}
where $T$ is the gas temperature, $a_r$ the radiation constant, and $\kappa_P$, $\kappa_E$ and $\kappa_F$ are three means of the comoving-frame opacity per unit mass over the spectrum: the Planck mean, which sets the emission; the energy mean, which sets the absorption of $E$; and the flux mean, which sets the exchange of momentum. For a gray material the three means are equal.

Following \citet{Lowrie99a}, it is helpful to non-dimensionalise the equations. For the matter quantities we let $\ell$ be the characteristic size of the system, $a_\infty$ be the characteristic isothermal sound speed, and $\rho_\infty$ be the characteristic density, and define
\begin{eqnarray}
\hvecx = \vecx/\ell & \hatt = t a_\infty/\ell & \hrho = \rho/\rho_\infty
\nonumber \\
\hvecv = \mathbf{v}/a_\infty & \hp = p/(\rho_\infty a_\infty^2) & \he = e/a_{\infty}^2
\end{eqnarray}
Here the quantities with carets are dimensionless versions of the dimensional quantities. For the radiation we also define a reference temperature $T_\infty$ and a reference opacity $\kappa_\infty$, with the corresponding photon mean free path $\lambda_\infty = 1/(\rho_\infty\kappa_\infty)$, and
\begin{equation}
\hE = \frac{E}{a_r T_\infty^4} \qquad
\hvecF = \frac{\vecF}{c\, a_r T_\infty^4} \qquad
\htenP = \frac{\tenP}{a_r T_\infty^4} \qquad
\hT = \frac{T}{T_\infty} \qquad
\hat{\kappa} = \frac{\kappa}{\kappa_\infty},
\end{equation}
the last for each of the three means. The four-force is then $cG^0 = (c a_r T_\infty^4/\lambda_\infty)\,\hat{G}^0$ and $\vecG = (a_r T_\infty^4/\lambda_\infty)\,\hvecG$, with
\begin{eqnarray}
\hat{G}^0 & = & \hrho\left(\hat{\kappa}_E\hE - \hat{\kappa}_P\hT^4\right) + \hrho\left(\hat{\kappa}_F - 2\hat{\kappa}_E\right)\frac{\hvecv\cdot\hvecF}{\calC},
\label{eq:G0_int_nondim}
\\
\hvecG & = & \hrho\hat{\kappa}_F\hvecF - \frac{\hrho}{\calC}\left[\hat{\kappa}_P\hT^4\,\hvecv + \left(\hat{\kappa}_F - \hat{\kappa}_E\right)\hE\,\hvecv + \hat{\kappa}_F\,\hvecv\cdot\htenP\right],
\label{eq:G_int_nondim}
\end{eqnarray}
where
\begin{equation}
\calC = \frac{c}{a_\infty}
\end{equation}
is the dimensionless speed of light. Substituting these definitions into the moment equations gives
\begin{eqnarray}
\frac{\partial}{\partial \hatt}\hrho + \hnabla\cdot\left(\hrho\hvecv\right) & = & 0 \\
\label{eq:gas_momentum}
\frac{\partial}{\partial \hatt}\left(\hrho \hvecv\right) + \hnabla\cdot\left(\hrho \hvecv\otimes\hvecv\right) + \hnabla \hat{p} & = & \calP \calL \hvecG, \\
\label{eq:gas_energy}
\frac{\partial}{\partial \hatt}\left(\hrho \he\right) +  \hnabla \cdot\left[\left(\hrho \he + \hat{p}\right)\hvecv\right]  & = & \calP \calC \calL \hat{G}^0 \\
\label{eq:rad_energy}
\frac{\partial}{\partial\hatt} \hE + \calC \hnabla\cdot\hvecF & = & -\calC \calL \hat{G}^0 \\
\label{eq:rad_momentum}
\frac{\partial}{\partial \hatt}\hvecF + \calC \hnabla \cdot \htenP & = & -\calC \calL \hvecG ,
\end{eqnarray}
where $\hnabla$ indicates a divergence with respect to the dimensionless position $\hvecx$, and we have defined
\begin{equation}
\calL = \frac{\ell}{\lambda_\infty} \qquad \calP = \frac{a_r T_\infty^4}{\rho_\infty a_\infty^2}.
\end{equation}

We therefore see that the equations contain three dimensionless quantities: $\calC$ is the dimensionless speed of light, $\calL$ is the ratio of the system size to the photon mean free path (and thus is equal to the characteristic optical depth), and $\calP$ is (up to factors of order unity) the ratio of radiation pressure to gas pressure at the characteristic temperature and sound speed of the system, for radiation in thermodynamic equilibrium with the matter, $E = a_r T^4$. Note that in optically thin systems the radiation energy density is generally much smaller than $a_r T^4$, and the actual ratio of radiation to gas pressure is then much smaller than $\calP$.

\subsection{The Standard Reduced Speed of Light Approximation}
\label{ssec:rsla}
\label{ssec:heat_capacity}

Since $\calC \gg 1$ for a non-relativistic problem, in a system that is even moderately optically thick ($\calL \gtrsim 1$) the exchange terms on the right-hand sides of \autoref{eq:rad_energy} and \autoref{eq:rad_momentum} are large compared to unity. They are therefore usually evaluated with an implicit step, which presents no great practical difficulty, because they are purely local. The transport terms, $\calC\hnabla\cdot\hvecF$ and $\calC\hnabla\cdot\htenP$, are more problematic. For reasons discussed in \autoref{sec:intro} it is often preferable on modern hardware to update them explicitly, but an explicit update must then satisfy a CFL condition, $\Delta\hatt \lesssim \Delta\hx/\calC$. Since $\calC$ is a large number for non-relativistic flows, this is a severe time step constraint.

The RSLA addresses this problem by multiplying the transport and exchange terms of the radiation equations, \autoref{eq:rad_energy} and \autoref{eq:rad_momentum}, by a constant factor $\phi \ll 1$, and leaving the matter equations unchanged:
\begin{eqnarray}
\frac{\partial}{\partial\hatt} \hE & = & -\phi\left(\calC\hnabla\cdot\hvecF + \calC\calL\hat{G}^0\right),
\label{eq:rad_energy_rsl}
\\
\frac{\partial}{\partial \hatt}\hvecF & = & -\phi\left(\calC\hnabla\cdot\htenP + \calC\calL\hvecG \right).
\label{eq:rad_momentum_rsl}
\end{eqnarray}
Here $\phi$ is a dimensionless number that obeys $\calC^{-1} \ll \phi \ll 1$. This reduces the signal speed from $\calC$ to $\phi\calC$, allowing a correspondingly larger time step. The logic behind this strategy is that on a fluid flow timescale the time derivatives on the left-hand sides are of order $\calM$, where $\calM = |\hvecv|$ is the characteristic Mach number of the flow, so as long as $\phi\calC \gg \calM$ then the right-hand sides remain large on a fluid flow timescale, and the solution on a fluid flow timescale is the same as the true solution computed without the extra factor of $\phi$. The RSLA therefore accepts errors on the radiation transport timescale, as long as these become small on the matter flow timescale.

This logic can fail in two regimes. One is where the two terms in parentheses on the right-hand sides nearly cancel, so that the full RHS does not remain $\gg 1$ even though each of the individual terms does. This can occur when $\calL \gg 1$. In this regime, \citet{Lowrie99a} show that to leading order $\hat{\kappa}_E\hE = \hat{\kappa}_P\hT^4$, $\htenP = \hE\tenI/3$, and
\begin{equation}
    \hvecF = -\frac{\hnabla\hE}{3\calL\hrho\hat{\kappa}_F} + \frac{4\hvecv\hE}{3\calC} .
    \label{eq:flux_next_order}
\end{equation}
The first term is the diffusion flux, of order $1/\calL$. The second is the radiation carried with the gas, of order $\calM/\calC$. Which term dominates depends on the dimensionless ratio
\begin{equation}
    \calD = \frac{\calM\calL}{\calC},
\end{equation}
where we refer to $\calD$ as the dynamic diffusion number; $\calD \ll 1$ is the static diffusion limit, where radiation diffuses across the system faster than the flow crosses it, and $\calD \gg 1$ is the dynamic diffusion limit where the reverse is true. Inserting these limiting forms into the terms on the RHS of \autoref{eq:rad_energy_rsl} shows that it is of order $\phi\calC/\calL$ in the static diffusion limit and of order $\phi\calM$ in the dynamic one. For either case, the condition that the RHS remain large is then satisfied only if $\phi \gg \calD$, which is equivalent to the criterion of \citet{Skinner13a}.

The second way in which the RSLA can fail is due to the mismatch that it introduces between the RHS of the radiation equations, \autoref{eq:rad_energy_rsl} and \autoref{eq:rad_momentum_rsl}, which are multiplied by the reduction factor $\phi$, and the corresponding matter equations, \autoref{eq:gas_energy} and \autoref{eq:gas_momentum}, which are not. For simplicity we analyze the effects of this mismatch for the case of a gray material at rest, with the opacity held at its reference value, $\hrho\hat{\kappa} = 1$, and an ideal gas of constant adiabatic index $\gamma$, $p = \rho k_B T/\mu$ with $\mu$ the mean particle mass. We choose the reference temperature $T_\infty = \mu a_\infty^2/k_B$, so that the internal energy of the gas at rest is $\hrho\he = \hrho\hT/(\gamma - 1)$.

To understand the effects of the RSLA, it is helpful to compare the heat capacities of the gas and the radiation. In equilibrium the radiation energy density is $E = a_r T^4$, so a change $\delta T$ of the common temperature changes the radiation energy by $4 a_r T^3\,\delta T$ and the gas energy by $\rho c_v\,\delta T$, with $c_v$ the specific heat at constant volume.\footnote{Alternately we could replace $c_v$ with the heat capacity at constant pressure $c_p$; for the order-of-magnitude argument we make here the distinction is unimportant.} Their ratio is
\begin{equation}
    \hat{h} \equiv \frac{4 a_r T^3}{\rho c_v} = 4(\gamma-1)\calP \frac{\hT^3}{\hrho}.
    \label{eq:h_ratio}
\end{equation}
The RSLA effectively changes this ratio. Without transport, \autoref{eq:gas_energy} and  \autoref{eq:rad_energy_rsl} reduce to
\begin{equation}
    \frac{\partial(\hrho\he)}{\partial\hatt} = \calP\calC\calL\left(\hE - \hT^4\right),
    \qquad
    \frac{\partial\hE}{\partial\hatt} = -\phi\,\calC\calL\left(\hE - \hT^4\right),
    \label{eq:rsla_exchange}
\end{equation}
so the gas exchanges energy with the radiation at the full rate, $\propto \calC$, while the radiation exchanges energy with the gas at the reduced rate, $\propto \phi\calC$, so that the gas loses energy $1/\phi$ times faster than the radiation gains it. Adding the first equation to $\calP/\phi$ times the second shows that the RSLA conserves $\hrho\he + \calP\hE/\phi$ rather than the total energy, $\hrho\he + \calP\hE$. To the gas, the radiation therefore acts as if its heat capacity were $4\calP\hT^3/\phi$, and the heat-capacity ratio becomes $\hat{h}/\phi$.

This has two consequences. First, the gas responds to heating with the wrong temperature change. Consider an opaque region, where $\hE = \hT^4$, heated by compression or by any other source in the gas energy equation. The heating raises $\hrho\he + \calP\hE/\phi$ rather than the total energy, so that the gas temperature changes by $\delta\hT_\phi$ instead of the true $\delta\hT_1$, with
\begin{equation}
    \frac{\delta\hT_\phi}{\delta\hT_1} = \frac{1 + \hat{h}}{1 + \hat{h}/\phi} .
    \label{eq:heating_error}
\end{equation}
This factor is close to unity only if $\phi \gg \hat{h}$. The same weighting makes a closed system relax to the wrong common temperature, so the RSLA never produces the correct answer for a closed system.

Second, temperature structures diffuse too slowly. The radiation still carries the true diffusion flux, $\hvecF = -\hnabla\hE/(3\calL)$ (\autoref{eq:flux_next_order}), but the temperature responds to it with the heat capacity of \autoref{eq:heating_error}, so that the diffusion time is longer under the RSLA, $t_\phi$, than without it, $t_1$, by the inverse factor,
\begin{equation}
    \frac{t_\phi}{t_1} = \frac{1 + \hat{h}/\phi}{1 + \hat{h}} .
    \label{eq:slowdown}
\end{equation}
Taking the ratio of the diffusion timescale, $t_\phi = 3\calL(1/\hat{h} + 1/\phi)/\calC$, to the flow timescale, $1/\calM$, gives
\begin{equation}
    \mathrm{Pe}_\phi = 3\calD\left(\frac{1}{\hat{h}} + \frac{1}{\phi}\right),
    \label{eq:Pe_phi}
\end{equation}
where Pe$_\phi$ is the P\'eclet number evaluated with reduction factor $\phi$, and $\mathrm{Pe} = \mathrm{Pe}_1$ is its true value. The RSLA need not preserve the diffusion time itself. It is accurate as long as it does not move the P\'eclet number away from its regime, $\mathrm{Pe} \ll 1$ or $\mathrm{Pe} \gg 1$: a structure that diffuses much faster than the flow moves it is smoothed out on the flow timescale either way, and one that diffuses much more slowly is carried unchanged either way. Since $\mathrm{Pe}_\phi - \mathrm{Pe} = 3\calD(1/\phi - 1) < 3\calD/\phi$, this holds whenever $\phi \gg \calD$, the condition that we found above. The slower diffusion therefore adds no condition of its own.

Combining $\phi \gg \hat{h}$ with $\phi \gg \calD$, and using $\hat{h} \sim \calP$ (\autoref{eq:h_ratio}), we find that the RSLA is accurate in an optically thick system only if
\begin{equation}
    \phi \gg \max(\calP, \calD).
    \label{eq:rsla_valid}
\end{equation}

\input{tab_media}

\begin{figure}[t]
    \centerline{\includegraphics[width=0.85\textwidth]{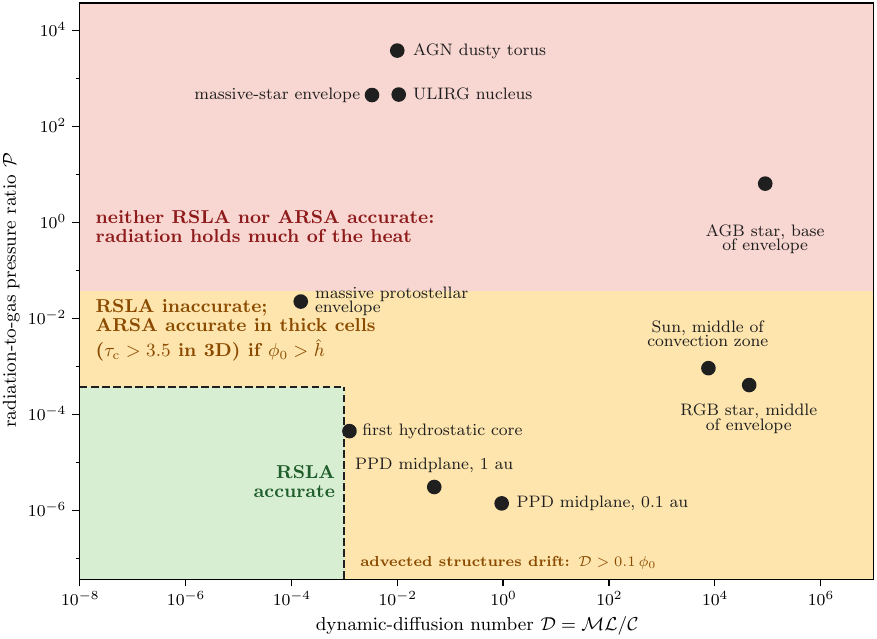}}
    \caption{The parameter space of radiation to gas pressure ratio $\calP$ versus dynamic diffusion number $\calD$, for optically thick systems. $\calP$ is the pressure ratio for radiation in thermodynamic equilibrium with the matter, $E = a_r T^4$. Circles indicate typical astrophysical system parameters (see \autoref{tab:media}). Lines and shading delineate different parameter regimes, computed for a speed of light reduction factor $\phi_0 = 10^{-2}$. The regions are set in $\hat{h} = 4(\gamma - 1)\calP$, which we convert to $\calP$ with $\gamma = 5/3$: $\hat{h} < 0.1\,\phi_0$ and $\calD < 0.1\,\phi_0$ (green); $0.1\,\phi_0 < \hat{h} < 0.1$ at any $\calD$, or $\hat{h} < 0.1\,\phi_0$ and $\calD > 0.1\,\phi_0$ (amber); and $\hat{h} > 0.1$ at any $\calD$ (red). The standard RSLA is accurate in the green region. In the amber region the RSLA either slows the thermal diffusion ($\hat{h} > 0.1\,\phi_0$, \autoref{eq:slowdown}) or lets the structures that the flow carries drift behind it ($\calD > 0.1\,\phi_0$), and the ARSA removes the error in cells of optical depth $\tau_{\rm c} \gtrsim 2\sqrt{D}$ in $D$ dimensions ($3.46$ in three dimensions), provided that $\phi_0 > \hat{h}$. Neither method is accurate in the red region.
    \label{fig:accuracy}
    }
\end{figure}

\autoref{tab:media} lists representative parameters of selected optically thick astrophysical systems, sorted by the heat-capacity ratio $\hat{h}$ and good to a factor of 2--3 (for the stars, $\calL$ and $v$ to an order of magnitude). The columns are, from left to right, the system (PPD = protoplanetary disk, AGN = active galactic nucleus, ULIRG = ultraluminous infrared galaxy, RGB = red giant branch, AGB = asymptotic giant branch), its characteristic density $\rho$, temperature $T$, specific opacity $\kappa$, velocity $v$, and optical depth across the system $\calL$; the two numbers that set the smallest reduction factor $\phi$ that the RSLA can safely use by \autoref{eq:rsla_valid}, which are the dynamic diffusion number $\calD = \calM\calL/\calC = \calL v/c$ and the heat-capacity ratio $\hat{h} = 4a_rT^3/(\rho c_v)$, which is $\calP$ up to a factor of order unity; and the gain $\phi/\phi_0$ of the ARSA over the standard RSLA, which is defined and discussed below. We adopt $c_v = 2\times10^8$~erg~g$^{-1}$~K$^{-1}$ for ionized hydrogen and helium, and $8\times10^7$ and $5\times10^7$ for warm and cold molecular gas. The stellar rows are halfway in radius between the base of the convective envelope and the surface, in the present Sun \citep{ChristensenDalsgaard96a} and in a $1\,M_\odot$ star on the lower RGB, and at the base of the envelope of an AGB star with hot bottom burning \citep{Karakas14b, Doherty17a}; their $\calL$ is across a pressure scale height. \autoref{fig:accuracy} plots these systems in the plane of $\calD$ versus $\calP$, and indicates where the standard RSLA is acceptable for the case of a reduction factor $\phi = 10^{-2}$, corresponding to $\hat{c} = 3000$ km s$^{-1}$. We omit optically thin systems, where the RSLA has other problems \citep{Gnedin25a}.

\subsection{Linear Wave Analysis}
\label{ssec:linear_waves}

We have now seen that the RSLA fails where $\mathcal{D}$ or $\mathcal{P}$ are too large, but we can gain some insight into how to fix this problem by analyzing the wave structure of the equations in these regimes. Such linear analyses have been conducted before by a number of authors \citep[e.g.,][]{Mihalas83a, Bogdan96a, Lowrie99a, Mihalas99a, Skinner13a}, so we discuss the calculation only briefly. For simplicity we consider a one-dimensional problem and orient our coordinate system so that $\hvecv$ and $\hvecF$ are purely in the $x$ direction, we take the gas to be ideal with adiabatic index $\gamma$, we again work in the gray regime $\kappa_{\rm E} = \kappa_{\rm P} = \kappa_{\rm F} = \mathrm{constant}$, we choose our non-dimensionalization so that $\hrho = \hT =  \hat{\kappa} = 1$ for the background state, and adopt the P1 approximation whereby $\htenP = \hE \tenI / 3$; this final approximation causes the wave speed to be $c/\sqrt{3}$ rather than $c$ in the optically thin limit, but otherwise does not affect the dispersion relations for RHD waves. With these assumptions, the RHD system reduces to
\begin{equation}
\frac{\partial}{\partial\hatt} \hvecU + \frac{\partial}{\partial\hx} \hvecFU = \hvecS_{\hvecU},
\label{eq:rhd_reduced}
\end{equation}
where $\hvecU = (\hrho, \hrho\hv, \hrho\he, \hE, \hF)$ is the vector of conserved quantities, and
\begin{equation}
\hvecFU =
\left(
\begin{array}{c}
\hrho\hv \\
\hrho\left(\hv^2 + \hT\right) \\
\left(\frac{1}{2}\hv^2+\frac{\gamma}{\gamma-1}\hT\right) \hrho\hv \\
\phi \calC \hF \\
\phi \calC \hE/3
\end{array}
\right)
\quad
\hvecS_{\hvecU} = \calL\left(
\begin{array}{c}
0 \\
\calP\left[\hF - \left(\hT^4 + \hE/3\right)\hv/\calC \right] \\
\calP \left[\calC\left(\hE-\hT^4\right) - \hv\hF \right] \\
-\phi \left[\calC\left(\hE-\hT^4\right) - \hv\hF \right] \\
-\phi \left[\calC \hF - (\hT^4+\hE/3)\hv\right]
\end{array}
\right)
\label{eq:flux_and_source}
\end{equation}
are the advection and source terms, respectively.

The background state is\footnote{The equilibrium state we give here differs at order $1/\calC^2$ from that given by \citet{Lowrie99a}, because in their analysis they selectively discard some terms of order $1/\calC^2$ while retaining others. Since \citet{Krumholz07b} show that terms of this order are necessary in the dynamic diffusion regime, we keep every term of \autoref{eq:G0_int_nondim} and \autoref{eq:G_int_nondim}, including the work terms, and solve for the equilibrium exactly. The second-order terms that we omit (\autoref{ssec:nondim}) would change $\hE_{\rm bg}$ only at order $\calC^{-4}$ and $\hF_{\rm bg}$ at order $\calC^{-3}$, so the state is complete to order $\calC^{-2}$; they would change the entries of the matrices below by relative amounts of order $\calM^2/\calC^2$, at most $4\times10^{-7}$ for the parameters we explore below.}
\begin{equation}
    \hvecU_{\mathrm{bg}} = \left(1, \calM, \frac{1}{\gamma-1} + \frac{\calM^2}{2}, \frac{3\calC^2 + 3\calM^2}{3\calC^2-\calM^2}, \frac{4\calC\calM}{3\calC^2-\calM^2}\right),
\end{equation}
or equivalently in terms of the primitive variables
\begin{equation}
    \hvecW_\mathrm{bg} = \left(\hrho, \hv, \hT, \hE, \hF\right) = \left(1, \calM, 1, \frac{3\calC^2 + 3\calM^2}{3\calC^2-\calM^2}, \frac{4\calC\calM}{3\calC^2-\calM^2}\right).
\end{equation}
Perturbing this background state with a small perturbation $\delta\hvecW$ and performing the usual linearization yields
\begin{equation}
    \frac{\partial}{\partial\hatt}\delta\hvecW + \tenA\,\frac{\partial}{\partial\hx}\delta\hvecW = \tenM\,\delta\hvecW,
    \label{eq:linear}
\end{equation}
with
\begin{eqnarray}
    \tenJ & \equiv & \frac{\partial\hvecU}{\partial\hvecW} = \left(\begin{array}{ccccc}
    1 & 0 & 0 & 0 & 0 \\
    \hv & \hrho & 0 & 0 & 0 \\
    \frac{\hT}{\gamma_1} + \frac{\hv^2}{2} & \hrho\hv & \frac{\hrho}{\gamma_1} & 0 & 0 \\
    0 & 0 & 0 & 1 & 0 \\
    0 & 0 & 0 & 0 & 1
    \end{array}\right)
    \\
    \tenA & \equiv & \tenJ^{-1}\frac{\partial\hvecFU}{\partial\hvecW} = \left(\begin{array}{ccccc}
    \calM & 1 & 0 & 0 & 0 \\
    1 & \calM & 1 & 0 & 0 \\
    0 & \gamma_1 & \calM & 0 & 0 \\
    0 & 0 & 0 & 0 & \phi\calC \\
    0 & 0 & 0 & \frac{\phi\calC}{3} & 0
    \end{array}\right)
    \\
    \tenM & \equiv & \tenJ^{-1}\frac{\partial\hvecS_{\hvecU}}{\partial\hvecW} =
    \calL \left(\begin{array}{ccccc}
    0 & 0 & 0 & 0 & 0 \\
    0 & -\frac{4\calC \calP}{3\calC^2-\calM^2} & -\frac{4\calP\calM}{\calC} & -\frac{\calP\calM}{3\calC} & \calP \\
    0 & 0 & -\frac{4\left(\calC^2-\calM^2\right)\calP\gamma_1}{\calC} & \frac{\left(3\calC^2+\calM^2\right)\calP\gamma_1}{3\calC} & -2\calP\calM\gamma_1 \\
    0 & \frac{4\phi \calC\calM}{3\calC^2-\calM^2} & 4\phi\calC & -\phi\calC & \phi\calM \\
    0 & \frac{4\phi\calC^2}{3\calC^2-\calM^2} & 4\phi\calM & \frac{\phi\calM}{3} & -\phi\calC
    \end{array}\right),
    \label{eq:M_matrix}
\end{eqnarray}
and $\gamma_1 \equiv \gamma-1$.

For a Fourier mode perturbation $\delta\hvecW \propto e^{i(\hk\hx-\homega\hatt)}$, this gives the eigenvalue problem
\begin{equation}
    -i\homega\,\delta\hvecW + i\hk\,\tenA\,\delta\hvecW = \tenM\,\delta\hvecW,
    \label{eq:perturb}
\end{equation}
which, divided by $i\hk$, is an eigenvalue problem for the complex phase speed $\hOmega \equiv \homega/\hk$, whose real part $\mathrm{Re}(\hOmega)$ is the phase speed and whose negative imaginary part $-\mathrm{Im}(\hOmega)$ the decay rate per unit wavenumber. We solve this eigenvalue problem numerically, and the solution depends only on the dimensionless coefficients $\calC$, $\calM$, and $\calP$, the adiabatic index $\gamma$, the speed of light reduction factor $\phi$, and the optical depth per radian of phase $\tau = \calL/\hk$. Following \citet{Lowrie99a}, the eigenvalues can be classified into three families based on their behavior as $\tau\to 0$ and radiation and matter decouple: there are two radiation modes whose propagation speeds approach $\calC/\sqrt{3}$ (or $\phi \calC/\sqrt{3}$ in the RSLA), two acoustic modes whose propagation speeds approach $\calM \pm \sqrt{\gamma}$, and one entropy mode whose propagation speed approaches $\calM$.

\begin{figure}[t]
    \centerline{\includegraphics{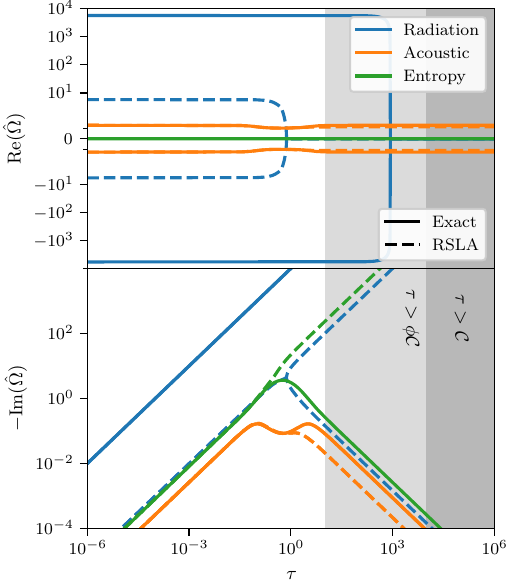}}
    \caption{Dimensionless complex phase speeds $\hOmega = \homega/\hk$ of linear waves for $\calC = 10^4$, $\calM = 0$, $\calP = 5\times10^{-4}$, $\gamma=5/3$ and $\phi = 10^{-3}$. Top: phase speeds, $\mathrm{Re}(\hOmega)$; bottom: decay rates, $-\mathrm{Im}(\hOmega)$. Colors: type of mode; solid: unmodified equations; dashed: RSLA. Light gray band: $\tau > \phi\calC$, where $\calD/\phi \gtrsim 1$ and the RSLA fails; dark gray band: dynamic diffusion, $\tau > \calC$, where $\calD \gtrsim 1$.
    \label{fig:waves_RSLA}
    }
\end{figure}

We plot $\hat{\Omega}$ as a function of $\tau$ for both the true RHD equations ($\phi = 1$) and the RSLA for the example set of parameters $\calC = 10^4$, $\calM = 0$, $\calP = 5\times10^{-4}$, $\gamma=5/3$, and $\phi = 10^{-3}$ in \autoref{fig:waves_RSLA}. We choose these parameters to illustrate the regime of interest to us: as we vary $\tau$ we move from the transparent regime, $\tau \ll 1$, to the static diffusion regime ($1 \lesssim \tau \lesssim \calC$, $\calD \lesssim 1$) to the dynamic diffusion regime ($\tau \gtrsim \calC$, $\calD \gtrsim 1$), and the chosen values of $\gamma$, $\calP$, and $\phi$ give $\hat{h} = 4(\gamma - 1)\calP = 1.3\times10^{-3}$ (\autoref{eq:h_ratio}), so that the gas holds most of the heat capacity, but $\hat{h}/\phi = 1.3$, so that under the RSLA the radiation holds more than half of the apparent heat capacity. The qualitative results we discuss below are the same for any parameter values with $\hat{h} \ll 1$ and $\hat{h}/\phi$ not small.

As expected based on our analysis from \autoref{ssec:rsla}, the RSLA behaves well at low $\tau$ and thus low $\calD$, leaving the acoustic and entropy modes unchanged from their true values and only modifying the radiation modes. This is precisely the goal of the RSLA, where we are interested in obtaining the correct behavior for the matter modes, and are willing to accept errors in the radiation modes as long as the matter modes are correct. However, as soon as we reach a regime where $\calD \ll 1$ but $\calD/\phi \gtrsim 1$, indicated by the light gray band in the figure, problems appear. The true solution for the acoustic modes (orange) shows a phase speed that switches from 1 at intermediate $\tau$ to $\sqrt{5/3}$ at high $\tau$, and a decay rate that reaches a maximum where this change in the phase speed occurs; both changes are associated with moving from the static diffusion regime where the radiation diffusion time across a mode is smaller than the oscillation period to the dynamic diffusion regime where it is larger. The RSLA misses this effect because its slowed diffusion suppresses the escape of radiation, and its enhanced radiation heat capacity artificially changes the compressibility of the gas -- indeed the phase speed that the RSLA recovers at high $\tau$ is 1.13, which is exactly the adiabatic sound speed, $\gamma_{\rm eff}^{1/2}$ with $\gamma_{\rm eff} = 1 + (\gamma - 1)/(1 + \hat{h}/\phi) = 9/7$, of a mixed gas-radiation fluid for which the ratio of radiation to gas heat capacity is $\hat{h}/\phi = 1.3$. A similar effect is visible for the entropy mode (which exchanges with the radiation mode for the RSLA), which decays too slowly in the RSLA due to the extra heat capacity of radiation.

However, the figure also points towards a possible resolution: the regime in which the RSLA fails is one in which the phase speed of the radiation modes is dropping sharply to zero. That is, in the high $\calD$ regime where the RSLA fails, it should not be necessary to suppress the propagation speed of the radiation modes, because the true phase speed -- once we account for the source terms on the RHS of \autoref{eq:rhd_reduced} as well as the propagation terms on the LHS -- is already $\ll \calC$. Conversely, in the regime where the phase speed is large and a reduced speed of light is required to simulate with a reasonable time step, the RSLA causes no errors. Thus we can obtain better performance by replacing the fixed speed of light reduction factor $\phi$ with a local quantity $\phi(\vecx,t)$ that is small only where it needs to be. This is the key insight that motivates our approach.

\section{An Adaptive Reduced Speed of Light Approximation}
\label{sec:irsla}

We now seek to design a numerical scheme in which, as in the standard RSLA, the time step is set by the CFL condition computed using a nominal reduction factor $\phi_0$, with $\calC^{-1} \ll \phi_0 \ll 1$, but the reduction factor $\phi$ that is actually used in solving the equations is set to a value in the range $[\phi_0,1]$, chosen cell-by-cell to be as large as possible while keeping the scheme numerically stable. We call this scheme the adaptive RSLA (ARSA), and our goal in this section is to derive this stability limit (\autoref{ssec:von_neumann}), use it to choose the adaptive reduction factor $\phi$ (\autoref{ssec:arsa}), and understand the qualitative behavior of this function (\autoref{ssec:irsla_phi}).
For this purpose we rewrite the RSLA radiation equations (\autoref{eq:rad_energy_rsl} and \autoref{eq:rad_momentum_rsl}) as
\begin{equation}
    \frac{1}{\phi}\frac{\partial \hvecU_{\rm r}}{\partial \hatt} = -\calC\,\hnabla\cdot\hvecF_{\rm r}\left(\hvecU_{\rm r}\right) + \hvecS_{\rm r},
    \label{eq:rsla_time}
\end{equation}
where $\hvecU_{\rm r} = (\hE, \hvecF)$ is the vector of radiation variables, $\hvecF_{\rm r}$ their flux, and $\hvecS_{\rm r}$ the exchange terms of the unmodified equations.
This choice does have a price: the quantity that \autoref{eq:rsla_time} conserves exactly, for a reduction factor that is constant in time, is $\rho e + E/\phi$ (or, for a scheme with multiple radiation groups, $\rho e + \sum_g E_g/\phi_g$ where the sum is over frequency groups $g$, each with its own reduction factor).

\subsection{Von Neumann Stability Analysis}
\label{ssec:von_neumann}

We now examine the class of update schemes used by modern two-moment codes: an explicit, upwind transport update of the radiation moments combined with an implicit update of the matter-radiation exchange terms. We specifically analyze the second-order implicit-explicit (IMEX) scheme PD-ARS \citep{Chu19a}, the scheme that QUOKKA uses for its radiation subsystem \citep{He24a, He24b}; an extension to other time integration schemes is provided in \autoref{sssec:other_schemes}. Writing $\mathcal{T}$ for the explicit transport operator and $\mathcal{S}$ for the exchange operator, one step of size $\Delta t$ with the PD-ARS scheme is
\begin{eqnarray}
    \vecU^{(1)} & = & \vecU^n + \Delta t\, \mathcal{T}(\vecU^n) + \Delta t\, \mathcal{S}(\vecU^{(1)}), \nonumber \\
    \vecU^{n+1} & = & \tfrac{1}{2}\vecU^n + \tfrac{1}{2}\left[\vecU^{(1)} + \Delta t\,\mathcal{T}(\vecU^{(1)})\right] + \tfrac{1}{2}\Delta t\,\mathcal{S}(\vecU^{n+1}),
    \label{eq:pdars}
\end{eqnarray}
so that each explicit stage is immediately followed by an implicit solve. Our goal is to determine for this scheme the largest value of $\phi$ that will be stable if we choose $\Delta t$ such that the CFL number computed using the reduced speed of light,
\begin{equation}
    \nu_0 \equiv \phi_0 c\frac{\Delta t}{\Delta x} \equiv \phi_0 \nu_c
\end{equation}
is of order unity, and therefore the CFL number computed using the \textit{true} speed of light $\nu_c = c\Delta t/\Delta x$ is much larger than unity.

\subsubsection{The amplification matrix}
\label{sssec:ampl}

We linearize \autoref{eq:pdars} about a uniform state, as in \autoref{ssec:linear_waves}, with coefficients and opacities frozen at their background values, and work with the primitive perturbations: since $\delta\hvecU = \tenJ\,\delta\hvecW$ with $\tenJ$ constant, we multiply \autoref{eq:pdars} by $\tenJ^{-1}$, as in \autoref{eq:linear}. We consider a uniform, one-dimensional grid in the $x$ direction, let $\delta\hvecW_j^n$ be the perturbation about the background in cell $j$ at time step $n$, and express the cell perturbations as a Fourier series, $\delta\hvecW^n_j = \sum_\theta \delta\hvecW^n_\theta e^{ij\theta}$, where $\theta = \hk\,\Delta\hx \in (0, \pi]$ is the phase change per cell and $\delta\hvecW^n_\theta$ is a five-element vector giving the amplitude of the Fourier mode for each primitive variable at time step $n$. Since we are working with the linearized equations each Fourier mode evolves independently, and since the $\mathcal{T}$ and $\mathcal{S}$ operators are linear with constant coefficients in the linearized system, we can represent each as a $5\times 5$ matrix that multiplies $\delta\hvecW_\theta$; we will derive the full expressions for these matrices below. Rewriting the first stage of \autoref{eq:pdars} in this way, it reads
\begin{equation}
    \delta\hvecW_\theta^{(1)} = \mathsf{I} \, \delta\hvecW_\theta^n + \tenZ_\theta \, \delta \hvecW_\theta^n + \tenS \, \delta\hvecW_\theta^{(1)},
\end{equation}
where $\tenI$ is the identity matrix, $\tenZ_\theta$ is the matrix for the transport step, and $\tenS$ is the matrix for the exchange step. Note that $\tenZ_\theta$ depends on the wavenumber of the Fourier mode $\theta$, since it couples different cells, but the exchange operator $\mathsf{S}$ does not because it is local to each cell. Both $\mathsf{Z}_\theta$ and $\mathsf{S}$ are linearly proportional to the time step $\Delta \hatt$.

Rearranging our expression for the first stage, we obtain
\begin{equation}
    \delta\hvecW_\theta^{(1)} = \left(\tenI - \tenS\right)^{-1}\left(\tenI + \tenZ_\theta\right)\delta\hvecW_\theta^n,
    \label{eq:stage1}
\end{equation}
and doing the same operations for the second stage gives
\begin{equation}
    \left(\tenI - \tfrac{1}{2}\tenS\right)\delta\hvecW_\theta^{n+1} = \tfrac{1}{2}\,\delta\hvecW_\theta^n + \tfrac{1}{2}\left(\tenI + \tenZ_\theta\right)\delta\hvecW_\theta^{(1)} .
    \label{eq:stage2}
\end{equation}
Combining the two stages, we can write the full update as a single matrix operation
\begin{equation}
    \delta\hvecW_\theta^{n+1} = \mathsf{G}_\theta \,\delta\hvecW_\theta^n, \qquad
    \tenG_\theta = \left(\tenI - \tfrac{1}{2}\tenS\right)^{-1}\left[\tfrac{1}{2}\tenI + \tfrac{1}{2}\left(\tenI + \tenZ_\theta\right)\left(\tenI - \tenS\right)^{-1}\left(\tenI + \tenZ_\theta\right)\right].
    \label{eq:ampl}
\end{equation}
The matrix $\tenG_\theta$ describes the amplification that a particular Fourier mode undergoes per time step. The scheme is stable only if the magnitude of the largest eigenvalue of $\tenG_\theta$ is $\leq 1$ for all $\theta \in (0,\pi]$. (Strictly, two coinciding eigenvalues of unit magnitude without independent eigenvectors would still let the mode grow linearly in time; \ref{app:theta} shows that this does not happen here.) The problem of analyzing the stability of the system therefore reduces to the problem of writing down the matrices $\tenS$ and $\tenZ_\theta$, using them to compute $\tenG_\theta$, and setting limits on its largest eigenvalue.

The exchange matrix $\tenS$ is straightforward to write down, since the calculation is completely analogous to the linearization carried out in \autoref{ssec:linear_waves}. For the case of equal mean opacities, the P1 closure, and an equilibrium background state, the result is $\tenS = \Delta\hatt \, \tenM$, where $\tenM$ is the exchange matrix derived from our linear wave analysis (\autoref{eq:M_matrix}), and with $\calL$ replaced by $\tau_{\rm c}/\Delta\hx$, where $\tau_{\rm c} = \rho\kappa\Delta x$ is the optical depth of a cell. We therefore have
\begin{equation}
    \tenS =
    \nu \tau_c
    \left(\begin{array}{ccccc}
    0 & 0 & 0 & 0 & 0 \\
    0 & -\frac{4\calP}{3\calC^2-\calM^2}\phi^{-1} & -\frac{4\calP\calM}{\calC^2}\phi^{-1} & -\frac{\calP\calM}{3\calC^2}\phi^{-1} & \calP\phi^{-1} \\
    0 & 0 & -\frac{4\left(\calC^2-\calM^2\right)\calP\gamma_1}{\calC^2}\phi^{-1} & \frac{\left(3\calC^2+\calM^2\right)\calP\gamma_1}{3\calC^2}\phi^{-1} & -\frac{2\calP\calM\gamma_1}{\calC}\phi^{-1} \\
    0 & \frac{4\calM}{3\calC^2-\calM^2} & 4 & -1 & \frac{\calM}{\calC} \\
    0 & \frac{4\calC}{3\calC^2-\calM^2} & 4\frac{\calM}{\calC} & \frac{\calM}{3\calC} & -1
    \end{array}\right),
    \label{eq:S_matrix}
\end{equation}
where the dimensionless numbers $\calC$, $\calP$, and $\calM$ and the cell opacity $\tau_c$ are all evaluated for the unperturbed background state, and we have defined
\begin{equation}
    \nu = \frac{\phi}{\phi_0} \nu_0 = \phi \nu_c
\end{equation}
as the CFL number computed at the local reduced speed of light $\phi$ (as opposed to at the maximum reduction factor $\phi_0$). The analysis is straightforward to generalize to the case of non-gray opacities and an M1 rather than a P1 closure, but we refrain from writing out the expression in full since it is lengthy and there is not much insight to be gained by doing so; we handle the fully general case in \ref{app:m1}.

Since the transport step transports only radiation (hydrodynamic transport is operator-split in QUOKKA and similar codes), the transport matrix $\tenZ_\theta$ affects only the final two elements of $\delta\hvecW^n_\theta$, $\mathbf{w} \equiv (\delta\hE, \delta\hF_x)$. Its first three rows and first three columns are therefore all 0, and we need only compute its lower right $2\times 2$ matrix. The final two rows of the flux vector $\hvecFU$, for one-dimensional transport in the $x$ direction and a general closure, are $\phi \calC\hF_x$ and $\phi\calC\hP_{xx}$ (cf.~\autoref{eq:rad_energy_rsl} and \autoref{eq:rad_momentum_rsl}), where $\hP_{xx} = T_{xx} \hE$ is the $xx$ component of the radiation pressure tensor and $\mathsf{T}$ is the Eddington tensor. To evaluate $\tenZ_\theta$, we insert $\hvecW = \hvecW_{\rm bg} + \delta\hvecW$, where $\hvecW_{\rm bg}$ is the background state, and linearize by dropping terms above first order, which gives
\begin{equation}
    \left(\begin{array}{c} \phi\calC\hF_x \\ \phi\calC\hat{P}_{xx} \end{array}\right)
    = \left(\begin{array}{c} \phi\calC\hF_{x,\mathrm{bg}} \\ \phi\calC\hat{P}_{xx,\mathrm{bg}} \end{array}\right)
    + \tenA_{\rm r}\,\mathbf{w},
    \label{eq:flux_von_neumann}
\end{equation}
with
\begin{equation}
    \tenA_{\rm r} \equiv \phi\calC\,\tilde{\tenA},\qquad \tilde{\tenA} \equiv \frac{\partial(\hF_x, \hat{P}_{xx})}{\partial(\hE, \hF_x)}\bigg|_{\rm bg}.
    \label{eq:flux_split}
\end{equation}
The value of $\tilde{\tenA}$ depends on the closure: for the P1 closure, and for M1 where the background state is isotropic, we have $T_{xx} = 1/3$ and thus $\tilde{\tenA} = \bigl(\begin{smallmatrix} 0 & 1 \\ 1/3 & 0 \end{smallmatrix}\bigr)$, but for the general case of the M1 closure $\tilde{\tenA}$ depends on the transverse components of $\hvecF$ as well as the $x$ component. We focus on the isotropic case in the main text, for which we will be able to obtain a rigorous stability bound, and defer the anisotropic case to \ref{app:theta}, where we treat it numerically.

The first term in \autoref{eq:flux_von_neumann} is the flux of the background state, and the second is the flux due to the perturbation. Since the background flux is spatially uniform, it causes no change to any cell, so the change that appears in $\tenZ_\theta$ depends only on the second term, $\tenA_{\rm r}\mathbf{w}$. To find the contribution from this term, we use the HLL flux \citep{Toro09a} with wave speeds $\pm s$, $s \equiv \sigma\phi\calC$ (or equivalently in physical units $\pm\sigma\hat{c}$), where $\sigma = T_{xx}^{1/2} \leq 1$ is the signal speed as calculated in the frozen Eddington factor approximation of \citet{Balsara_1999}: $\sigma = 1/\sqrt{3}$ for an isotropic radiation field, rising to $\sigma = 1$ where the radiation free-streams along $x$ \citep{Levermore84a}. For simplicity we take the face states to be the cell averages, i.e., piecewise-constant reconstruction; all the tests of \autoref{sec:tests} use PPM instead, and \autoref{sssec:validity} discusses this. To leading order the signal speeds depend only on the uniform background state, and thus are the same on the two faces, so that the HLL flux reduces to the Rusanov flux:
\begin{equation}
    \mathbf{f}_{j+1/2} = \tfrac{1}{2}\tenA_{\rm r}\left(\mathbf{w}_j + \mathbf{w}_{j+1}\right) - \tfrac{1}{2}s\left(\mathbf{w}_{j+1} - \mathbf{w}_j\right),
    \label{eq:rusanov}
\end{equation}
on the $j+1/2$ face, and similarly for the $j-1/2$ face. The difference in the fluxes on the two faces is
\begin{equation}
    \mathbf{f}_{j+1/2} - \mathbf{f}_{j-1/2} = \tfrac{1}{2}\tenA_{\rm r}\left(\mathbf{w}_{j+1} - \mathbf{w}_{j-1}\right) - \tfrac{1}{2}s\left(\mathbf{w}_{j+1} - 2\mathbf{w}_j + \mathbf{w}_{j-1}\right),
    \label{eq:flux_diff}
\end{equation}
and one transport stage changes cell $j$ by $\Delta t\,\mathcal{T}(\mathbf{w})_j = -(\Delta\hatt/\Delta\hx)(\mathbf{f}_{j+1/2} - \mathbf{f}_{j-1/2})$. For the Fourier mode, $\mathbf{w}_{j\pm1} = e^{\pm i\theta}\mathbf{w}_j$, so that
\begin{equation}
    \mathbf{w}_{j+1} - \mathbf{w}_{j-1} = 2i\sin\theta\,\mathbf{w}_j,
    \qquad
    \mathbf{w}_{j+1} - 2\mathbf{w}_j + \mathbf{w}_{j-1} = -2\left(1 - \cos\theta\right)\mathbf{w}_j .
    \label{eq:fourier_diff}
\end{equation}
Inserting these into \autoref{eq:flux_diff} gives $\Delta t\,\mathcal{T}(\mathbf{w})_j = \tenZ_\theta\,\mathbf{w}_j$, and we then arrive at our final result for the lower right $2\times 2$ corner of the transport matrix,
\begin{equation}
    (\tenZ_\theta)_{4,5}= -\nu\left[i\sin\theta\,\tilde{\tenA} + \sigma\left(1-\cos\theta\right)\tenI\right]
    \label{eq:hll_symbol}
\end{equation}
where the subscript (4,5) indicates that we are giving the $2\times 2$ elements in rows and columns 4 and 5 (with the other elements of $\tenZ_\theta$ all zero).
The imaginary term of \autoref{eq:hll_symbol}, from the average flux, shifts the phase of the mode (transport); the real term, from the jump, damps it (numerical diffusion).

\subsubsection{The checkerboard perturbation}
\label{sssec:checkerboard}

At this point we have written down exact expressions for $\tenS$ and $\tenZ_\theta$, and therefore for the amplification matrix $\tenG_\theta$, and shown that it depends only on the choice of closure scheme, the dimensionless numbers for the system $\calC$, $\calM$, and $\calP$, the cell optical depth $\tau_c$ (or for non-gray opacities the three optical depths $\tau_P$, $\tau_E$, and $\tau_F$), the gas adiabatic index $\gamma$, and the CFL at the local reduced speed of light $\nu = (\phi/\phi_0) \nu_0$. It is therefore straightforward to use numerical methods to find the largest stable CFL number $\nu$, and thus the largest stable reduction factor $\phi$, for any combination of parameters of interest: we simply compute $\tenG_\theta$ for these parameters, find the eigenvalues at each $\theta \in (0,\pi]$, and adjust $\nu$ until the magnitude of the largest eigenvalue at any $\theta$ drops below unity. We will indeed resort to this strategy to handle the most general case of an arbitrary anisotropic radiation field on a moving background. However, we can derive an analytic expression for $\nu$ in the special case $\theta = \pi$, the checkerboard perturbation in which neighboring cells have phase factors of opposite sign, on an equilibrium background at rest, $\calM = 0$. We focus on this case for both physical and practical reasons.

The physical reason to focus on the checkerboard is that, intuitively, it is the mode we expect to be the least stable. This is because the ultimate source of instability in our scheme, as with any other upwind scheme, is the numerical diffusion term $\sigma(1-\cos\theta)$ in $\tenZ_\theta$ (\autoref{eq:hll_symbol}). This term is largest, $2\sigma$, at $\theta = \pi$, and since the other term in $\tenZ_\theta$ vanishes at $\theta=\pi$, the eigenvalue with the largest magnitude is $-2\nu\sigma$. When $\nu>1/2\sigma$, we have the usual instability found in an explicit scheme for the diffusion equation with a time step that is too large: a grid-scale mode whose sign inverts and whose amplitude grows at every time step. The exchange term $\tenS$ can damp this growth -- indeed, that is the effect upon which the ARSA rests -- but its effect is independent of $\theta$, and thus we expect that the $\theta = \pi$ mode for which the growth is strongest will be the first to become unstable as we increase $\nu$. We formally prove that this is indeed the case for $\calM = 0$ and an isotropic radiation field in \ref{app:proof}, and our numerical tests in \ref{app:numerics} show that, while this intuition does not strictly hold for $\calM \neq 0$ or for an anisotropic radiation state, the violation is by at most a few percent over a broad parameter range.

The fact that the checkerboard with $\calM = 0$ is, if not always the most unstable mode, at least within a few percent of it, supplies the first practical reason to focus on it: this result means that we can in practice implement our scheme by using the analytic stability limit we derive below for the checkerboard case and then applying a small additional safety factor. The other practical reason to focus on this case is that it is by far the easiest to analyze, which is why we can derive an exact result for it in the first place. The property that makes this case easier is that it reduces $\tenG_\theta$ from a $4\times 4$ matrix to a pair of independent $2\times 2$ matrices. To see this, first note that for $\calM = 0$, the exchange matrix $\tenS$ (\autoref{eq:S_matrix}) reduces to
\begin{equation}
\tenS\big|_{\calM = 0} = \nu
    \left(\begin{array}{ccccc}
    0 & 0 & 0 & 0 & 0 \\
    0 & -\phi^{-1}\frac{\calP}{\calC^2}\frac{\kappa_P}{\kappa_E} (1+T_{xx})\tau_F & 0 & 0 & \phi^{-1}\frac{\calP}{\calC}\tau_F \\
    0 & 0 & -4\phi^{-1}\calP\gamma_1\tau_P & \phi^{-1}\calP\gamma_1\tau_E & 0 \\
    0 & 0 & 4\tau_P & -\tau_E & 0 \\
    0 & \frac{1}{\calC}\frac{\kappa_P}{\kappa_E}(1+T_{xx})\tau_F & 0 & 0 & -\tau_F
    \end{array}\right),
    \label{eq:S_static}
\end{equation}
so that $\delta\hv_x$ and $\delta\hF_x$ (the second and fifth rows) are coupled and $\delta\hT$ and $\delta\hE$ are coupled (the third and fourth rows), but there is no coupling across these pairs. Similarly, the only off-diagonal terms in the transport matrix $\tenZ_\theta$ (\autoref{eq:hll_symbol}) appear in the term $i\sin\theta \,\tilde{\tenA}$, which vanishes when $\theta = \pi$, leaving $\tenZ_\pi$ a pure diagonal matrix. In this case the transport stage of the update, which multiplies $\delta\hvecW$ by $\tenI + \tenZ_\theta$, reduces to simply multiplying $\delta\hE$ and $\delta\hF_x$ by a factor $1-2\nu\sigma$. Thus $\tenG_\pi$ consists of one two-by-two block that couples $\delta\hT$ and $\delta\hE$ -- to which we refer as the energy block -- and one two-by-two block that couples $\delta\hv_x$ and $\delta\hF_x$ -- to which we refer as the momentum block.

\subsubsection{Stability boundaries for the checkerboard}
\label{sssec:stabilty_checkerboard}

Having decomposed $\tenG_\theta$ to a pair of $2\times 2$ blocks, we are now in a position to calculate its stability boundaries analytically.

\paragraph{The energy block} For the energy block we have the source matrix,
\begin{equation}
    \tenS^{(E)} = \nu\left(\begin{array}{cc} -\hat{h}\tau_P/\phi & \hat{h}\tau_E/(4\phi) \\ 4\tau_P & -\tau_E \end{array}\right).
    \label{eq:SE_nongray}
\end{equation}
acting on $(\delta\hT, \delta\hE)$, where we have used the relation $\hat{h} = 4\gamma_1 \calP$ (\autoref{eq:h_ratio}); the $1/\phi$ difference between the two rows is the heat capacity mismatch of the RSLA. If we rescale the radiation perturbation from $\delta\hE$ to $\delta\hE' = (\kappa_E/\kappa_P)\delta\hE$, which does not change the eigenvalues, we can rewrite this in the more compact form
\begin{equation}
    \tenS^{(E)} = q\left(\begin{array}{cc} -\lambda & \lambda/4 \\ 4 & -1 \end{array}\right),
    \qquad
    q \equiv \nu\,\tau_E,
    \qquad
    \lambda \equiv \frac{\hat{h}_{\rm eff}}{\phi},
    \qquad
    \hat{h}_{\rm eff} \equiv \hat{h}\,\frac{\kappa_P}{\kappa_E} .
    \label{eq:SE}
\end{equation}
Here $q$ is the optical depth that the reduced light crosses in one step, and $\hat{h}_{\rm eff}$ is the effective heat-capacity ratio of a non-gray material, reducing to $\hat{h}_{\rm eff} = \hat{h}$ for gray material. The corresponding matrix for the transport stage for $\theta = \pi$ is
\begin{equation}
    \tenD \equiv \left(\tenI + \tenZ_\pi\right)\big|_{(\delta\hT, \delta\hE)} = \left(\begin{array}{cc} 1 & 0 \\ 0 & 1 - a \end{array}\right).
    \label{eq:D_energy}
\end{equation}
where $a \equiv 2\sigma\nu$. Inserting these two into the definition of $\tenG_\theta$ (\autoref{eq:ampl}), we arrive at our final expression for the amplification matrix of the energy block for our special case of non-moving background, isotropic radiation, and the checkerboard mode:
\begin{equation}
    \tenG_E = \left(\tenI - \tfrac{1}{2}\tenS^{(E)}\right)^{-1}\left[\tfrac{1}{2}\tenI + \tfrac{1}{2}\,\tenD\,\tenB^{-1}\tenD\right],
    \qquad
    \tenB = \left(\begin{array}{cc} 1 + \lambda q & -\lambda q/4 \\ -4q & 1 + q \end{array}\right).
    \label{eq:G_energy}
\end{equation}
Since this is a $2\times 2$ matrix its eigenvalues are straightforward to find, and in \ref{app:proofs} we show that all eigenvalues of $\tenG_E$ have magnitude $\leq 1$ if and only if $a \leq a_* \equiv 2(2 + q + \lambda q)/(2 + \lambda q)$. At $a=a_*$ exactly, one of the eigenvalues is exactly unity, and this case corresponds to one where the growth of the unstable mode during the transport stage $\tenI + \tenZ_\pi$ is exactly damped by the exchange $\tenS$, so that a checkerboard perturbation is neutrally stable.

To obtain the corresponding limit on $\phi$, note that $\lambda q = (\hat{h}_{\rm eff}/\phi)\,\nu\,\tau_E = \hat{h}_{\rm eff}\,\nu_c\,\tau_E$ does not depend on $\phi$: the factor $1/\phi$ in $\lambda$ cancels the factor $\phi$ in $\nu$. Writing $K \equiv \lambda q$ and inserting $a = 2\nu\sigma$ and $q = \nu\tau_E$, the condition $a \leq a_*$ becomes linear in $\nu$, $\nu[\sigma(2 + K) - \tau_E] \leq 2 + K$, and dividing by $2 + K$ gives the stability condition
\begin{equation}
    \nu\left[\sigma - \frac{\tau_E}{2 + K}\right] \leq 1,
    \qquad
    K \equiv \lambda q = \hat{h}\,\nu_c\,\tau_P = \hat{h}\,\rho\kappa_P\, c\,\Delta t .
    \label{eq:numax}
\end{equation}
Where the term in square brackets is positive, this condition is $\nu \leq \nu_{\max} \equiv [\sigma - \tau_E/(2 + K)]^{-1}$; where it is not, there is no bound at all and the checkerboard is stable even for $\phi = 1$. The dimensionless number $K$ that appears here is the time step divided by the time in which the gas exchanges its thermal energy with the radiation at the true speed of light.

\paragraph{The momentum block} The momentum block $(\delta\hv, \delta\hF)$ is treated in the same way. The $2\times 2$ block in the exchange matrix for this pair is
\begin{equation}
    \tenS^{(F)} = q_F\left(\begin{array}{cc} -\lambda_F & \lambda_F/r \\ r & -1 \end{array}\right),
    \qquad
    q_F \equiv \nu\,\tau_F,
    \qquad
    \lambda_F \equiv \frac{(1 + T_{xx})E}{\phi\,\rho c^2},
    \qquad
    r \equiv \frac{(1 + T_{xx})\hE}{\calC},
    \label{eq:SF}
\end{equation}
with $E$ and $\hE$ the radiation energy density of the background in physical and dimensionless units, respectively; for the equilibrium state this is $\hE=\kappa_P/\kappa_E$. This form is identical to that of $\tenS^{(E)}$ (\autoref{eq:SE}), just with $q$ replaced by $q_F$, $\lambda$ by $\lambda_F$ and the factor 4 by $r$. The factor $(1 + T_{xx})E/(\rho c^2)$ in $\lambda_F$ is the ratio of the inertia of the radiation, $(1 + T_{xx})E/c^2$, to that of the gas, $\rho$, and the extra factor $1/\phi$ is the same mismatch of the RSLA as in $\lambda$. Since the transport block $\tenI + \tenZ_\pi$ is also the same for the momentum block as for the energy block (i.e., a diagonal matrix with a value of 1 on the diagonal for the matter row, and $1-a$ for the radiation row), we can make exactly the same argument as for the energy block and arrive at the stability condition
\begin{equation}
    \nu\left[\sigma - \frac{\tau_F}{2 + K_F}\right] \leq 1,
    \qquad
    K_F \equiv \lambda_F\,q_F = \frac{(1 + T_{xx})E\,\nu_c\,\tau_F}{\rho c^2} = \frac{1 + T_{xx}}{4\gamma_1\calC^2}\,\frac{\kappa_F}{\kappa_E}\,K ,
    \label{eq:KF}
\end{equation}
Since $K_F/K \sim \calC^{-2} (\kappa_F/\kappa_E)$, the momentum stability condition is never relevant in gray material, since then $\tau_F = \tau_E$ and $K_F < K$; this implies that if a given CFL number $\nu$ obeys the condition \autoref{eq:numax}, it automatically obeys \autoref{eq:KF} as well. For non-gray material the momentum condition becomes relevant only when the flux mean opacity is much smaller than the energy mean.

Note that in the conditions we have derived the pressure ratio $\calP$ and adiabatic index appear only via $\hat{h}$. Thus $\hat{h}$ alone sets the limit on the value of $\phi$ allowed in the ARSA.

\subsubsection{Extension to multiple dimensions}
\label{sssec:md}

On a grid with $D$ active dimensions and cubic cells of width $\Delta x$, the analysis carries over with one change: the checkerboard is amplified by the transport in every direction at once. A Fourier mode now has a phase change $\theta_d$ per cell in each direction $d$, and the transport operator \autoref{eq:hll_symbol} becomes a sum of one such term per direction, each with the same CFL number $\nu$ and its own Jacobian $\tilde{\tenA}_d$ and signal speed $\sigma_d$. As in one dimension, the least stable mode is the checkerboard in every direction, $\theta_d = \pi$ for all $d$ (Proposition~\ref{prop:checkerboard}). There the transport stage is again $(1 - a)\,\tenI$ on the radiation variables, now with
\begin{equation}
    a = 2\nu\sum_{d=1}^{D}\sigma_d .
    \label{eq:a_md}
\end{equation}
The exchange in multiple dimensions changes only in that the Eddington tensor element $T_{xx}$ appearing in $K_F$ is different in different directions, so to be conservative we replace it with $\chi$, the largest diagonal element of $\mathsf{T}$, when computing $K_F$. Nothing else changes, and each component of $\vecF$ forms its own momentum block with the matching velocity component. The multi-dimensional bound is therefore the pair of conditions of \autoref{sssec:stabilty_checkerboard} with $\sigma$ replaced by $\sum_d\sigma_d$,
\begin{equation}
    \nu\left[\sum_{d=1}^{D}\sigma_d - \frac{\tau_E}{2 + K}\right] \leq 1,
    \qquad
    \nu\left[\sum_{d=1}^{D}\sigma_d - \frac{\tau_F}{2 + K_F}\right] \leq 1 .
    \label{eq:bound_md}
\end{equation}
For non-cubic grids the only change is that $\nu\sum_d\sigma_d$ is replaced by $\sum_d\nu_d\sigma_d$, where $\nu_d = \phi c\,\Delta t/\Delta x_d$; the damping terms are unchanged because $\nu\tau = \phi c\,\Delta t\,\rho\kappa$ does not depend on the cell width.

Note that in \autoref{eq:bound_md} the transport term grows with the number of dimensions, but the damping term does not, and for anisotropic radiation may slightly decrease since we must use the largest diagonal element of $\mathsf{T}$ in $K_F$. Thus with the same $\sigma$ in every direction the transport term is $D\sigma$, so a cell of a given optical depth $\tau_E$ provides $D$ times less damping relative to the transport in $D$ dimensions than in one.

\subsubsection{The stability limit for the standard RSLA}
\label{sssec:phi0}

An immediate corollary of our analysis, which to our knowledge has not appeared in the literature to date, is that we can derive the stability bound for the standard RSLA, $\phi = \phi_0$. In this case we have $\nu = \nu_0$, and for optically thin cells, $\tau_E \to 0$ and $\tau_F \to 0$, the maximum allowed value of $\nu_0$ is determined solely by the maximum possible value of $\sum_d \sigma_d$. The HLL flux uses in direction $d$ the signal speed $\sigma_d = T_{dd}^{1/2}$, with $T_{dd}$ the diagonal element of the Eddington tensor in that direction \citep{Balsara_1999}, which allows us to prove the following proposition:

\begin{prop}
\label{prop:rsla}
The standard RSLA, $\phi = \phi_0$, satisfies the checkerboard bound \autoref{eq:bound_md} for every radiation state and every (frozen) opacity and heat capacity, if $\nu_0 \leq 1/\sqrt{D}$. For the M1 closure this condition is also necessary.
\end{prop}

\begin{pf}
The Eddington tensor has unit trace, so $\sum_d T_{dd} \leq 1$ over the $D \leq 3$ active directions. By the Cauchy--Schwarz inequality, $\sum_d\sigma_d = \sum_d T_{dd}^{1/2} \leq (D\sum_d T_{dd})^{1/2} \leq \sqrt{D}$, with equality when $T_{dd} = 1/D$ in every active direction. At $\nu = \nu_0$ the damping terms of both conditions of \autoref{eq:bound_md} are non-negative, so both conditions hold if $\nu_0\sum_d\sigma_d \leq 1$, which is true for every radiation state if $\nu_0 \leq 1/\sqrt{D}$. Conversely, in an optically thin cell the damping terms vanish, and an M1 state with $T_{dd} = 1/D$, such as isotropic radiation in three dimensions, then requires $\nu_0 \leq 1/\sqrt{D}$. \qed
\end{pf}

Proposition~\ref{prop:rsla} holds for the M1 closure in any radiation state: its proof uses only the unit trace of the Eddington tensor and the transport factor of the checkerboard, which depends on the closure only through $\sigma_d$ (\autoref{sssec:checkerboard}). The bound itself also assumes that the checkerboard is the least stable mode, which \ref{app:theta} proves for the Eddington closure and for M1 in its isotropic limit.

The proof also shows why a large $\hat{h}$ does not matter here: the exchange never destabilizes the scheme. Its damping terms are non-negative for every $K$, so the bound is never stricter than the CFL condition of the transport alone. A large $\hat{h}$ only makes the exchange less helpful, by capping its damping once $K \gg 2$; it never makes it harmful. Hence the standard RSLA needs no time step restriction beyond $\nu_0 \leq 1/\sqrt{D}$, and the value $\phi = \phi_0$ is stable whenever the RSLA itself is.

\subsubsection{Stability analysis for alternative time integration schemes}
\label{sssec:other_schemes}

The analysis presented above for the PD-ARS scheme can be extended straightforwardly to other time integration schemes. We consider two here, again limiting ourselves to analyzing the checkerboard perturbation, $\theta=\pi$, for a static, equilibrium background, $\calM = 0$, so that the exchange stage again factorizes into $2\times2$ matrices.

The first alternative is a single forward Euler transport stage followed by an implicit exchange update. Its amplification matrix for the energy block is
\begin{equation}
    \tenG_{\rm FE} = \tenB^{-1}\tenD,
    \qquad
    \det\left(\tenI - \tenG_{\rm FE}\right) = \frac{\lambda q\,a}{\det\tenB},
    \qquad
    \det\left(\tenI + \tenG_{\rm FE}\right) = \frac{\left(2 + \lambda q\right)\left(a_* - a\right)}{\det\tenB},
    \label{eq:G_FE}
\end{equation}
where the determinants follow from $\det(\tenB - \tenD) = \lambda q a$ and $\det(\tenB + \tenD) = (2 + \lambda q)(a_* - a)$ (\ref{app:proofs}). Of the three conditions of the Schur--Cohn--Jury criterion (Lemma~\ref{lem:jury}), the first, $\det(\tenI - \tenG_{\rm FE}) \geq 0$, now always holds, and the third, $|\det\tenG_{\rm FE}| = |1 - a|/\det\tenB \leq 1$, holds for $a \leq a_*$ because $a_* \leq 2 + q + \lambda q = 1 + \det\tenB$. The second, $\det(\tenI + \tenG_{\rm FE}) \geq 0$, however, fails for $a > a_*$, so this scheme has exactly the same boundary as PD-ARS, \autoref{eq:numax}, although an eigenvalue leaves the unit circle through $g = -1$ instead of $g = +1$. This is no coincidence, since this scheme is essentially a single-stage version of PD-ARS, and in both schemes the boundary is set by the factor $\det(\tenB + \tenD)$.

The second alternative is a second-order Runge-Kutta update in which the two explicit transport stages are taken back to back and the exchange terms are updated only once, at the end. At $\theta = \pi$ the two transport stages multiply the radiation by $1 - a + a^2/2$, so that, with $\tenD_2 \equiv \mathrm{diag}(1, 1 - a + a^2/2)$ and the same steps as in \autoref{eq:G_FE},
\begin{equation}
    \tenG_{\rm RK2} = \tenB^{-1}\tenD_2,
    \qquad
    \det\left(\tenI - \tenG_{\rm RK2}\right) = \frac{\lambda q\left(a - a^2/2\right)}{\det\tenB} .
    \label{eq:G_RK2}
\end{equation}
For $a > 2$ the factor $1 - a + a^2/2$ exceeds unity, and the first condition fails whatever the optical depth. The boundary is therefore $a \leq 2$, i.e., $\nu \leq 1/\sigma$, and nothing can be gained over the standard RSLA. The reason is physical. The growing mode is close to the equilibrium $\delta\hE = 4\,\delta\hT$, in which the gas and radiation checkerboards have the same sign. Since $1 - a + a^2/2 > 0$, the transport amplifies the radiation checkerboard without flipping its sign, and so it adds energy to this mode. The exchange conserves the total energy of the gas and the radiation, so it can only share this gain between them, not remove it. In PD-ARS and in the forward Euler scheme, by contrast, the transport factor $1 - a$ is negative for $a > 1$. The radiation checkerboard then has the opposite sign to the gas checkerboard, far from equilibrium, and this departure is what the exchange damps.

A local $\phi$ therefore works best with an upwind transport update and an implicit exchange update after every explicit stage such as PD-ARS. A scheme
that exchanges energy only after multiple explicit transport stages cannot benefit from it.

\subsubsection{Stability analysis for wave-speed correction schemes}
\label{sssec:wave_speed}

In addition to alternative time integration schemes, some codes also modify the HLL wave speeds to reduce numerical diffusion in optically thick cells; for example FORNAX multiplies the wave-speed for energy by a factor $\varepsilon = \min(1, 1/\tau_F)$ \citep{Skinner19a}, which at $\theta = \pi$ multiplies $a$ by $\varepsilon$ in the energy block and leaves the momentum block unchanged. The checkerboard condition is therefore the first of \autoref{eq:bound_md} with $\sigma_d$ replaced by $\varepsilon\sigma_d$, which in thick cells is much weaker.

This condition is, however, no longer a stability bound, for two reasons. First, the checkerboard is no longer always the least stable mode. With the energy dissipation lowered, modes near $\theta = \pi/2$ can grow while the checkerboard is stable, where the inertia of the radiation is not negligible against that of the gas: at $\tau_{\rm c} = 10^6$, $\nu_c = 3000$, $\calP = 10$ and $\calC = 10^4$, for instance, the corrected condition allows $\phi = 1$, but the spectral radius at $\theta \approx \pi/2$ is $1.017$. The growth disappears as $\calC$ increases, i.e., as $K_F$ of \autoref{eq:KF} decreases, but its onset depends strongly on the Eddington factor. A scan of the amplification matrix over all $\theta$, for $\tau_{\rm c} = 1$--$10^6$, $\nu_c = 30$--$3000$, $\calP = 10^{-2}$--$30$ and $\calC = 10^3$--$10^5$, finds unstable cases from $K_F \approx 18$ for $\chi = 1/3$ but from $K_F \approx 0.6$ for $\chi = 1$, and 110 out of 936 in all when the corrected condition is used everywhere. Second, the correction at a face is set by both cells beside it, so a thick cell adjacent to a thin one is then bounded as if it were barely dissipated. We therefore do not combine the wave-speed correction with a local $\phi$.

\subsubsection{Caveats}
\label{sssec:validity}

For a static background and the Eddington closure, or M1 about an isotropic background, \autoref{eq:bound_md} is the exact stability boundary of the linearized scheme in any number of dimensions: the checkerboard is the least stable mode (Proposition~\ref{prop:checkerboard}), and every other step of its derivation, the separation into blocks and Proposition~\ref{prop:energy_block}, is exact. Beyond this case the bound rests on five approximations.
\begin{enumerate}
    \item \emph{Moving background.} For $\calM \neq 0$ the blocks of the checkerboard no longer separate. For $\calM$ up to 3 the bound differs from the stability boundary computed numerically from the amplification matrix on all five variables by at most $6\times10^{-3}$ relative to it (\ref{app:theta}), in cells with $\tau_{\rm c}$ just below $2\sigma$, where the bound slightly overestimates the stable $\nu$.
    \item \emph{Anisotropic radiation.} If the Eddington tensor is highly anisotropic, the checkerboard need not be the least stable mode, and the exchange couples the radiation energy to the gas velocity through the background flux. In the diffusion regime, which is where ARSA raises $\phi$, we show in \ref{app:m1} that for the M1 closure the result is unchanged: a time step sufficient to keep the checkerboard stable keeps all modes stable. Outside of this regime we rely on numerical scans as we do for the moving background case, which we present in \ref{app:theta}. These numerical experiments find that our stability limit for the checkerboard also guarantees the stability of all modes, for every direction of the flux that a simulation can carry. For strongly directed radiation in optically thick cells, the linearized equations about such a background can themselves have growing modes (\ref{app:m1}), a property of the equations rather than of the scheme.
    \item \emph{Choice of wave speeds.} The analysis we have presented assumes HLL wave speeds that are symmetric about a uniform state, $\mp\sigma_d\hat{c}$, as in QUOKKA and Honghu.jl. Codes that take the extreme eigenvalues of the M1 Jacobian do not satisfy this assumption (e.g., \citealt{Skinner13a}).
    \item \emph{Lagged opacities.} The analysis omits the temperature derivatives of the opacities from the exchange.
    \item \emph{Discretization.} The analysis assumes piecewise-constant reconstruction, while codes such as QUOKKA and Honghu.jl reconstruct to higher order. A reconstruction with a limiter drops to first order at a local extremum, which the checkerboard has in every cell, so we expect the bound to carry over. We have not proven this, but every test of \autoref{sec:tests} uses PPM for both the matter and the radiation and every run is stable. 
    The analysis also omits the matter transport within a radiation step, as codes do; including it in $\tenZ_\theta$, as an upper estimate of its effect, shifts the boundary by at most $2.4\%$ in our linear wave tests.
\end{enumerate}

Where the checkerboard is not the least stable mode (items 1 and 2), the stable limit lies at most $0.6\%$ below \autoref{eq:bound_md} in every case that we have scanned. The safety factor of the ARSA, $\beta = 0.9$ (\autoref{ssec:arsa}), sets a margin of $10\%$ below the bound, which covers this and the shift of up to $2.4\%$ from the matter transport (item 5), and so guarantees stability.

\subsection{The Adaptive Reduced Speed of Light Approximation}
\label{ssec:arsa}

The ARSA sets the reduction factor of every cell to a fraction $\beta \leq 1$ of the largest value that \autoref{eq:bound_md} permits at the step being taken, and keeps it between $\phi_0$ and 1. Since $K$ and $K_F$ do not depend on $\phi$, both conditions of \autoref{eq:bound_md} are linear in $\phi$. With $\nu = \phi c\,\Delta t/\Delta x$, $\tau_E = \rho\kappa_E\Delta x$, $\tau_F = \rho\kappa_F\Delta x$, $K = \hat{h}\,\rho\kappa_P\,c\,\Delta t$ and $K_F = (1 + \chi)\,(E/\rho c^2)\,\rho\kappa_F\,c\,\Delta t$, with $\chi$ the largest diagonal element of the Eddington tensor, every cell $i$ evaluates
\begin{eqnarray}
    \phi_i & = & \min\left[1,\; \max\left(\phi_0,\; \frac{\beta}{\Lambda_i}\frac{\Delta x}{c\Delta t}\right)\right],
    \nonumber \\
    \Lambda_i & = & \max\left(\sum_d\sigma_{d,i} - \frac{\tau_{E,i}}{2 + K_i},\; \sum_d\sigma_{d,i} - \frac{\tau_{F,i}}{2 + K_{F,i}}\right),
    \label{eq:phi_sb}
\end{eqnarray}
with $\phi_i = 1$ where $\Lambda_i \leq 0$, i.e., where no reduction is needed at any time step. In words, the transport term, less the weaker of the two damping terms, sets the largest stable $\phi c\,\Delta t$, and $\phi$ is $\beta$ times that value, kept between $\phi_0$ and 1. On non-cubic grids, $\Delta x$ becomes the smallest grid spacing $\Delta x_{\rm min}$, and $\sum_d\sigma_{d,i}$ becomes $\sum_d\sigma_{d,i}\Delta x_{\rm min}/\Delta x_d$. All quantities are local properties of the cell and the step: its opacities, temperature and heat capacity, the signal speed of the radiation field it contains, the cell width, and $\Delta t$.

The safety factor $\beta$ keeps every cell below the boundary itself, where the checkerboard would be neutral, and covers the small errors of the bound listed in \autoref{sssec:validity}. It is a fixed safety margin rather than a parameter to tune; we use $\beta = 0.9$ throughout. The lower limit $\phi_0$ ensures that no cell uses a timestep reduction factor less than that of the standard RSLA, which is stable at the same step by Proposition~\ref{prop:rsla}. At the upper limit the cap is 1, not $\beta$: where $\Lambda_i \leq 0$ no time step is unstable, so no safety margin is needed, and cells can safely be updated using the unmodified RHD equations.

In a multigroup code each group carries its own $\phi$, from \autoref{eq:phi_sb} with the mean opacities of the group, and with the heat-capacity ratio $\hat{h}$ in $K$ and the energy density $E$ in $K_F$ of all the radiation: group $g$ of cell $i$ uses $K_{g,i} = \hat{h}_i\,\rho_i\kappa_{P,g,i}\,c\,\Delta t$, with $\hat{h}_i = 4a_rT_i^3/(\rho_i c_v)$. Since a group holds only part of the radiation heat capacity, this $K$ is an upper bound, which lowers the damping term and so keeps the bound conservative. We have checked the checkerboard of the coupled multigroup energy block, on a static background and with all groups sharing one gas temperature, in $4000$ random cases with two to five groups, optical depths $10^{-3}$--$10^4$ and $\calP \in [10^{-6}, 1]$; none is unstable. That the checkerboard is also the least stable mode of this coupled system is not proven: Proposition~\ref{prop:checkerboard} pairs each radiation variable with its own gas variable.

\subsection{Regimes of the Reduction Factor}
\label{ssec:irsla_phi}

The limiter has three regimes. First, in optically thin cells where $\tau_{E,i}\to 0$ and $\tau_{F,i}\to 0$, we have $\Lambda_i \to \sum_d \sigma_{d,i}$, which is of order unity, so the factor $(\beta/\Lambda_i)(\Delta x/c\Delta t) = \beta\phi_0/(\nu_0\sum_d\sigma_{d,i})$ is about $\phi_0$, since $\Delta x/(c\Delta t) = \phi_0/\nu_0$ and the RSLA uses a time step corresponding to a CFL number $\nu_0\sum_d\sigma_d \leq \beta$. Consequently, the limiter selects $\phi \approx \phi_0$, and reduces to the standard RSLA.

In optically thick cells the behavior of the ARSA depends on $K$ (or $K_F$ in the rare case where the second branch for $\Lambda_i$ in \autoref{eq:phi_sb} is activated). For $K \ll 2$, the time step is short compared with the time in which the gas exchanges its thermal energy, so the gas temperature hardly changes within a step, and the exchange simply relaxes the radiation towards a fixed gas temperature. Since $\sum_d \sigma_d$ is bounded above by $\sqrt{D}$, as long as $\tau_{E,i} > 2\sqrt{D}$ then $\Lambda_i$ is negative, and $\phi_i = 1$, meaning that even a simulation run with no reduction in the speed of light whatsoever will still remain stable, because damping by exchange fully suppresses unstable modes.

For $K \gg 2$, on the other hand, the damping term saturates at $\tau_E/K = 1/(\hat{h}_{\rm eff}\nu_c)$, independent of the cell optical depth. The reason for this limit is that when $K$ is large the gas and radiation reach a common temperature within a single time step, and in this regime the amount of damping that exchange with the gas can provide is limited by the effective ratio of specific heats $\hat{h}_{\rm eff}$ (\autoref{eq:SE}). We can then ask two closely related questions: (1) for what value of $\hat{h}_{\rm eff}$ will the ARSA produce $\phi = 1$ in very optically thick cells, $\tau\to\infty$, and (2) if this condition is met, how opaque does a cell have to be to reach $\phi = 1$?  The answers to these two questions will determine under what circumstances ARSA recovers the correct solution to the true RHD equations with no reduction in the speed of light.

These questions are straightforward to answer: in the limit $\tau \to \infty$, \autoref{eq:phi_sb} gives $\phi = 1$ only if the nominal reduction factor $\phi_0$ satisfies
\begin{equation}
    \phi_0 \geq \phi_* \equiv \frac{\hat{h}_{\rm eff}\,\nu_0\sum_d\sigma_d}{1 +\beta\hat{h}_{\rm eff}} = \frac{\hat{h}_{\rm eff}}{1 + \beta\hat{h}_{\rm eff}},
    \label{eq:phi_star}
\end{equation}
where the second equality holds assuming that the time step is set to the largest stable step for the standard RSLA, corresponding to $\nu_0 \sum_d \sigma_d = 1$. If this condition is met, then $\phi = 1$ is reached in any cell satisfying
\begin{equation}
    \tau_E \geq \tau_{\rm thresh} \equiv \frac{2\left(\sum_d\sigma_d - \beta\phi_0/\nu_0\right)\phi_0}{(1 + \beta \hat{h}_{\rm eff})(\phi_0 - \phi_*)},
    \label{eq:tau_one}
\end{equation}
assuming we are in the much more common case where the first term in $\Lambda_i$ in \autoref{eq:phi_sb} is larger. When $\phi_0 \gg \phi_*$, this condition reduces to $\tau_{\rm thresh} = 2 \sum_d \sigma_d\leq 2\sqrt{D}$, so as long as we have $\phi_0$ significantly larger than $\phi_*$, ARSA recovers the exact RHD equations in any cell with optical depth $\gtrsim 3.46$ in three dimensions. Even in cases where $\phi_0 < \phi_*$ and thus ARSA does not reach the value $\phi = 1$ required to exactly reproduce the full RHD equations, it still improves over the classic RSLA by moving the solution closer to them; for $\phi_0 < \phi_*$ and $\tau\to\infty$, the reduction factor will be set to
\begin{equation}
    \phi_{\max} = \frac{\beta\phi_0}{\nu_0\sum_d\sigma_d - \phi_0/\hat{h}_{\rm eff}},
    \label{eq:phi_sat}
\end{equation}
which exceeds $\phi_0$ when the RSLA runs at $\nu_0\sum_d\sigma_d \leq \beta$, as in our tests, though only slightly when $\phi_0 / \hat{h}_{\rm eff} \ll 1$. Where we quote values of $\phi_*$ and $\tau_{\rm thresh}$, we evaluate them at the bound itself, $\beta = 1$ and $\nu_0\sum_d\sigma_d = 1$; at the values of our runs, $\beta = \nu_0\sum_d\sigma_d = 0.9$, both are slightly smaller, so the quoted values are conservative.

We illustrate the condition for the ARSA to restore the exact solutions to the RHD equation in \autoref{fig:accuracy} for a gray material with $\hat{h}_{\rm eff} = \hat{h}$. In the amber region, $0.1\,\phi_0 < \hat{h} < 0.1$, the ARSA run with $\phi_0$ above $\phi_*$ removes the error of the RSLA in thick cells, at a local speedup of up to $1/\phi_* \approx 1/\hat{h}$. At its upper edge, $\hat{h} = 0.1$, this speedup has fallen to about 10, and in the red region above it no affordable $\phi_0$ lets a cell reach $\phi = 1$. Where $\hat{h} < 0.1\,\phi_0$, thick cells also reach $\phi = 1$. In the green part of this band, $\calD < 0.1\,\phi_0$, the RSLA is already accurate; in the amber part, $\calD > 0.1\,\phi_0$, the ARSA removes the drift of the structures that the flow carries (\autoref{ssec:heat_capacity}), which the RSLA leaves however small $\hat{h}$ is.

The last column of \autoref{tab:media} reports $\phi/\phi_0$, with $\phi$ set to the maximum allowed, for an example simulation of each of the given systems run with $\phi_0 = 10^{-2}$ and 100 cells across: it is the reduction factor that the limiter of the ARSA sets in a cell of optical depth $\calL/100$, for isotropic radiation in three dimensions at the RSLA step $\nu_0 = 0.9/\sqrt{3}$, divided by $\phi_0$. For this value of $\phi_0$, $\phi/\phi_0 = 100$ means that the ARSA runs the unmodified RHD equations, and we see that this is achieved in the first five example systems. In all other systems either the cells are optically thin or the radiation holds a large share of the heat capacity, and $\phi \approx \phi_0$, so the ARSA provides no gain over the standard RSLA.

\subsection{Linear Waves with ARSA}
\label{ssec:arsa_waves}

\begin{figure}[t]
    \centerline{\includegraphics[width=0.75\textwidth]{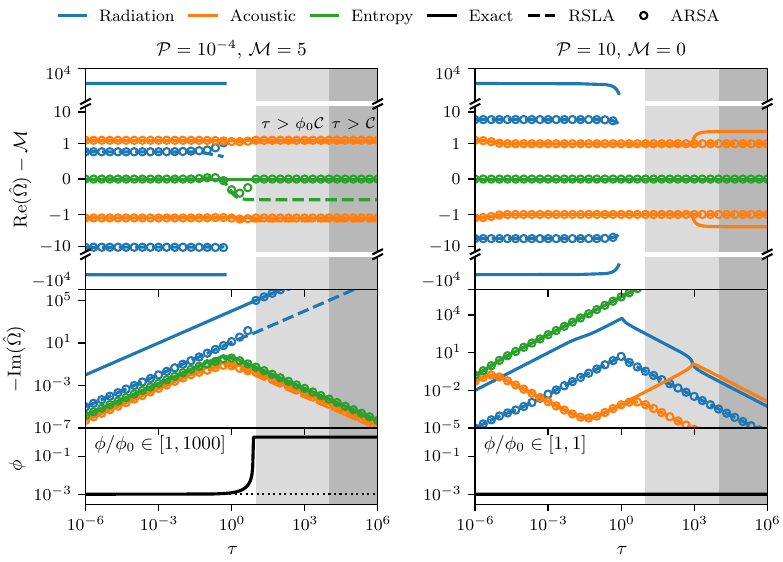}}
    \caption{Linear waves computed for the exact RHD equations (solid lines), the RSLA (dashed lines), and the ARSA (circles); colors indicate the type of mode. The RSLA calculation uses $\phi = 10^{-3}$, and the ARSA one uses $\phi_0 = 10^{-3}$, $N = 32$ cells per wavelength, and a time step corresponding to $\nu_0 = 0.9\sqrt{3}$. All calculations are for a gas with $\calC = 10^4$ and $\gamma = 5/3$. Panels show, from top to bottom, the wave phase speed relative to the background medium velocity $\mathrm{Re}(\hOmega) - \calM$, the wave decay rate $-\mathrm{Im}(\hOmega)$, and the value of the speed of light reduction factor $\phi$ computed by the ARSA, all as a function of wave optical depth per radian $\tau$; annotations in the bottom panel give the range of $\phi/\phi_0$. The left column shows the case $\calP = 10^{-4}$, $\calM = 5$ ($\phi_* \ll \phi_0$), and the right shows $\calP = 10$ ($\phi_* \gg \phi_0$) at $\calM = 0$. Gray bands show the approach to the dynamic diffusion regime, $\tau>\phi_0\calC$ and $\tau>\calC$, where the classic RSLA fails.
    \label{fig:waves_IRSLA}
    }
\end{figure}

We can gain more insight into how the ARSA performs by repeating the linear wave analysis of \autoref{ssec:linear_waves} using the ARSA. We therefore repeat that calculation for a gas with $\gamma = 5/3$ and $\mathcal{C} = 10^4$, using both the classic RSLA with $\phi = 10^{-3}$ and the ARSA with $\phi_0 = 10^{-3}$ and $\phi$ given by \autoref{eq:phi_sb}; we evaluate this equation for the case of $N = 32$ cells per wavelength, $\beta = 0.9$, and a time step $\Delta t$ chosen to give CFL ratio $\nu_0 = 0.9\sqrt{3}$.

We show the results for two example cases in \autoref{fig:waves_IRSLA}: $\calM = 5$ and $\calP = 10^{-4}$ and $\calM = 0$ and $\calP = 10$. The first of these cases gives $\phi_* = 2.7\times10^{-4}$, placing it in the green and amber regions of \autoref{fig:accuracy} where $\phi_0 > \phi_*$, and the second case has $\phi_* = 0.96$, placing it in the red region. As in \autoref{fig:waves_RSLA}, to which it can be compared, we plot the wave phase speed $\mathrm{Re}(\hOmega)$ in the top panels (compensated by the background flow speed $\calM$), and the damping rate $-\mathrm{Im}(\hOmega)$ below this; the bottom panel shows the value of $\phi$ for the ARSA.

The left panel shows the case where $\phi_* < \phi_0$; in this case $\phi$ rises with $\tau$ and reaches unity at $\tau \approx 8$, i.e., at $\tau_{\rm c} \approx 1.5$, the threshold of \autoref{eq:tau_one}. Beyond this point the ARSA uses the unmodified equations and reproduces their phase speeds and decay rates exactly, in the moving background as in a static one. Non-negligible errors in acoustic and entropy modes, which are the significant ones for matter, are present only over a very narrow range of wavelength near $\tau = 1$. The RSLA, by contrast, gets the matter modes wrong because it raises the heat capacity of the radiation to $\hat{h}/\phi_0 = 0.27$ of that of the gas. The right panel shows the case where $\phi_*$ is almost unity, where the RSLA does not help. In this regime the ARSA cannot stably use $\phi$ significantly larger than $\phi_0$, and thus the RSLA and ARSA results are essentially identical, and both are incorrect at high $\tau$.

\subsection{Extension to Decoupled Dust and Gas}

So far the radiation has exchanged energy with the gas directly, at the gas temperature $T$. In much of the interstellar medium, however, the radiation is absorbed and emitted by dust grains, whose temperature $T_{\rm d}$ can differ from $T$ by orders of magnitude, while the heat is held by the gas. We now extend the ARSA to this case. Our central assumption in doing so is that the dust holds a negligible share of the heat, so that its temperature is a function of the local state, $T_{\rm d} = T_{\rm d}(T, E)$, which we take to be known exactly.

For a system with decoupled dust and gas temperatures, the term $\rho c (\kappa_E E - \kappa_P a_r T^4)$ in the time-like component of $cG^0$ (\autoref{eq:G0_int}) is replaced by $-N$, where $N = \rho c\,(\kappa_P\, a_r T_{\rm d}^4 - \kappa_E E)$ is the net emission of the dust. Its perturbation, $\delta N = (\partial N/\partial T)_E\,\delta T + (\partial N/\partial E)_T\,\delta E$, can be written as
\begin{equation}
    \delta N = \rho c\left(\kappa_P\, 4 a_r T_{\rm d}^3\,\psi\,\delta T - \alpha\,\kappa_E\,\delta E\right),
    \label{eq:two_temp_exchange}
\end{equation}
with two dimensionless parameters,
\begin{equation}
    \psi \equiv \left(\frac{\partial T_{\rm d}}{\partial T}\right)_E,
    \qquad
    \alpha \equiv \frac{(\partial N/\partial E)_T}{(\partial N/\partial E)_{T_{\rm d}}} .
    \label{eq:psi_alpha}
\end{equation}
Intuitively, $\psi$ measures how much the dust warms per unit change in the gas temperature with a fixed radiation field, and $\alpha$ compares how the net matter cooling rate responds at fixed gas temperature versus at fixed dust temperature. When the dust is tightly coupled to the gas, $T_{\rm d} = T$, we have $\psi = \alpha = 1$ and \autoref{eq:two_temp_exchange} reduces to the case we have already analyzed. When the dust is completely decoupled from the gas, it is in radiative equilibrium, $\kappa_P\, a_r T_{\rm d}^4 = \kappa_E E$, and we have $\psi = \alpha = 0$. In the common case where the dust temperature is set by a balance between radiative heating and cooling and dust-gas collisions, we have \citep[e.g.,][]{Goldsmith01a}
\begin{equation}
    N = \alpha_{\rm gd} n^2 \sqrt{T} (T - T_{\rm d}),
\end{equation}
where $\alpha_{\rm gd}$ is a constant dust-gas coupling coefficient and $n$ is the number density of free particles. This gives
\begin{equation}
    \alpha = \left(\frac{4 a_r T_{\rm d}^3 \rho \kappa_P c}{\alpha_{\rm gd} n^2 \sqrt{T}} + 1 \right)^{-1},
    \qquad
    \psi = \left(\frac{3}{2} - \frac{T_{\rm d}}{2T}\right)\alpha.
\end{equation}
Here $\alpha$ lies in $[0,1]$ and $\psi \leq 1$, and $\psi \geq 0$ where $T_{\rm d} \leq 3T$, which is generally true for typical interstellar media.

\autoref{eq:two_temp_exchange} has the form of the non-gray exchange of \autoref{sssec:checkerboard}: the gas loses what the radiation gains, the derivative of the emission with respect to $T$ is $4 a_r T_{\rm d}^3\psi$ instead of $4 a_r T^3$, and the absorption coefficient is $\alpha\rho\kappa_E$. In the radiation row of \autoref{eq:SE_nongray}, this amounts to $\tau_E \to \alpha\tau_E$ and $\tau_P \to \psi\,(T_{\rm d}/T)^3\,\tau_P$. The analysis of \autoref{ssec:von_neumann} therefore carries over unchanged, Propositions~\ref{prop:energy_block} and \ref{prop:checkerboard} included. The two-temperature bound is \autoref{eq:bound_md} with $\tau_E$ replaced by $\alpha\tau_E$ in the energy condition, and with $K = \hat{h}\,\nu_c\,\tau_P$ computed with the heat-capacity ratio
\begin{equation}
    \hat{h} = \frac{4 a_r T_{\rm d}^3\,\psi}{\rho c_v} .
    \label{eq:h_2T}
\end{equation}
In practice dust and gas tend to become well-coupled, $\psi \to 1$, only at densities $\gtrsim 10^5$ cm$^{-3}$ \citep[e.g.,][]{Goldsmith01a}, and in this regime we have $\hat{h} = 0.15 n_6^{-1} T_1^3$, where $n_6$ is the gas number density in units of $10^6$ cm$^{-3}$, $T_1$ is $T_d$ in units of 10 K, and we have computed $c_v$ for a monatomic ideal gas, appropriate for H$_2$ molecules at temperatures too low to excite their rotational degrees of freedom.
The momentum condition is unchanged, provided that the dust passes the force of the radiation on to the gas at once. Note that $\hat{h}$ is a ratio of heat capacities, not of energies: it is not 4 times $a_r T_{\rm d}^4/(\rho c_v T)$, from which it differs by the factor $\psi\,T/T_{\rm d}$. In the limiter \autoref{eq:phi_sb}, the energy condition uses $\alpha_i\rho_i\kappa_{E,i}$ in place of $\rho_i\kappa_{E,i}$, and $K_i = \hat{h}_i\,\rho_i\kappa_{P,i}\,c\,\Delta t$ with $\hat{h}_i$ of \autoref{eq:h_2T}. In the notation of \autoref{eq:SE}, the two-temperature bound is the one-temperature bound with $q$ replaced by $\alpha q$ and with the effective heat-capacity ratio $\hat{h}_{\rm eff} = \hat{h}\kappa_P/(\alpha\kappa_E)$.

\section{A Numerical Algorithm for Grid-Based Two-Moment Codes}
\label{sec:implementation}

We now describe how the ARSA can be implemented in a finite-volume two-moment code that uses the IMEX scheme of \autoref{eq:pdars}; for generality we write the algorithm for the case of multiple radiation groups $g$. Relative to an RSLA code, the ARSA requires three changes, all local: a reduction factor in every cell and group, a transport step that multiplies the flux divergence of each cell by its own reduction factor, and an exchange step with the radiation source terms scaled by it. \autoref{alg:arsa} summarizes the operations for one radiation sub-step, for which we explain the rationale in detail below.

\begin{algorithm}[t]
\caption{One radiation sub-step of the ARSA.}
\label{alg:arsa}
\begin{enumerate}
    \item Fill the ghost cells of the gas and radiation states.
    \item In every cell $i$ and group, compute $\sigma_{d,i}$, $\rho_i\kappa_{E,i}$, $\rho_i\kappa_{F,i}$, $K_i$ and $K_{F,i}$, then $\Lambda_i$ and $\phi_i$ from \autoref{eq:phi_sb}, using the sub-step $\Delta t$ being taken.
    \item In each of the two stages of \autoref{eq:pdars}, with the same $\phi_i$:
    \begin{enumerate}
        \item compute the HLL fluxes of the unmodified radiation system and update each cell by $\phi_i$ times its flux divergence (\autoref{eq:transport_discrete});
        \item floor $E$ only in a cell where $\rho e + \sum_g E_g/\phi_g \leq 0$;
        \item perform the implicit matter-radiation exchange update with the radiation source terms scaled by $\phi_i$;
        \item repair the state, $E > 0$ and $|\vecF| < cE$.
    \end{enumerate}
\end{enumerate}
\end{algorithm}

\paragraph{Steps 1 and 2} An important part of the algorithm is that we evaluate \autoref{eq:phi_sb} once per radiation sub-step, in every cell and every group, including ghost cells, from the state at the start of the sub-step, and using the actual time step $\Delta t$ that will be used for that step, and the actual maximum signal speed that will enter the Riemann solver -- $\sigma_d = T_{dd}^{1/2}$ for the M1 closure, where $T_{dd}$ is the diagonal element of the Eddington tensor in direction $d$. We use this $\phi$ in both stages of \autoref{eq:pdars}. This is the scheme that the stability analysis describes, and so for the stability guarantee to hold we must follow it.

\paragraph{Step 3a} Following \autoref{eq:rsla_time}, $\phi$ multiplies the time derivative, not the flux. We therefore compute the HLL fluxes of the \emph{unmodified} radiation moment system, with wave speeds $\pm\sigma c$ at every face, and then scale the update of each cell by its own reduction factor,
\begin{equation}
    \frac{\mathrm{d}\vecU_{{\rm r},i}}{\mathrm{d}t} = -\phi_i\sum_d \frac{\vecF^*_{{\rm r},i+\frac{1}{2}\hat{e}_d} - \vecF^*_{{\rm r},i-\frac{1}{2}\hat{e}_d}}{\Delta x_d},
    \label{eq:transport_discrete}
\end{equation}
group by group. For a uniform $\phi$ this is identical to the standard RSLA, because the HLL flux at reduced speed is exactly $\phi$ times the HLL flux at the true speed, term by term. Where $\phi$ varies, \autoref{eq:transport_discrete} keeps the steady states of the unmodified scheme cell-by-cell, which is the key property required of a reduced speed of light scheme -- the alternative, a single flux $\phi_{i+1/2}\vecF^*$ at each face with, say, the minimum of the two adjacent reduction factors, conserves the radiation energy in the transport step but not the set of steady states, and it can be badly wrong in situations where the value of $\phi$ changes rapidly over a small number of cells due to the presence of a discontinuity. The supercritical shock test presented in \autoref{ssec:test_ensman} provides an example of this: in this test at $\phi_0 = 0.03$, $\phi$ jumps by a factor of 65 between neighboring cells on either side of the shock, and for this case \autoref{eq:transport_discrete} achieves a $3.8\%$ error relative to the reference solution, while a scheme using the minimum reduction factor of two adjacent cells makes a 52\% error.

\new{Moreover, the discontinuities in $\phi$ do not create, amplify, or reflect radiation. To see this, consider a cell with $\phi=\phi_0$ adjacent to a cell with $\phi=1$. Through their common face the first cell loses only $\phi_0$ times the radiation energy that the neighboring cell gains; however, the radiation in the first cell is worth $E_g/\phi_0$ to its gas, so the gas of the second cell absorbs exactly the energy emitted by the gas of the first. While $\phi$ is held fixed, transport and exchange together conserve $\sum_i (\rho e + \sum_g E_g/\phi_g)_i$. In steady state the radiative flux and the energy density are continuous across the discontinuity and assume their true values, and since $\phi$ multiplies both radiation equations identically, the discontinuity reflects no radiation.}

\paragraph{Step 3b} After the transport step the realizability condition, $E > 0$ and $|\vecF| < cE$, may not hold; indeed, for optically thick cells where the ARSA sets $\phi = 1$, we will in general have $|\vecF| \gg cE$. In the ARSA scheme, this condition should be repaired by the exchange step, and it is important to let the exchange step act before applying floors. Limiting $|\vecF|$ at this stage would artificially reduce the flux and lead to the scheme failing to recover the diffusion limit, while limiting $E$ artificially adds energy to the system. We therefore apply only one floor after the transport step, which is to floor $E$ if the \emph{total} energy $\rho e + \sum_g E_g/\phi_g$ -- which is conserved during the exchange step -- is not positive. This never occurs in the runs of the radiative shocks (\autoref{ssec:test_ensman}), where $E$ most often becomes negative after the transport. We apply floors to enforce realizability only after the exchange is complete, in \emph{Step 3d.}

The remaining details of the IMEX PD-ARS update, including the second stage, are as in \citet{He24a}.

\section{Numerical Tests}
\label{sec:tests}

We test the algorithm described above in Honghu.jl \citep{He26a}, a three-dimensional radiation-magnetohydrodynamics code written in Julia for the development and testing of algorithms intended for QUOKKA \citep{Wibking22a, He24a, He24b}. Honghu.jl is open source (\url{https://github.com/chongchonghe/Honghu.jl}), and its release v0.1.0 reproduces every numerical result of this paper. The scripts and data that make every figure and quoted number of this paper are at \url{https://github.com/chongchonghe/paper-ARSA}. Honghu.jl follows QUOKKA's discretization closely, so that a kernel written in one transliterates to the other: the hydrodynamics uses a method-of-lines update with piecewise parabolic reconstruction, an HLLC Riemann solver and a third-order strong-stability-preserving Runge-Kutta integrator \citep{Shu88a}; the radiation uses the gray or multigroup M1 moment system with the Levermore closure \citep{Levermore84a}, the HLL Riemann solver with the wave speeds of \citet{Balsara_1999}, piecewise parabolic reconstruction of $E$ and $\vecF/(cE)$, and the IMEX PD-ARS scheme of \autoref{eq:pdars} for transport and matter-radiation exchange, sub-cycled within each hydrodynamic step. See \citet{Wibking22a} and \citet{He24b, He24a} for a full description of the algorithm.

All the tests below use piecewise parabolic reconstruction (PPM) for both the matter and the radiation. The stability analysis of \autoref{sec:irsla} assumes piecewise-constant reconstruction (\autoref{sssec:validity}), so the tests also check that the bound carries over to PPM.

In every test the RSLA runs at $0.9$ times its largest stable step, $\nu_0\sum_d\sigma_d = 0.9$: $\nu_0 = 0.9/\sqrt{3}$ in the three-dimensional tests (Proposition~\ref{prop:rsla}) and $0.9\sqrt{3}$ for the one-dimensional linear waves with the P1 closure, and the ARSA, with $\beta = 0.9$, at the same radiation step, so that the ARSA is compared with the RSLA at its best. Unless stated otherwise, an error is the relative $L_1$ error of the gas temperature, $\sum_i |T_i - T_i^{\rm ref}|/\sum_i T_i^{\rm ref}$, against a reference at the same resolution: the exact solution where one exists, and otherwise a run with $\hat{c} = c$ at $\nu = 0.4$.

The linear waves use a one-dimensional grid. The pulse, the radiative shocks and the Marshak wave are planar problems, which we run in three dimensions as slabs of $N \times 2 \times 2$ cubic cells, periodic in $y$ and $z$, in which every column is the same; the expanding sphere is a three-dimensional box. In every test the three mean opacities are equal, within each group for the Marshak wave, so that $\tau_E = \tau_{\rm c}$ and $\hat{h}_{\rm eff} = \hat{h}$.

\subsection{Linear Waves}
\label{ssec:test_waves}

We first test the performance of ARSA on linear waves in order to verify that our numerical implementation reproduces the results calculated in \autoref{ssec:arsa_waves}. For this test we work in dimensionless units with fixed $\calC = 10^3$ and $\calP = 10^{-2}$. We use a one-dimensional grid of $N=64$ cells, and a box of size 1. On this grid we simulate a gas with adiabatic index $\gamma=5/3$ and a constant gray opacity $\hat{\kappa}$, which is initialized as a uniform background state plus a sinusoidal perturbation, $\hvecW_{\rm bg} + \delta\hvecW e^{i\hk\hx}$, with wavenumber $\hk = 2\pi$ (i.e., one wavelength in the simulation domain), background state
\begin{equation}
    \hvecW_{\rm bg} = (\hrho, \hv, \hT, \hE, \hF) =  (1,0,1,1,0)
\end{equation}
and perturbation $\delta\hvecW$ set equal to the eigenvector for an RHD mode -- either a right-going acoustic mode or an entropy mode -- with optical depth per radian $\tau = [0.1, 1, 10, 100, 1000]$, scaled so that the density perturbation is $\delta\hrho =10^{-5}$. We carry out simulations for each value of $\tau$ using the RSLA, the ARSA, and using no reduction in the speed of light. For the simulation with no reduction in the speed of light we solve for the eigenvector of the unmodified RHD equations, for the RSLA we find the eigenvector for a reduction factor $\phi = 0.03$, and for the ARSA we use a reduction factor computed by adopting a nominal reduction factor $\phi_0 = 0.03$ and solving for the local reduction factor $\phi$ using \autoref{eq:phi_sb}.

We simulate the system for a time $\hatt = \min[2, 3/(\hk\,|\mathrm{Im}(\hOmega)|)]$, where $|\mathrm{Im}(\hOmega)|$ is the analytically computed decay rate per unit wavenumber of the mode we have initialized; thus the simulation runs for either two sound crossing times or three decay times, whichever is smaller. The simulations using the RSLA and ARSA use the same time step, set so $\nu_0 = 0.9\sqrt{3}$, while the simulation using the unmodified speed of light uses a time such that $\nu_0 = 0.4$; note that this makes the time step for the unmodified equations a factor of $\approx 100$ smaller than for the RSLA or ARSA. For this test only we use the P1 closure (i.e., the Eddington tensor is fixed to $T_{ij} = (1/3)\delta_{ij}$) to match the assumptions of the analytic wave mode calculation. We run the simulation using no subcycling of radiation, so both radiation and hydrodynamics are advanced in each time step.

To calculate the numerical phase speed $\mathrm{Re}(\hOmega)$ and decay rate $-\mathrm{Im}(\hOmega)$ from the simulations, we sample the output at 40 times evenly spaced from $\hatt = 0$ to the final time. For each output we compute the Fourier coefficient of the density perturbation over all cells, $c(\hatt) = N^{-1}\sum_i(\hrho_i - 1)e^{-i\hk\hx_i}$; we then do a least-squares linear fit of $\ln|c|$ versus $\hatt$ to measure the decay rate, and a least-squares linear fit of the phase of $c$ versus $\hatt$ to measure the phase speed.

\begin{figure}[t]
    \centerline{\includegraphics[width=\textwidth]{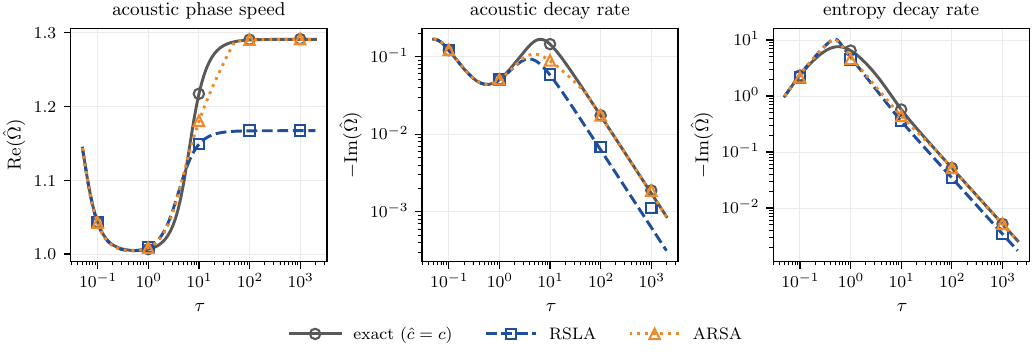}}
    \caption{Phase speeds and decay rates for linear RHD waves as a function of optical depth per radian $\tau$, as measured in Honghu.jl (symbols) and predicted by linear theory (lines). Panels show, from left to right, the phase speed and decay rate of the acoustic mode, and the decay rate of the entropy mode, and colors show the exact solution with no reduction in the speed of light (black), the result using the RSLA (blue), and the result using ARSA (orange). See \autoref{ssec:test_waves} for details on the simulation setup.
    \label{fig:linear_waves_code}
    }
\end{figure}

We plot the results of this procedure in \autoref{fig:linear_waves_code}; lines show the analytically-computed phase speed and decay rate (omitting the phase speed of the entropy mode, which is exactly zero), and points show the numerical measurements. We see that the code reproduces the phase speeds and decay rates predicted by linear theory very accurately, confirming our implementation; the only visible difference between the linear theory predictions and the measurements is for the RSLA at $\tau = 1000$, where the numerical decay rate somewhat exceeds the (very small) linear theory prediction due to numerical dissipation. We also see that the ARSA achieves its goal of removing almost all of the error in the RSLA and matching the exact solutions, despite taking the same time step as the RSLA. The only visible differences between the ARSA and exact solutions are for the acoustic mode at $\tau = 10$ and the entropy mode at $\tau = 1$, where the ARSA is not quite able to reach $\phi = 1$ -- but even in these cases the ARSA is considerably closer to the true answer than the RSLA.

\subsection{Advecting Radiation Pulse}
\label{ssec:test_thick}

We next carry out a version of the advecting pulse from \citet{Krumholz07b}. In the first two versions of this test we initialize an ideal gas with mean mass per particle $\mu = 2.33$ (in units of amu), adiabatic index $\gamma=5/3$, and a temperature as a function of position
\begin{equation}
    T = T_{\rm lo}\left[1 + e^{-x^2/(2w^2)}\right],
    \label{eq:pulse}
\end{equation}
with $w = 24$~cm and $T_{\rm lo}$ chosen as described below. We set the density as a function of position so $\rho = 1.2$~g~cm$^{-3}$ as $x\to \pm\infty$, and with constant $\rho T$ so that the system is everywhere in pressure balance. The simulation domain is 1024~cm wide, periodic, and resolved by 512 cells; the (gray) opacity is $\kappa = 500$~cm$^2$~g$^{-1}$, so every cell is optically thick, $\tau_{\rm c} = 600$--$1200$. The entire medium advects uniformly at $v = 3\times10^7$~cm~s$^{-1}$, chosen so that the dynamic diffusion number $\calD = \kappa\rho w\,v/c = 14$, placing the system in the dynamic diffusion regime.

We simulate this system with both the RSLA and the ARSA at $\phi_0 = 0.01$, and compare both with the exact solution: the pulse at rest, diffusing, shifted by $vt$, which we compute from the diffusion equation for the enthalpy of gas and radiation on a fine grid in column mass. For these runs we choose $T_{\rm lo}$ so that $\calP/\phi_0 = 0.33$ and $0.0033$ at the peak of the pulse. The first run lies at the top right of the amber region in \autoref{fig:accuracy}, where the RSLA both slows the diffusion and lets the pulse drift, and the second at its lower right, where only the drift remains. In both cases $\hat{h} < \phi_0$ and the cells are thick, so the ARSA runs $\phi = 1$ everywhere.

We also carry out a third version of this test using a weak pulse at rest, $\calD = 0$, with amplitude $0.1$ instead of 1, i.e., $T = T_{\rm lo}[1 + 0.1\,e^{-x^2/(2w^2)}]$, and $T_{\rm lo}$ chosen again so that $\calP/\phi_0 = 0.33$ at the peak. This case lies at the top left of the amber region, where the RSLA slows the diffusion and nothing drifts. In this run we also lower the opacity to $\kappa = 30$~cm$^2$~g$^{-1}$, which shortens the diffusion time but still keeps the cells thick enough for $\phi = 1$ ($\tau_{\rm c} = 65$--$72 > \tau_{\rm thresh}$). The domain for this test is 768~cm wide and is resolved by 384 cells.

\begin{figure}[t]
    \centering
    \includegraphics[width=0.9\textwidth]{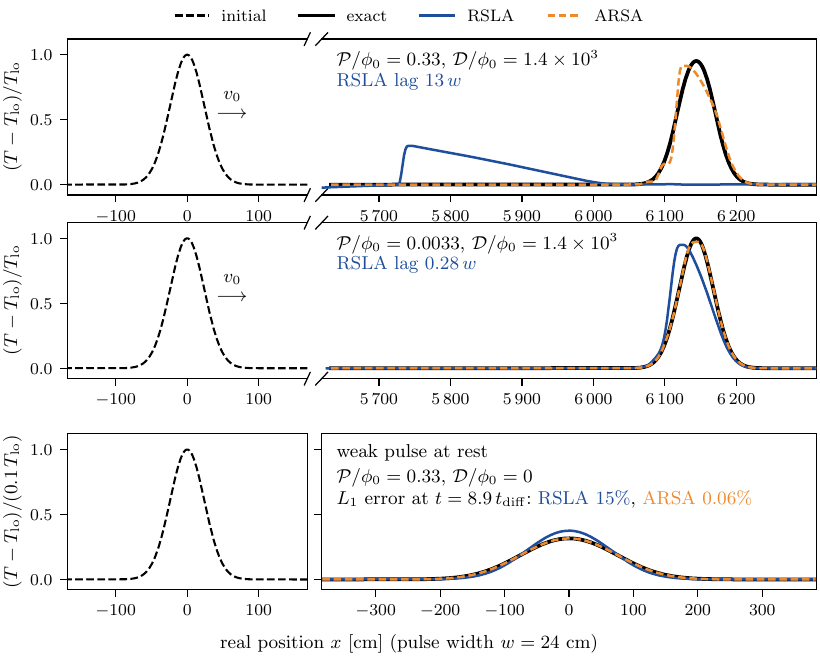}
    \caption{Fractional temperature deviation $(T - T_{\rm lo})/T_{\rm lo}$ as a function of position $x$ in the pulse test, at $\phi_0 = 0.01$. Top two rows: the moving pulse, $\calD = 14$, at $\calP/\phi_0 = 0.33$ and $0.0033$, after 256 widths traveled, with the lag of the RSLA in pulse widths $w$. Bottom row: a weak pulse at rest, of amplitude $0.1$ (shown divided by it), at $\calP/\phi_0 = 0.33$, after $t = 8.9\,t_{\rm diff}$ (2.1~ms), with $t_{\rm diff} = 3\rho\kappa w^2/(\hat{h}c)$ at the peak, with the $L_1$ errors of $T - T_{\rm lo}$. In each row the black dashed line (left) shows the initial condition, the black solid line shows the exact solution, and the blue solid and orange dashed lines show results for the RSLA and ARSA, respectively. For the moving pulse the horizontal axis is the real position, broken so that the pulse has the same width on the page in every window.}
    \label{fig:pulse_drift}
\end{figure}

We run the two moving pulses for $t = 256\,w/v$. Since the gas holds the heat, the pulse diffuses only on $t_{\rm diff} \sim 3\rho\kappa w^2/(\hat{h}c)$, longer than this, so the exact pulse advects with little change of shape. The top two panels of \autoref{fig:pulse_drift} show the temperature distribution at the final times of both the ARSA and RSLA simulations of these cases. We see that for the RSLA the pulse position (defined numerically by $\sum_i x_i (T_i - T_{\rm lo}) / \sum_i (T_i - T_{\rm lo})$) falls behind by 13 and 0.28 times $w$ in the two simulations; the pulse shape is also deformed. This is the expected failure of the RSLA in the dynamic diffusion regime. By contrast, the ARSA result is essentially perfect for all cases except $\calP/\phi_0 = 0.33$; in all cases the pulse centroid stays within $0.04w$ of the exact position in every run. We note that even the small distortion in the ARSA is not due to a failure of the ARSA \textit{per se}, since the ARSA gives $\phi = 1$ in every cell for this test; the distortion is simply because we are advecting the radiation at a high CFL number, and is completely analogous to the distortion that an entropy mode advected across the grid at high Mach number might undergo in a purely hydrodynamic simulation. Using a smaller CFL number removes the distortion in the ARSA, but does not repair the RSLA.

We show the results of the third run in the bottom panel, run to time $t = 8.9\,t_{\rm diff} = 2.1$~ms, when the peak temperature has fallen to a third. The panel shows that the RSLA diffuses the pulse too slowly, since it raises the heat capacity of the radiation to $\hat{h}/\phi_0$ of that of the gas (\autoref{ssec:heat_capacity}): its peak is $19\%$ too high, and its $L_1$ error is $15\%$, against $0.064\%$ for the ARSA.

We can understand the results of all these tests as follows: in the amber zone where $\hat{h} < \phi_0$ and the ARSA runs $\phi = 1$, \autoref{eq:slowdown} gives $t_\phi/t_1 < 2$: the RSLA at most halves the diffusion rate, so the amplitude of a spreading pulse, $\propto (w_0^2 + 2D_{\rm th} t)^{-1/2}$ with $D_{\rm th}$ the thermal diffusivity, is at most $\sqrt{2}$ too high.
The drift, in contrast, grows for as long as the flow carries a structure, by about $3\calD/\phi_0$ widths per diffusion time, since the structure lags at about $\hat{h}/\phi_0$ of the flow speed. There the drift is therefore the larger error of the RSLA, unless the flow moves a structure much less than its own size before it diffuses.

\new{We can also use these runs as a test of energy conservation in the ARSA: since the box is periodic, the unmodified RHD equations conserve the total energy $E_{\rm gas} + E_{\rm rad}$, and for this problem the ARSA, which runs $\phi = 1$, should as well. Our numerical tests confirm this expectation: the ARSA runs conserve total energy to round-off in all three runs, to a relative change below $2\times10^{-9}$. The RSLA conserves $E_{\rm gas} + E_{\rm rad}/\phi_0$ instead (\autoref{ssec:heat_capacity}), and changes the total energy by $6.2\times10^{-4}$, $7.8\times10^{-8}$ and $3.2\times10^{-4}$ of its value; in the first run this is 23 times the radiation energy in the box.}

\subsection{Opaque Expanding Sphere}
\label{ssec:test_sphere}

The pulse tests the transport of radiation by the flow. The opaque expanding sphere of \citet{Roth15a}, which we examine next, tests the work that an expanding flow does on the radiation. In this test a uniform sphere expands homologously, $\vecv = \vecx/(t + t_0)$, with outer radius $r_{\rm out} = r_{\rm out,i}(1 + t/t_0)$, from $r_{\rm out,i} = 10^{13}$~cm to $11\,r_{\rm out,i}$. As in \citet{Roth15a}, the gas starts at density $\rho = 4.75\times10^{-7}$~g~cm$^{-3}$ and temperature $T = 10^4$~K, and has mean particle mass $\mu = 0.5$, adiabatic index $\gamma = 5/3$ and specific opacity $\kappa = 0.4$~cm$^2$~g$^{-1}$. The radiation starts in equilibrium with the gas.

Compared to the published setup of \citet{Roth15a}, we change two things. First, we make the opacity absorptive, $\kappa_P = \kappa_E = \kappa_F$, so that the gas and the radiation share one temperature, as opposed to scattering in the original version. In that case nothing damps the radiation energy, the damping term of \autoref{eq:phi_sb} vanishes however thick the cell, and the ARSA sets $\phi \approx \phi_0$, as in an optically thin cell: for scattering-only opacities the ARSA is the RSLA. Second, we lower the expansion speed from $10^4$ to $10$~km~s$^{-1}$, so that $\hat{c} \geq 9\,v_{\rm out}$ at the $\phi_0$ we use, and the flow is subsonic; the expansion speed in the original test is close enough to $c$ that there is little need to reduce the speed of light.

Since the medium stays uniform, we compute only its interior: a cube $[-r_{\rm out,i}, r_{\rm out,i}]^3$ of $32^3$ cells, in which the velocity is homologous throughout. The ghost cells copy the nearest cell of the domain and take the homologous velocity at their own position, so that each run sees only its own solution. The cells are very thick, $\tau_{\rm c} = 1.2\times10^5$ at the start and $89$ at the end, and $\calD = \kappa\rho\,r_{\rm out,i}v_{\rm out}/c = 63$, so the radiation is trapped in the flow.

\begin{figure}[t]
    \centerline{\includegraphics[width=0.55\textwidth]{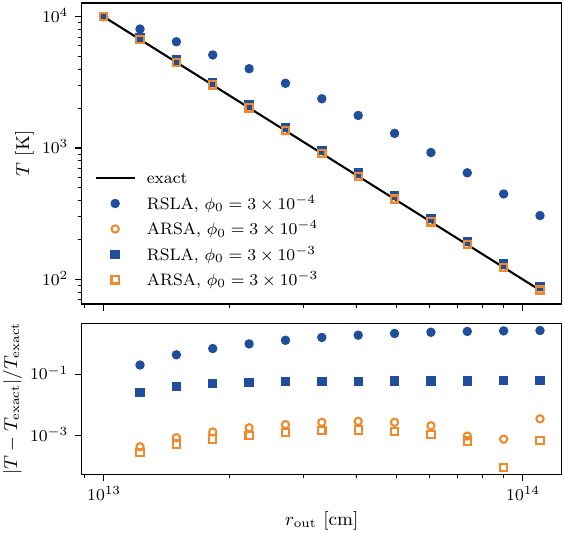}}
    \caption{Temperature as a function of outer radius in the opaque expanding sphere of \citet{Roth15a}, modified to use absorption rather than scattering opacity. The top panel shows the gas temperature, averaged over the domain, for the RSLA (filled) and the ARSA (open) at $\phi_0 = 3\times10^{-4}$ (circles) and $3\times10^{-3}$ (squares); the black line shows the exact solution (\autoref{eq:sphere_adiabat}). The bottom panel shows the corresponding relative error between the numerical and exact solutions.}
    \label{fig:expanding_sphere}
\end{figure}

The exact solution for the relationship between the gas temperature and the radius of the expanding sphere is the adiabat of gas and radiation: in a uniform flow the entropy per unit mass of the mixture is constant,
\begin{equation}
    \frac{3}{2}\ln T - \ln\rho + \frac{4 a_r T^3\mu m_{\rm H}}{3\rho k_B} = \mathrm{constant},
    \label{eq:sphere_adiabat}
\end{equation}
with $\rho \propto (t + t_0)^{-3}$. \autoref{fig:expanding_sphere} compares the exact result to the result of simulations using both the RSLA and the ARSA at reduction factors $\phi_0 = 3\times10^{-4}$ and $3\times10^{-3}$, giving $\hat{h}/\phi_0 = 0.85$ and $0.085$, respectively, at the start. Both values of $\phi_0$ lie above $\phi_* = 2.6\times10^{-4}$, and the cells lie far above $\tau_{\rm thresh} = 23$ and $3.8$, so the ARSA runs $\phi = 1$ in every cell at every step. The figure shows that at $r_{\rm out} = 11\,r_{\rm out,i}$ the RSLA leaves the gas $270\%$ and $6.2\%$ too hot, at $306$~K and $88$~K instead of the exact answer of $83$~K; this failure occurs because the RSLA raises the heat capacity of the radiation, which then reduces the rate at which the gas-radiation mixture cools adiabatically. The ARSA, by contrast, produces results within $0.35\%$ and $0.067\%$ of the exact solution, using the same time step as the RSLA; this remaining error, which the RSLA shares, comes from the discretization of the hydrodynamics.

\subsection{Radiative Shocks}
\label{ssec:test_ensman}

Radiative shocks test the coupling of matter and radiation where both change over a few cells. We use the subcritical and supercritical shocks of \citet{Ensman94a} as set up by \citet{MelonFuksman21a}: gas at $T = 10$~K with radiation in equilibrium, $\rho = 7.78\times10^{-10}\,\mathrm{g\,cm^{-3}}$, $\mu = 1$ and $\gamma = 7/5$, flowing onto a reflecting wall in a domain of $7\times10^{10}$~cm, at $6\,\mathrm{km\,s^{-1}}$ until $t = 3.8\times10^4$~s (subcritical) or at $20\,\mathrm{km\,s^{-1}}$ until $t = 7.5\times10^3$~s (supercritical). In the subcritical case we use their opacity, $\rho\kappa = 3.1\times10^{-10}\,\mathrm{cm^{-1}}$ everywhere. In the supercritical case we use a constant mass opacity ten times theirs, so that $\rho\kappa = 3.1\times10^{-9}\,(\rho/\rho_0)\,\mathrm{cm^{-1}}$, with $\rho_0$ the upstream density; this makes the upstream cells thick enough for the ARSA to raise $\phi$. Production runs resolve a shock by a few cells, so we use 64 cells along the flow. This gives cell optical depths $\tau_{\rm c} = 0.34$ and $3.4$ in the upstream regions of the two tests, respectively. In the dimensionless numbers of \autoref{sec:analysis}, the upstream gas has $\calC = c/a_\infty = 1.0\times10^6$ and $\calP = 1.2\times10^{-10}$, the Mach numbers are 18 and 59, and the domain is $\calL = \rho\kappa L = 22$ photon mean free paths long for the subcritical shock and 220 for the supercritical one.

We again run simulations using the full speed of light, and the RSLA and ARSA with $\phi_0 = 10^{-3}$ for the subcritical shock and $\phi_0 = 0.03$ for the supercritical one. In the subcritical shock $\phi_* = \hat{h}/(1 + \hat{h})$ lies far below $\phi_0$, so whether a cell reaches $\phi = 1$ is set by its optical depth; the threshold $\tau_{\rm thresh}$ of \autoref{eq:tau_one} is then close to $2\sqrt{3}$, which no cell reaches. In the supercritical shock the hottest gas of the precursor has $\phi_*$ above $\phi_0 = 0.03$, and there $\phi$ saturates by \autoref{eq:phi_sat}. The other cells have $\phi_* < \phi_0$. The upstream cells, the thinnest, lie just above $\tau_{\rm thresh}$ at $\phi_0 = 0.03$, so that 53 of the 64 cells satisfy both conditions for $\phi = 1$, but just below it at $\phi_0 = 0.011$, where almost none does.

\begin{figure}[t]
    \centerline{\includegraphics[width=0.85\textwidth]{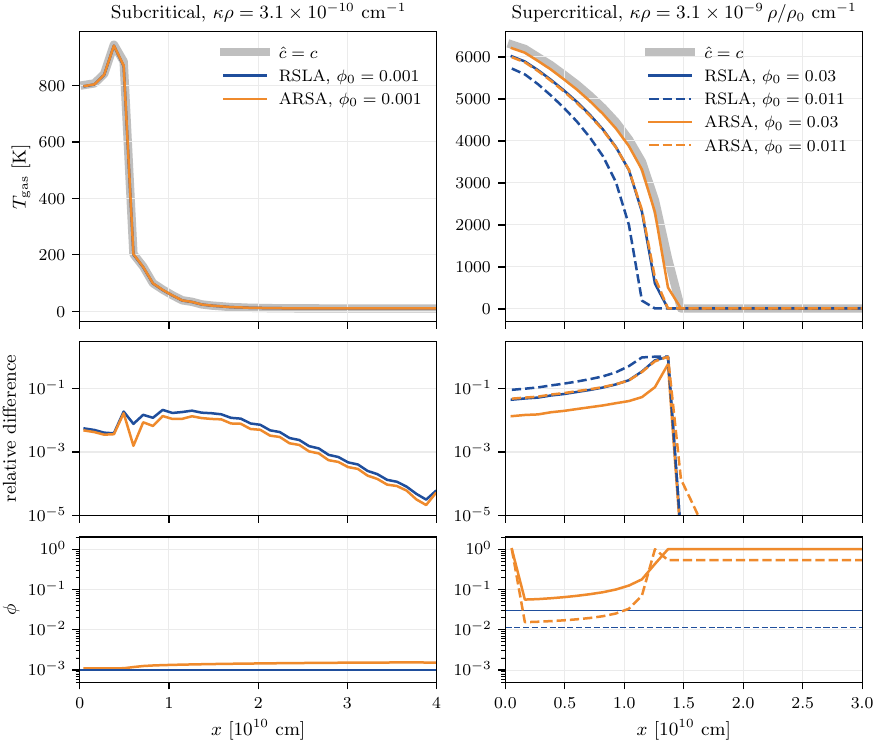}}
    \caption{Gas temperature versus position in the two radiative shock tests of \citet{MelonFuksman21a}; the left column shows the subcritical case, and the right the supercritical case.  The top row shows temperature, with gray showing the solution computed with no reduction in the speed of light, and blue and orange showing the classic RSLA and ARSA, respectively; solid lines show
    the RSLA and ARSA with $\phi_0 = 10^{-3}$ (left) and $\phi_0 = 0.03$ (right), and in the right panel dashed lines show $\phi_0 = 0.011$. The middle row shows the relative difference between the ARSA/RSLA cases and the reference solution with no reduction. The bottom panel shows the value of $\phi$ in each cell in the ARSA/RSLA cases.
    \label{fig:shocks_Tgas}
    }
\end{figure}

\begin{figure}[t]
    \centerline{\includegraphics[width=0.85\textwidth]{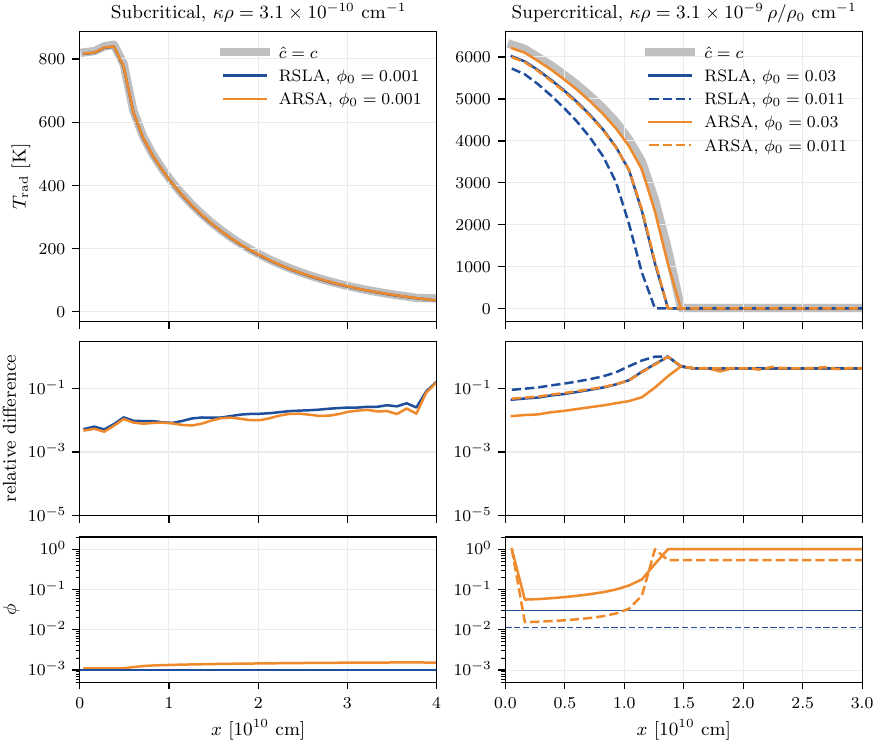}}
    \caption{Same as \autoref{fig:shocks_Tgas}, but for the radiation temperature.
    \label{fig:shocks_Trad}
    }
\end{figure}

\autoref{fig:shocks_Tgas} and \autoref{fig:shocks_Trad} show the results. In the subcritical shock the RSLA is already accurate, with an $L_1$ error of $0.78\%$ in the gas temperature at $\phi_0 = 10^{-3}$. The cells are too thin for $\phi = 1$, so the ARSA raises $\phi$ only to $0.0011$--$0.0015$, and its error is $0.62\%$. In the supercritical shock at $\phi_0 = 0.03$ the RSLA leaves the precursor behind, with an error of $14\%$. The ARSA has most of the domain at $\phi = 1$, so that the precursor is nearly right, and the $L_1$ error is $3.8\%$ for the same number of time steps. \new{The box gains energy through the inflow, so the total energy $E_{\rm gas} + E_{\rm rad}$ is not conserved. Therefore, we compare the total energy at the end with that of the reference: the RSLA misses $7.9\%$ of the energy that the box gains over the run, and the ARSA $2.3\%$.} The advantage of the ARSA can also be spent on speed: a bisection in $\phi_0$ finds that the ARSA at $\phi_0 = 0.011$ produces the same error as the RSLA at $\phi_0 = 0.03$, achieving the same accuracy with 2.6 times fewer steps.

\subsection{Multigroup Marshak Wave}
\label{ssec:test_marshak}

Our final test is the non-equilibrium multigroup Marshak wave of \citet{Vaytet11a}, in the version of \citet{He24b}: radiation from a wall at $1000$~K enters a static slab of density $10^{-3}$~g~cm$^{-3}$, initial temperature $300$~K and volumetric heat capacity $10^{-3}$~erg~cm$^{-3}$~K$^{-1}$, with opacity $\kappa = 2000\,(\nu/4\times10^{13}\,{\rm Hz})^{-2}$~cm$^2$~g$^{-1}$. Radiation is discretized into four logarithmically spaced frequency groups from $6\times10^{10}$ to $6\times10^{14}$~Hz. The problem domain is 20~cm in length and is resolved by 64 cells. We run the simulation to $t = 1.36\times10^{-7}$~s. The cell optical depth ranges from $\sim 10^4$ in the lowest group to $0.03$ in the highest, so every regime of \autoref{eq:phi_sb} occurs at once: the highest frequency group free streams and the lowest frequency group diffuses.

\begin{figure}[t]
    \centerline{\includegraphics[width=\textwidth]{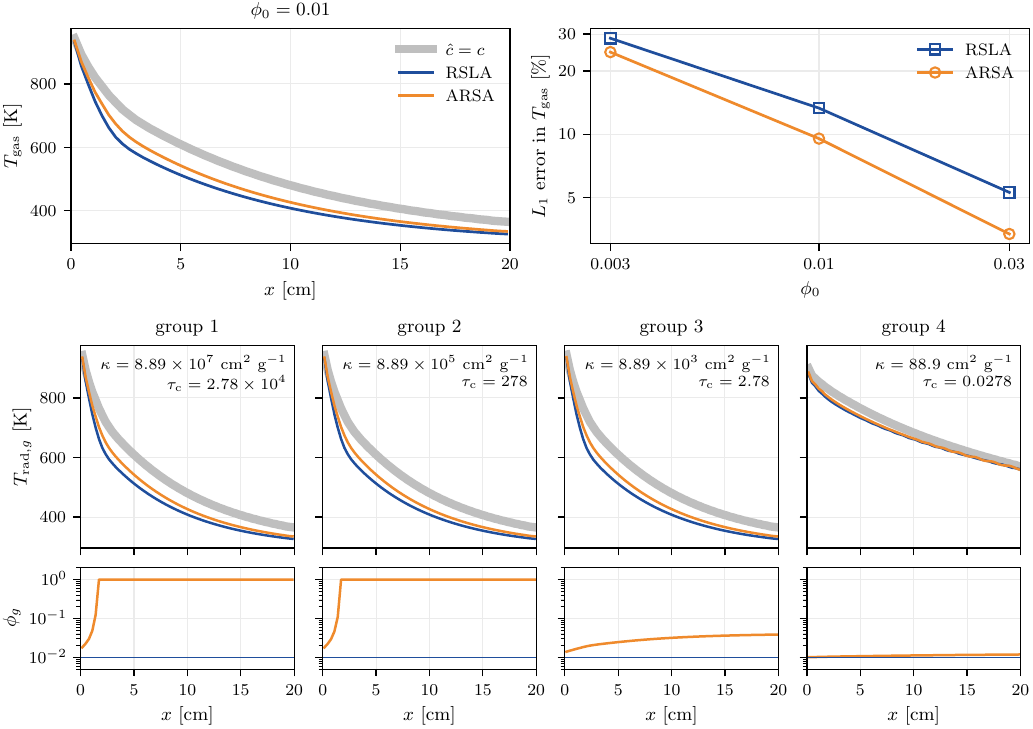}}
    \caption{Multigroup Marshak wave at $t = 1.36\times10^{-7}$~s. Top left: gas temperature of the reference with $\hat{c} = c$ (wide gray), the RSLA (blue) and the ARSA (orange), both at $\phi_0 = 0.01$ and the same time step. Top right: error of the gas temperature against $\phi_0$. Bottom: for each group, labeled with its opacity $\kappa$ and cell optical depth $\tau_{\rm c}$, the radiation temperature of the group, $a_r T_{{\rm rad},g}^4 f_g(T_{{\rm rad},g}) = E_g$ with $f_g$ the fraction of the Planck function in the group (upper panel) and the reduction factor of the ARSA (orange) and $\phi_0$ (blue) (lower panel), at $\phi_0 = 0.01$.
    \label{fig:marshak_vaytet}
    }
\end{figure}

We run the RSLA and the ARSA at $\phi_0 = 0.003$, $0.01$ and $0.03$, against a reference with $\hat{c} = c$ (\autoref{fig:marshak_vaytet}). We see that the RSLA retards the wave, with $L_1$ errors of $29\%$, $13\%$ and $5.3\%$. The ARSA reduces them to $25\%$, $9.5\%$ and $3.3\%$ with the same steps; to match them, the RSLA needs $1.4$, $1.5$ and $1.6$ times as many steps.

The gain is modest because the thin group 4 ($\tau_{\rm c} = 0.028$) carries over $96\%$ of the flux and $88\%$ of the heating of the gas, and there the stability bound keeps $\phi$ at $0.010$--$0.012$. Reducing $\hat{c}$ to $0.01\,c$ in this group alone, at the step of the reference, gives an error of $9.3\%$, close to that of the ARSA.

The thick groups carry little flux, but they stay at the gas temperature, so the RSLA adds their energy, raised by $1/\phi_0$, to the heat capacity of the gas; reducing $\hat{c}$ in groups 1--3 alone gives $5.9\%$. The ARSA removes most of this error: groups 1 and 2 run at $\phi = 1$ except in the five cells next to the wall, and group 3, whose cells ($\tau_{\rm c} = 2.8$) lie below $\tau_{\rm thresh}$, at $\phi = 0.014$--$0.039$. No local choice of $\phi$ can remove the error of the thin group at this step.

\section{Summary and Conclusion}
\label{sec:summary}

The reduced speed of light approximation (RSLA), in which the equations of RHD are simulated using a speed of light reduced by some factor $\phi \ll 1$ compared to the true one, fails in optically thick media because it scales the radiation exchange terms but not the matter ones. This raises the effective heat capacity of the radiation by $1/\phi$, which slows thermal relaxation, corrupts wave properties, and makes a flow carry temperature structures too slowly. In a system where gas holds most of the heat capacity, these failures occur where no reduction of the signal speed is needed for stability, because there the exchange with the gas damps the radiation modes. Here we present an adaptive reduced speed of light approximation (ARSA) that exploits this by setting a reduction factor $\phi$ in every cell close to the largest value that the scheme supports stably at the time step of the standard RSLA. Our main results are as follows.

\begin{enumerate}
    \item A von~Neumann analysis of the IMEX PD-ARS update with HLL transport allows us to express the local $\phi$ in closed form, \autoref{eq:bound_md}. The bound we derive is exact for the linearized scheme on a static background with diffusive radiation and piecewise-constant reconstruction, and all our tests, which use PPM, confirm that it carries over to higher-order reconstruction. We verify numerically that it changes by less than $1\%$ for a moving background and not at all for beamed radiation, so that a safety margin of $10\%$ guarantees stability.
    \item The bound we derive reduces to the CFL condition in transparent cells and permits $\phi = 1$, recovering the unmodified RHD equations, in cells more than a few photon mean free paths thick ($2\sqrt{D}$ in $D$ dimensions), provided that the gas holds most of the heat capacity. The standard RSLA satisfies this bound at a CFL number $\nu_0 = 1/\sqrt{D}$ in $D$ dimensions for every radiation state, opacity and heat capacity.
    \item We provide a numerical implementation that evaluates the maximum safe speed of light reduction factor in every cell (\autoref{eq:phi_sb}), and demonstrate that, when coupled to an upwind transport update with an implicit exchange update after every explicit stage, this method allows simulations to run at the same time step as the traditional RSLA while producing much more accurate results. We verify this improvement for a wide range of test problems.
\end{enumerate}

Our new scheme can be used as a drop-in replacement for the RSLA in any RHD code where radiation is handled by upwind transport with symmetric HLL wave speeds and an implicit exchange after every explicit stage, as in PD-ARS, with piecewise-constant or higher-order reconstruction: all our tests use PPM. 
Where the ARSA cannot raise $\phi$ it reduces to the RSLA, so it never performs worse, and evaluating $\phi$ is a local operation per cell and group, with no communication; replacing the RSLA with the ARSA is therefore an essentially cost-free gain in accuracy. In the tests where the ARSA reaches $\phi = 1$ in every opaque cell, it reaches an accuracy for which the RSLA would need a $\hat{c}$ up to 100 times larger. We therefore anticipate that the scheme will be widely useful in astrophysical applications.

\section*{Acknowledgements}

The authors acknowledge support from the Australian Research Council through Laureate Fellowship FL220100020 (to MRK). This research was supported by the Australian Government's National Collaborative Research Infrastructure Strategy (NCRIS) via the National Computational Merit Allocation Scheme (NCMAS) and ANU Merit Allocation Scheme (ANUMAS), award jh2.

We used Claude Code (Anthropic) to implement code from the authors' design, and Claude (Anthropic) and ChatGPT (OpenAI) to assist with mathematical derivations and proofs, including the formal proofs in Lean.
All code and data are public (\url{https://github.com/chongchonghe/Honghu.jl} and \url{https://github.com/chongchonghe/paper-ARSA}), and a single command rebuilds every figure and quoted number from them.
The authors reviewed and edited the code and the text and take full responsibility for the content of the published article.

\bibliographystyle{apsrev4-1}
\bibliography{main}

\begin{appendix}
\setcounter{prop}{0}
\setcounter{lem}{0}
\makeatletter
\@addtoreset{prop}{section}
\@addtoreset{lem}{section}
\makeatother
\renewcommand{\theprop}{\Alph{section}.\arabic{prop}}
\renewcommand{\thelem}{\Alph{section}.\arabic{lem}}

\section{Stability boundary of the energy block}
\label{app:proofs}

Here we derive the stability boundary of the energy block that \autoref{sssec:stabilty_checkerboard} uses, from the amplification matrix $\tenG_E$ of \autoref{eq:G_energy}. We first show that the scheme is stable at $a = 0$, i.e., without transport. Then $\tenD = \tenI$, and writing $\tfrac{1}{2}\tenI = \tfrac{1}{2}\tenB^{-1}\tenB$ in \autoref{eq:G_energy} gives
\begin{equation}
    \tenG_E\big|_{a = 0}
    = \left(\tenI - \tfrac{1}{2}\tenS^{(E)}\right)^{-1}\tfrac{1}{2}\tenB^{-1}\left(\tenB + \tenI\right)
    = \left(\tenI - \tfrac{1}{2}\tenS^{(E)}\right)^{-1}\tenB^{-1}\left(\tenI - \tfrac{1}{2}\tenS^{(E)}\right)
    = \tenB^{-1},
    \label{eq:G_a0}
\end{equation}
where the last step uses that $\tenB^{-1}$ and $\tenI - \tfrac{1}{2}\tenS^{(E)}$, both functions of $\tenS^{(E)}$, commute. Since $\mathrm{tr}\,\tenB = 2 + q + \lambda q$ and $\det\tenB = 1 + q + \lambda q$, the eigenvalues of $\tenB$ are $1$ and $1 + q + \lambda q$, so those of $\tenG_E$ are $g = 1$, the equilibrium $\delta\hE = 4\,\delta\hT$, and $g = 1/(1 + q + \lambda q) < 1$, the decaying departure from it. Neither exceeds unity, so the scheme is stable at $a = 0$: without transport, the two stages act as a single backward-Euler exchange step, which is unconditionally stable.

As $a$ increases from zero, an eigenvalue $g$ of $\tenG_E$ can leave the unit circle in three ways: through $g = +1$, through $g = -1$, or as a complex-conjugate pair with $|g| = 1$. The following proposition states where this happens.

\begin{prop}
\label{prop:energy_block}
Let $q, \lambda > 0$ and $a \geq 0$. All eigenvalues of $\tenG_E$ satisfy $|g| \leq 1$ if and only if $a \leq a_*$, where
\begin{equation}
    a_* \equiv \frac{2\left(2 + q + \lambda q\right)}{2 + \lambda q} .
    \label{eq:boundary_a}
\end{equation}
\end{prop}

We prove Proposition~\ref{prop:energy_block} in two steps. First, the Schur--Cohn--Jury criterion \citep{Jury64a} reduces $|g| \leq 1$ to three scalar conditions, one for each way out of the unit circle. Second, three lemmas check these conditions for $\tenG_E$: the second and third hold for every $a \leq a_*$, so neither a crossing of $g = -1$ nor a complex pair can come first, and the first fails exactly when $a$ exceeds $a_*$, where a real eigenvalue leaves the unit circle through $g = +1$.

\begin{lem}[Schur--Cohn--Jury criterion for $2\times2$ matrices]
\label{lem:jury}
Let $\tenG$ be a real $2 \times 2$ matrix. Both eigenvalues of $\tenG$ satisfy $|g| \leq 1$ if and only if
\begin{equation}
    \mathrm{(i)}\;\; \det\left(\tenI - \tenG\right) \geq 0,
    \qquad
    \mathrm{(ii)}\;\; \det\left(\tenI + \tenG\right) \geq 0,
    \qquad
    \mathrm{(iii)}\;\; \left|\det\tenG\right| \leq 1 .
    \label{eq:jury}
\end{equation}
\end{lem}

Condition (i) fails when a real eigenvalue crosses $g = +1$, condition (ii) when one crosses $g = -1$, and condition (iii) when a complex pair crosses $|g| = 1$, since then $|g|^2 = \det\tenG$.

We now evaluate the three conditions for $\tenG_E$. Two properties of the matrices in \autoref{eq:G_energy} are used throughout.

First, $\det\tenB = 1 + q + \lambda q > 0$, as found above.

Second, the matrix $\tenH \equiv \tenI - \tfrac{1}{2}\tenS^{(E)}$ is invertible, with $\det\tenH = 1 + \tfrac{1}{2}q(1 + \lambda) > 0$: the eigenvalues of $\tenS^{(E)}$ are $0$ and $-q(1 + \lambda)$ (its determinant vanishes and its trace is $-q(1 + \lambda)$), so those of $\tenH$ are $1$ and $1 + \tfrac{1}{2}q(1 + \lambda)$.

To evaluate condition (i), note that for a $2\times2$ matrix $\det(\tenI - \tenG_E) = \det(\tenG_E - \tenI)$. Multiplying $\tenG_E - \tenI$ from the left by $\tenH$ gives
\begin{equation}
    \tenH\left(\tenG_E - \tenI\right) = \tfrac{1}{2}\left(\tenD\,\tenB^{-1}\tenD - \tenB\right),
    \label{eq:GminusI}
\end{equation}
so that $\det(\tenG_E - \tenI) = \det(\tenD\,\tenB^{-1}\tenD - \tenB)/(4\det\tenH)$. The determinant on the right factorizes.

\begin{lem}
\label{lem:det_split}
Let $\tenB$ and $\tenD$ be $n \times n$ matrices with $\tenB$ invertible. Then
\begin{equation}
    \det\left(\tenD\,\tenB^{-1}\tenD - \tenB\right) = (-1)^n\,\frac{\det(\tenB - \tenD)\,\det(\tenB + \tenD)}{\det\tenB} .
    \label{eq:det_split}
\end{equation}
\end{lem}

\begin{pf}
Expanding the product shows that
\begin{equation}
    \left(\tenD - \tenB\right)\tenB^{-1}\left(\tenD + \tenB\right)
    = \tenD\,\tenB^{-1}\tenD + \tenD\,\tenB^{-1}\tenB - \tenB\,\tenB^{-1}\tenD - \tenB\,\tenB^{-1}\tenB
    = \tenD\,\tenB^{-1}\tenD - \tenB .
    \label{eq:factor_DB}
\end{equation}
Taking the determinant of both sides, and using $\det(\tenD - \tenB) = (-1)^n\det(\tenB - \tenD)$, gives \autoref{eq:det_split}. \qed
\end{pf}

\begin{lem}
\label{lem:cond_i}
Let $q, \lambda > 0$ and $a \geq 0$. Condition (i) of \autoref{eq:jury} holds for $\tenG_E$ if and only if $a \leq a_*$.
\end{lem}

\begin{pf}
From \autoref{eq:D_energy} and \autoref{eq:G_energy},
\begin{equation}
    \begin{split}
    \det\left(\tenB - \tenD\right) &= \lambda q\, a, \\
    \det\left(\tenB + \tenD\right) &= \left(2 + \lambda q\right)\left(2 + q - a\right) - \lambda q^2 = \left(2 + \lambda q\right)\left(a_* - a\right).
    \end{split}
    \label{eq:dets_pm}
\end{equation}
With \autoref{eq:GminusI} and Lemma~\ref{lem:det_split} for $n = 2$,
\begin{equation}
    \det\left(\tenI - \tenG_E\right) = \frac{\det\left(\tenD\,\tenB^{-1}\tenD - \tenB\right)}{4\det\tenH}
    = \frac{\lambda q\left(2 + \lambda q\right)}{4\det\tenB\,\det\tenH}\; a\left(a_* - a\right).
    \label{eq:detGminusI}
\end{equation}
The prefactor is positive, so $\det(\tenI - \tenG_E)$ has the sign of $a(a_* - a)$, which for $a \geq 0$ is non-negative if and only if $a \leq a_*$. \qed
\end{pf}

Condition (i) holds with equality at $a = 0$ and $a = a_*$, where $g = 1$ is an eigenvalue. At $a = 0$ this is the neutral mode of \autoref{eq:G_a0}: without transport, the equilibrium $\delta\hE = 4\,\delta\hT$, the null vector of \autoref{eq:SE}, is left unchanged by every stage. At $a = a_*$ a real eigenvalue leaves the unit circle through $g = +1$.

\begin{lem}
\label{lem:cond_ii}
Let $q, \lambda > 0$. Condition (ii) of \autoref{eq:jury} holds for $\tenG_E$ for every $a$, with strict inequality.
\end{lem}

\begin{pf}
Evaluating the determinant from \autoref{eq:G_energy} gives
\begin{equation}
    \begin{split}
    \det\left(\tenI + \tenG_E\right) &= \frac{N(a)}{4\det\tenB\,\det\tenH}, \\
    N(a) &= \left(2 + \lambda q\right)^2 a^2 - 2\left(\lambda^2 q^2 + \lambda q^2 + 4\lambda q + 4\right) a + 4\left(2 + q + \lambda q\right)^2 .
    \end{split}
    \label{eq:detGplusI}
\end{equation}
$N$ is a quadratic in $a$ with a positive leading coefficient and discriminant
\begin{equation}
    \tfrac{1}{4}\,\mathrm{disc}\,N = -\left(\lambda^2 q^2 + \lambda q^2 + 4\lambda q + 4q + 4\right)\left(3\lambda^2 q^2 + 3\lambda q^2 + 12\lambda q + 4q + 12\right) < 0 ,
    \label{eq:discN}
\end{equation}
so $N(a) > 0$ for every real $a$. The denominator is positive, so $\det(\tenI + \tenG_E) > 0$. \qed
\end{pf}

\begin{lem}
\label{lem:cond_iii}
Let $q, \lambda > 0$. For $0 \leq a \leq a_*$, $0 < \det\tenG_E < 1$, so condition (iii) of \autoref{eq:jury} holds for $\tenG_E$.
\end{lem}

\begin{pf}
Evaluating the determinant from \autoref{eq:G_energy} gives
\begin{equation}
    \det\tenG_E = \frac{\left(2 + \lambda q\right)\left(1 - a\right)^2 + 2 + 2q + \lambda q}{2\left(1 + q + \lambda q\right)\left(2 + q + \lambda q\right)} ,
    \label{eq:detG_energy}
\end{equation}
which is positive for every $a$ since its numerator and denominator are, and grows with $(1 - a)^2$. For $0 \leq a \leq a_*$, $1 - a$ lies between $1 - a_*$ and 1, and $a_* - 1 = (2 + 2q + \lambda q)/(2 + \lambda q)$ exceeds 1, so that $(1 - a)^2 \leq (a_* - 1)^2$: the determinant is largest at $a = a_*$, where it equals
\begin{equation}
    \det\tenG_E\big|_{a = a_*} = \frac{2 + 2q + \lambda q}{\left(2 + \lambda q\right)\left(1 + q + \lambda q\right)} < 1,
    \label{eq:detG_astar}
\end{equation}
since $(2 + \lambda q)(1 + q + \lambda q) - (2 + 2q + \lambda q) = \lambda q\,(2 + q + \lambda q) > 0$. \qed
\end{pf}

Condition (iii) can fail for $a > a_*$, but there condition (i) has already failed.

\begin{pfprop}
$\tenG_E$ is real, so Lemma~\ref{lem:jury} applies. If $0 \leq a \leq a_*$, conditions (i), (ii) and (iii) hold by Lemmas~\ref{lem:cond_i}, \ref{lem:cond_ii} and \ref{lem:cond_iii}, so $|g| \leq 1$. If $a > a_*$, condition (i) fails by Lemma~\ref{lem:cond_i}, so an eigenvalue has $|g| > 1$. \qed
\end{pfprop}

\section{Full M1 in the diffusion regime}
\label{app:m1}

In the main text we derive a stability limit for transport assuming a fixed Eddington tensor. However, under M1 the tensor depends on the flux, so that its derivatives enter the flux Jacobian, and about a background with a flux the exchange couples the radiation energy to the gas velocity. Here we show that the bound derived in the main text, \autoref{eq:bound_md}, nevertheless remains a stability condition for the complete M1 update where the radiation is diffusive, which is where the ARSA raises $\phi$ (\autoref{ssec:irsla_phi}). We state the result as Proposition~\ref{prop:m1} and outline its proof. The full proof, and its extension to unequal mean opacities, are machine-checked in Lean~4 \citep{deMoura21a, mathlib20a}, given the explicit change of variables that makes the isotropic problem symmetric; so are the results of \ref{app:proofs} and \ref{app:theta}, Lemma~\ref{lem:strict} in one dimension. The formal proofs are in the code and data repository of this paper, \url{https://github.com/chongchonghe/paper-ARSA}.

\paragraph{Setting} We linearize the radiation step about a uniform background at rest, with radiation energy density $E$, flux $\vecF$, reduced flux $f \equiv |\vecF|/(cE) < 1$ and the Eddington tensor of the M1 closure \citep{Levermore84a}. Unlike in \ref{app:proof}, we keep every term of the linearization: the derivatives of the Eddington tensor in the flux Jacobian, the coupling of the radiation energy to the gas velocity through the background flux, and the drag on the gas velocity. The gas need not be in exact energy balance with the radiation. The opacities are frozen, the HLL speeds are those of \autoref{eq:hll_symbol} with $\sigma_d = \max(0.1, T_{dd}^{1/2})$, and the amplification matrix $\tenG_{\boldsymbol{\theta}}$ is that of \autoref{eq:ampl}.

\paragraph{The premise: diffusive radiation} The result assumes that the radiation is diffusive, i.e., that its flux is set by a gradient over a macroscopic length. In the diffusion limit $\vecF \approx -c\nabla E/(3\rho\kappa_F)$, so that $\rho\kappa_F f \approx |\nabla E|/(3E)$. Accordingly, we assume that
\begin{equation}
    \rho\kappa_F f \leq \frac{1}{\ell},
    \qquad
    \rho\kappa_F\,\Delta x_d \geq \tau_* > 0
    \label{eq:diffusive}
\end{equation}
in every direction $d$, with a fixed macroscopic length $\ell$ and a fixed $\tau_*$. The first condition bounds the reduced flux by the inverse of the optical depth of the length $\ell$; the second excludes cells of vanishing optical depth. A cell can be optically thick and still violate the first condition, for example where a directed beam is absorbed; such cells lie outside the result (see the end of this appendix).

\paragraph{What stability means here} For a fixed Eddington tensor, \ref{app:proof} shows that the step never increases a fixed norm, so that the powers $\tenG^n$ stay bounded for all $n$. Under M1 no such bound holds in general, because the linearized equations about some anisotropic backgrounds have growing modes themselves (see below). The appropriate notion is the classical one of \citet{Lax56a}: a perturbation may grow, but at a rate per unit of physical time that does not depend on the grid,
\begin{equation}
    \left\|\tenG_{\boldsymbol{\theta}}^n\right\| \leq A\, e^{\Gamma\,\hat{c}\,n\Delta t/\ell} \quad \mbox{for every}\; n \;\mbox{and every}\; \boldsymbol{\theta},
    \label{eq:finite_time}
\end{equation}
with $\ell$ the length of \autoref{eq:diffusive} and constants $A$ and $\Gamma$ that do not depend on the cell width, the time step, the opacity, the mode or the direction of the flux. Over a fixed time, a perturbation then grows by at most a fixed factor, however fine the grid.

\paragraph{The result} With these definitions the result reads as follows.

\begin{prop}
\label{prop:m1}
Consider the radiation step \autoref{eq:pdars} with M1 transport and exchange, linearized about a uniform background at rest as described above, with frozen opacities. Assume that
\begin{enumerate}
    \item the radiation is diffusive, \autoref{eq:diffusive}, and its reduced flux satisfies $f \leq f_* < 1$;
    \item the ratios $\lambda$ and $\lambda_F$ of \autoref{eq:SE} and \autoref{eq:SF}, the ratios of the three mean opacities and the ratio $a_r T^4/E$ are fixed, and
    \begin{equation}
        \gamma_{\rm d} \equiv 1 + \frac{\kappa_P\,a_r T^4 - \kappa_E E}{\kappa_F E} \geq 0 ,
        \label{eq:gamma_drag}
    \end{equation}
    so that the drag on the gas does not change sign;
    \item the step satisfies both conditions of \autoref{eq:bound_md} with the isotropic signal speed, $\sigma_d = 1/\sqrt{3}$, in every direction, and with $\gamma_{\rm d} + 1/3$ in place of $1 + \chi$ in $K_F$:
    \begin{equation}
        \nu\left[\frac{D}{\sqrt{3}} - \frac{\tau_E}{2 + K}\right] \leq 1,
        \qquad
        \nu\left[\frac{D}{\sqrt{3}} - \frac{\tau_F}{2 + K_F}\right] \leq 1 .
        \label{eq:bound_iso}
    \end{equation}
\end{enumerate}
Then there are constants $A$ and $\Gamma$ such that \autoref{eq:finite_time} holds for every $n$ and every mode $\boldsymbol{\theta}$. They depend on the fixed quantities of assumptions 1 and 2 and on the number of dimensions $D$, but not on the cell width, the time step, the scale of the opacities, the mode or the direction of the flux.
\end{prop}

The quantity $\gamma_{\rm d}$ measures the balance of emission and absorption. In energy balance, $\kappa_P a_r T^4 = \kappa_E E$, it equals one, $\gamma_{\rm d} + 1/3 = 1 + \chi$ for $\chi = 1/3$, and \autoref{eq:bound_iso} is \autoref{eq:bound_md} for isotropic radiation; assumption 2 then only asks that the response of the gas be fixed. On non-cubic grids, $\nu D/\sqrt{3}$ becomes $\sum_d\nu_d/\sqrt{3}$.

The proposition uses the isotropic signal speed, whereas the limiter \autoref{eq:phi_sb} uses the actual $\sigma_d$. In the diffusion limit $T_{dd} = 1/3 + \epsilon_d$ with $\epsilon_d = O(f^2)$. In three dimensions the unit trace of the Eddington tensor gives $\sum_d\epsilon_d = 0$, so the first-order term cancels and $\sum_d\sigma_d = \sqrt{3} - (3\sqrt{3}/8)\sum_d\epsilon_d^2 + O(\epsilon^3)$: the two differ only at fourth order in $f$. In one or two dimensions they differ at second order. A code that wants the proposition to apply exactly can use $\max(\sum_d\sigma_d, D/\sqrt{3})$ in the limiter. This costs nothing in thick cells, where $\sum_d\sigma_d \approx D/\sqrt{3}$, and gives $\phi = \phi_0$ in thin cells, removing only the small rise of $\phi$ in thin beamed cells. Like the analysis of \autoref{ssec:von_neumann}, the proposition is a statement about a fixed $\phi$ and a frozen background; it does not cover changes of $\phi$ between cells or steps.

\begin{pfm1}
The proof compares the M1 step $\tenG$ with the step $\tenG_0$ of the isotropic problem at the same $\nu$ and $q$, in three steps.
\begin{enumerate}
    \item The isotropic step is bounded uniformly in the opacity. \ref{app:proof} gives a norm that $\tenG_0$ never increases, but this norm contains the optical depth that the reduced light crosses in a step, $q$, and is not uniform in it. Bounding the first step separately turns it into $\|\tenG_0^n\| \leq A_0$ in a fixed norm, with $A_0$ independent of $q$.
    \item M1 changes the exchange and the transport by amounts proportional to $f$: the coupling of the energy to the velocity is proportional to $f$, and the flux Jacobian and the HLL speeds differ from their isotropic values by $O(f)$. The implicit exchange solves stay bounded however large $q$ is, because the thermal exchange and the drag only damp. Hence $\|\tenG - \tenG_0\| \leq c_1\,q_F f + c_2\,f\sum_d\nu_d$, with $q_F = \hat{c}\,\Delta t\,\rho\kappa_F$ and fixed $c_1$ and $c_2$.
    \item By \autoref{eq:diffusive}, $q_F f \leq \hat{c}\,\Delta t/\ell$ and $f\nu_d \leq \hat{c}\,\Delta t/(\ell\tau_*)$: the difference per step is proportional to the physical time step. Writing $\tenG^n$ as $\tenG_0^n$ plus the differences accumulated over the steps then gives \autoref{eq:finite_time}. This is the discrete form of the statement that a bounded growth rate gives a bounded growth over a fixed time. \qed
\end{enumerate}
\end{pfm1}

The constants that the proof gives are far from sharp. It shows that $A$ and $\Gamma$ exist independently of the grid and the opacity, which is what stability requires, but not that they are small. The proposition is therefore a qualitative result: it does not by itself justify the margin $\beta = 0.9$, which rests on the scans of \ref{app:numerics}, including anisotropic M1 states, and on the tests of \autoref{sec:tests}.

\paragraph{Strongly directed radiation} Assumption 1 of Proposition~\ref{prop:m1} cannot simply be dropped. Consider a flux along a grid axis with $f > 2\sqrt{3}/5 \approx 0.69$, and a gas whose inertia responds weakly to the radiation compared with its heat capacity, as for any gas whose sound speed is much below $c$. The exchange then shares a long-wavelength perturbation between the gas and the radiation in a way that turns the two real characteristic speeds of M1 into a complex pair, and a mode of wavenumber $k$ grows at a rate proportional to $\hat{c}k$ in the linearized equations themselves, before any discretization. An example is $f = 80/91$ with $\hat{h} = \phi = 1$ and an isothermal sound speed of $c/61$. At a fixed phase per cell, the growth rate then scales as $1/\Delta x$ at a fixed cell optical depth, so that refining the grid admits ever faster growth and no condition on the time step gives \autoref{eq:finite_time}. The background of this example is not steady, since its flux pushes the gas, and the growth belongs to the linearization about a frozen state; whether it survives in a background that evolves is an open question. It shows, however, that for strongly directed radiation in thick cells no condition on the time step alone gives a result uniform in the opacity.

\section{The checkerboard sets the stability boundary}
\label{app:theta}

Here we show that the von Neumann stability of the update scheme described in \autoref{ssec:von_neumann} is determined by the stability of the checkerboard -- proving the result exactly for a static background with isotropic radiation (\ref{app:proof}) and demonstrating numerically that it holds to good approximation for moving background and anisotropic M1 states (\ref{app:numerics}).

\subsection{Proof for a static background and isotropic radiation}
\label{app:proof}

We consider the amplification matrix \autoref{eq:ampl} at $\calM = 0$ on a grid with $D$ active dimensions. The density perturbation is neither transported nor exchanged in the radiation step, so its eigenvalue is exactly unity, and we drop it. The remaining variables are the gas perturbations $(\delta\hT, \delta\hv_1, \ldots, \delta\hv_D)$ and the radiation perturbations $(\delta\hE, \delta\hF_1, \ldots, \delta\hF_D)$. As in \autoref{eq:a_md}, we take the Eddington tensor to be diagonal along the grid axes, with signal speed $\sigma_d$ along direction $d$, and the flux along $d$ to be that of a fixed Eddington tensor, $\hat{P}_{dd} = \sigma_d^2\hE$, as for the Eddington closure and for M1 about an isotropic state. The flux along $d$ then links $\delta\hE$ only with $\delta\hF_d$, with characteristic speeds $\pm\sigma_d$. A Fourier mode has the phase change $\theta_d$ per cell in direction $d$, and we write $\boldsymbol{\theta} = (\theta_1, \ldots, \theta_D)$.

At $\calM = 0$ the exchange splits into separate channels: an energy channel, which couples $\delta\hT$ with $\delta\hE$ through the block \autoref{eq:SE}, and one momentum channel per direction, which couples $\delta\hv_d$ with $\delta\hF_d$ through the block \autoref{eq:SF}. For each channel $k$ let $q_k$ and $K_k$ be its values of $q$ and $K$ (for the energy channel $q$ and $K$, for a momentum channel $q_F$ and $K_F$ of \autoref{eq:KF}, with the optical depths $\tau_E$, $\tau_P$ and $\tau_F$ for a non-gray material), and let
\begin{equation}
    a_*^k \equiv \frac{2\left(2 + q_k + K_k\right)}{2 + K_k}
    \label{eq:astar_channel}
\end{equation}
be the value for that channel. With $a_d \equiv 2\nu_d\sigma_d$, the transport factor of the checkerboard in every direction is $a \equiv \sum_d a_d$, as in \autoref{eq:a_md}. At the checkerboard, $\theta_d = \pi$ for all $d$, the amplification matrix splits into the blocks of the channels, each with transport factor $1 - a$, so that by Proposition~\ref{prop:energy_block}, applied to each channel, the checkerboard is stable if and only if $a \leq a_*^k$ for every $k$. The following proposition states that this is also the stability condition of every other mode.

\begin{prop}
\label{prop:checkerboard}
Let $\calM = 0$. All eigenvalues of $\tenG_{\boldsymbol{\theta}}$ satisfy $|g| \leq 1$ for every $\boldsymbol{\theta}$ if and only if $a \leq a_*^k$ for every channel $k$. If this holds, there is a weighted norm $\|\cdot\|_\star$, the same for every $\boldsymbol{\theta}$, that no mode can increase: $\|\tenG_{\boldsymbol{\theta}}\,\mathbf{x}\|_\star \leq \|\mathbf{x}\|_\star$ for every perturbation $\mathbf{x}$.
\end{prop}

\paragraph{Notation} The proof uses a few standard terms of linear algebra, which we collect here. For a complex vector or matrix, the asterisk denotes the complex conjugate of the transpose, so that $\mathbf{x}^*\mathbf{x} = \sum_i |x_i|^2$. A matrix $\mathsf{A}$ with $\mathsf{A}^* = \mathsf{A}$, which for a real matrix means a symmetric one, has real eigenvalues. It is positive semidefinite if these are all non-negative, which is the same as $\mathbf{x}^*\mathsf{A}\,\mathbf{x} \geq 0$ for every vector $\mathbf{x}$, and positive definite if they are all positive. We compare two such matrices through their quadratic forms: $\mathsf{A} \preceq \mathsf{B}$ means $\mathbf{x}^*\mathsf{A}\,\mathbf{x} \leq \mathbf{x}^*\mathsf{B}\,\mathbf{x}$ for every $\mathbf{x}$, i.e., that $\mathsf{B} - \mathsf{A}$ is positive semidefinite. A positive definite matrix $\mathsf{H}$ defines a weighted norm, $\|\mathbf{x}\|_{\mathsf{H}} \equiv (\mathbf{x}^*\mathsf{H}\,\mathbf{x})^{1/2}$, which plays the role of an energy of the perturbation $\mathbf{x}$. A step $\mathbf{x} \to \tenG\mathbf{x}$ does not increase it if $\|\tenG\mathbf{x}\|_{\mathsf{H}} \leq \|\mathbf{x}\|_{\mathsf{H}}$ for every $\mathbf{x}$, i.e., if $\tenG^*\mathsf{H}\,\tenG \preceq \mathsf{H}$. Every eigenvalue then has $|g| \leq 1$. The norm bound is in fact stronger than this: it also prevents the slow, linear-in-time growth that two coinciding eigenvalues on the unit circle can cause when they lack independent eigenvectors.

\paragraph{Symmetric form} We first rescale each variable by a positive constant: $\delta\hF_d$ by $1/\sigma_d$, $\delta\hT$ by $4/\lambda^{1/2}$, and $\delta\hv_d$ by $(1 + T_{dd})/(\sigma_d\calC\lambda_{F,d}^{1/2})$, with $\lambda$ from \autoref{eq:SE} and $\lambda_{F,d}$ the value of $\lambda_F$ of \autoref{eq:SF} for direction $d$. A rescaling by constants changes no eigenvalue. In the new variables the flux Jacobian along $d$ becomes $\sigma_d\tenX_d$, where $\tenX_d$ is the real symmetric matrix that swaps $\delta\hE$ and $\delta\hF_d$, and the exchange over one step becomes $\tenS = -\tenQ$, with
\begin{equation}
    \tenQ = \sum_k \mathbf{u}_k\mathbf{u}_k^{\rm T},
    \qquad
    \mathbf{u}_k = \left(K_k^{1/2},\; -q_k^{1/2}\right) \;\mbox{on the gas and radiation variable of channel}\; k .
    \label{eq:Q_sym}
\end{equation}
$\tenQ$ is real, symmetric and positive semidefinite, since $\mathbf{x}^*\tenQ\,\mathbf{x} = \sum_k |\mathbf{u}_k^{\rm T}\mathbf{x}|^2$: the exchange only removes energy in this norm. It vanishes exactly on the exchange equilibrium. The rescaling is thus a symmetrizer in the sense of \citet{Yong99a}: it makes the flux Jacobians symmetric and the exchange symmetric and negative semidefinite, with the equilibria as its kernel. The argument below uses these properties and the structure of the transport, which acts only on the radiation variables, with speeds $\pm\sigma_d$; it therefore applies unchanged to any linear relaxation system with the same structure. The transport over one step, \autoref{eq:hll_symbol} summed over directions, becomes $\tenZ_{\boldsymbol{\theta}} = -\tenW$, with
\begin{equation}
    \tenW(\boldsymbol{\theta}) = \sum_d \frac{a_d}{2}\left[\zeta_d\,\tenI_{\rm r} + i\sin\theta_d\,\tenX_d\right],
    \qquad \zeta_d \equiv 1 - \cos\theta_d \in [0, 2],
    \label{eq:W_sym}
\end{equation}
where $\tenI_{\rm r}$ is the identity on the radiation variables and zero on the gas variables. In one dimension, $\tenI_{\rm r} - (2/a)\tenW = \cos\theta\,\tenI_{\rm r} - i\sin\theta\,\tenX$ preserves the length $\mathbf{x}^*\mathbf{x}$ of the radiation part of every vector: it is the exact shift by one cell, i.e., upwinding at CFL number one. The transport stage $\tenI - \tenW$ is thus the average, with weights $1 - a/2$ and $a/2$, of the identity and an exact shift. With $\tenB \equiv \tenI + \tenQ = \tenI - \tenS$, as in \autoref{eq:G_energy}, $\tenD \equiv \tenI - \tenW$, and $\tenI - \tfrac{1}{2}\tenS = \tfrac{1}{2}(\tenI + \tenB)$, the amplification matrix \autoref{eq:ampl} reads
\begin{equation}
    \tenG = \left(\tenI + \tenB\right)^{-1}\left(\tenI + \tenD\,\tenB^{-1}\tenD\right).
    \label{eq:G_sym}
\end{equation}
Here $\tenB$ is positive definite, so it defines the norm $\|\mathbf{x}\|_{\tenB}$, and so is $\tenI + \tenB$.

\paragraph{Two lemmas} The proof has two steps. One stage of the scheme, \autoref{eq:stage1}, is a transport step followed by an implicit exchange solve, $\mathbf{x} \to \tenB^{-1}\tenD\,\mathbf{x}$. The first lemma shows that if one such stage does not increase the norm $\|\cdot\|_{\tenB}$, the full two-stage step does not increase the norm $\|\cdot\|_{\tenI + \tenB}$. The second lemma shows that the first condition holds for every $\boldsymbol{\theta}$ exactly when the checkerboard is stable.

\begin{lem}
\label{lem:two_stage}
If one stage, $\mathbf{x} \to \tenB^{-1}\tenD\,\mathbf{x}$, does not increase the norm $\|\cdot\|_{\tenB}$, then the full step $\mathbf{x} \to \tenG\,\mathbf{x}$ of \autoref{eq:G_sym} does not increase the norm $\|\cdot\|_{\tenI + \tenB}$.
\end{lem}

\begin{pf}
Let $\mathbf{y} \equiv \tenG\mathbf{x}$ be the result of the full step and $\mathbf{z} \equiv (\tenB^{-1}\tenD)^2\mathbf{x}$ that of two successive stages. Since $\tenD\,\tenB^{-1}\tenD = \tenB\,(\tenB^{-1}\tenD)^2$, \autoref{eq:G_sym} reads $(\tenI + \tenB)\,\mathbf{y} = \mathbf{x} + \tenB\,\mathbf{z}$. Expanding the squared norms and using this relation gives
\begin{equation}
    \|\mathbf{x}\|_{\tenI + \tenB}^2 - \|\mathbf{y}\|_{\tenI + \tenB}^2 = |\mathbf{x} - \mathbf{y}|^2 + \|\mathbf{z} - \mathbf{y}\|_{\tenB}^2 + \|\mathbf{x}\|_{\tenB}^2 - \|\mathbf{z}\|_{\tenB}^2 ,
    \label{eq:energy_identity}
\end{equation}
where $|\cdot|$ is the ordinary length: the cross terms on the right add up to $-2\,\mathrm{Re}[\mathbf{y}^*(\mathbf{x} + \tenB\,\mathbf{z})] = -2\|\mathbf{y}\|_{\tenI + \tenB}^2$. The hypothesis, applied to two successive stages, gives $\|\mathbf{z}\|_{\tenB} \leq \|\mathbf{x}\|_{\tenB}$, so every term on the right is non-negative, and $\|\mathbf{y}\|_{\tenI + \tenB} \leq \|\mathbf{x}\|_{\tenI + \tenB}$. \qed
\end{pf}

\autoref{eq:energy_identity} also shows that a mode whose norm the full step leaves unchanged is left unchanged itself: then every term on the right vanishes, the first one in particular, so that $\tenG\mathbf{x} = \mathbf{x}$.

\begin{lem}
\label{lem:stage}
One stage does not increase the norm $\|\cdot\|_{\tenB}$, for every $\boldsymbol{\theta}$, if and only if $a \leq a_*^k$ for every channel $k$.
\end{lem}

\begin{pf}
The condition is that $\tenB - \tenD^*\tenB^{-1}\tenD$ is positive semidefinite. Each channel couples one gas variable with one radiation variable, so in the partition into gas and radiation variables all four blocks of $\tenB$ and of $\tenB^{-1}$ are diagonal, and $\tenD$, the transport, acts only on the radiation variables. The gas block of $\tenB - \tenD^*\tenB^{-1}\tenD$ is then diagonal with the positive entries $K_k(2 + q_k + K_k)/(1 + q_k + K_k)$. Minimizing its quadratic form over the gas variables leaves a quadratic form on the radiation variables alone, with a matrix $\tenC$ (the Schur complement); since the gas block is positive definite, the matrix is positive semidefinite if and only if $\tenC$ is. Writing out $\tenB$ and $\tenB^{-1}$ channel by channel, a short calculation gives
\begin{equation}
    \tenC = \left(2\tenI - \tenV\right) - \tenD_{\rm r}^*\tenV\tenD_{\rm r} + \left(\tenV - \tenI\right)\tenD_{\rm r} + \tenD_{\rm r}^*\left(\tenV - \tenI\right)
    = \tenW + \tenW^* - \tenW^*\tenV\tenW,
    \qquad
    \tenV \equiv \mathrm{diag}_k\left(\frac{2}{a_*^k}\right),
    \label{eq:schur_stage}
\end{equation}
where $\tenD_{\rm r} = \tenI - \tenW$ is the radiation block of $\tenD$ and $\tenV$ multiplies the radiation variable of each channel by $2/a_*^k$. By \autoref{eq:W_sym}, $\tenW = \Sigma\,\tenI + i\tenY$, with $\Sigma \equiv \sum_d a_d\zeta_d/2$ and the real symmetric $\tenY \equiv \sum_d (a_d/2)\sin\theta_d\,\tenX_d$. Hence $\tenW + \tenW^* = 2\Sigma\,\tenI$ and $\tenW^*\tenW = \Sigma^2\,\tenI + \tenY^2$, where $\tenY^2$ has the largest eigenvalue $\sum_d (a_d/2)^2\sin^2\theta_d = \sum_d (a_d/2)^2\zeta_d(2 - \zeta_d)$. Let $a_* \equiv \min_k a_*^k$, so that $\tenV$ multiplies every variable by at most $2/a_*$. Then, for every radiation vector $\mathbf{x}$,
\begin{equation}
    \frac{\mathbf{x}^*\tenC\,\mathbf{x}}{\mathbf{x}^*\mathbf{x}} \geq 2\Sigma - \frac{2}{a_*}\left[\Sigma^2 + \sum_d \frac{a_d^2}{4}\zeta_d(2 - \zeta_d)\right]
    = 2\Sigma - \frac{2}{a_*}\left[\sum_{d \neq d'}\frac{a_d a_{d'}}{4}\zeta_d\zeta_{d'} + \sum_d \frac{a_d^2}{2}\zeta_d\right] .
    \label{eq:schur_bound}
\end{equation}
Since $\zeta_d, \zeta_{d'} \in [0, 2]$, $(\zeta_d - 1)(\zeta_{d'} - 1) \leq 1$, i.e., $\zeta_d\zeta_{d'} \leq \zeta_d + \zeta_{d'}$. The bracket is therefore at most $\sum_d (a_d\zeta_d/2)\left(\sum_{d' \neq d} a_{d'} + a_d\right) = a\,\Sigma$, and the right-hand side is at least $2\Sigma\,(1 - a/a_*)$, which is non-negative if $a \leq a_*$. Conversely, at the checkerboard $\zeta_d = 2$, $\tenY = 0$ and $\Sigma = a$, so that $\tenC = 2a\,\tenI - a^2\tenV$, which is positive semidefinite only if $a \leq a_*^k$ for every $k$. \qed
\end{pf}

\begin{pfcb}
If $a \leq a_*^k$ for every $k$, then by Lemma~\ref{lem:stage} the hypothesis of Lemma~\ref{lem:two_stage} holds for every $\boldsymbol{\theta}$, so that the full step does not increase the norm $\|\cdot\|_{\tenI + \tenB}$ of the rescaled variables, and $|g| \leq 1$ for every eigenvalue. Undoing the rescaling gives $\|\cdot\|_\star$. Conversely, if $a > a_*^k$ for some $k$, the checkerboard is unstable by Proposition~\ref{prop:energy_block}. \qed
\end{pfcb}

The matrix $\tenI + \tenB = 2(\tenI - \tfrac{1}{2}\tenS)$ is twice the matrix of the final implicit solve, $\tenH$, written for all variables. For a gray material in one dimension, in the original variables, the norm is
\begin{equation}
    \|\delta\hvecW\|_\star^2 = 2\left[\frac{16}{\lambda}\,\delta\hT^2 + \delta\hE^2 + \frac{(1 + \chi)^2}{\sigma^2\calC^2\lambda_F}\,\delta\hv^2 + \frac{\delta\hF^2}{\sigma^2}\right] + q\left(4\,\delta\hT - \delta\hE\right)^2 + \frac{q}{\sigma^2}\left(\frac{1 + \chi}{\calC}\,\delta\hv - \delta\hF\right)^2 .
    \label{eq:H_star}
\end{equation}
The bracket is a sum of squares in which the gas perturbations are weighted by the inverse of the heat-capacity ratio $\lambda$ and of its momentum analogue $\lambda_F$; the last two terms measure the departures from the exchange equilibria, $\delta\hE = 4\,\delta\hT$ and $\delta\hF = (1 + \chi)\,\delta\hv/\calC$, with the weight $q$.

\paragraph{Strict damping below the boundary} At and below the boundary, every mode other than the checkerboard and the uniform mode, $\boldsymbol{\theta} = 0$, is strictly damped, in any number of dimensions. With $\Sigma(\boldsymbol{\theta}) = \sum_d a_d\zeta_d/2$ the damping of \autoref{eq:W_sym}, these are the modes with $0 < \Sigma < a$.

\begin{lem}
\label{lem:strict}
Let $a_d > 0$ in every active direction $d$ and $a \leq a_*^k$ for every channel $k$. Then every eigenvalue of $\tenG_{\boldsymbol{\theta}}$ satisfies $|g| < 1$ for every $\boldsymbol{\theta}$ with $0 < \Sigma(\boldsymbol{\theta}) < a$.
\end{lem}

\begin{pf}
Let $\tenG\mathbf{x} = g\mathbf{x}$ with $|g| = 1$ and $\mathbf{x} \neq 0$. Such a mode keeps its norm, so $\tenG\mathbf{x} = \mathbf{x}$ by the remark after Lemma~\ref{lem:two_stage}, and $(\tenI + \tenB)\,\mathbf{y} = \mathbf{x} + \tenB\,\mathbf{z}$ with $\mathbf{y} = \mathbf{x}$ gives $\mathbf{z} = \mathbf{x}$: two stages return the mode unchanged. One stage, $\tenB^{-1}\tenD$, must then have an eigenvalue $+1$ or $-1$: if $(\tenB^{-1}\tenD + \tenI)\,\mathbf{x} \neq 0$, this vector is an eigenvector with $+1$; otherwise $\mathbf{x}$ is one with $-1$. We show that neither exists. Write the gas and radiation variables of channel $k$ as $x_{{\rm g},k}$ and $x_{{\rm r},k}$, and recall from \autoref{eq:W_sym} that the transport acts on the radiation variables as $\tenW = \Sigma\,\tenI + i\tenY$ with $\tenY$ real symmetric, so that the real part of $\mathbf{x}_{\rm r}^*\tenW\mathbf{x}_{\rm r}$ is $\Sigma\,|\mathbf{x}_{\rm r}|^2$.

An eigenvalue $+1$ means $\tenD\mathbf{x} = \tenB\mathbf{x}$. The gas row of channel $k$ gives $K_k^{1/2}x_{{\rm g},k} = q_k^{1/2}x_{{\rm r},k}$, the exchange equilibrium, and the radiation row then gives $(\tenW\mathbf{x}_{\rm r})_k = 0$. Hence $\Sigma\,|\mathbf{x}_{\rm r}|^2 = 0$, so $\mathbf{x}_{\rm r} = 0$, and then $\mathbf{x}_{\rm g} = 0$.

An eigenvalue $-1$ means $\tenD\mathbf{x} = -\tenB\mathbf{x}$. The gas row now gives $(2 + K_k)\,x_{{\rm g},k} = (K_kq_k)^{1/2}x_{{\rm r},k}$, and eliminating $x_{{\rm g},k}$ from the radiation row gives $(\tenW\mathbf{x}_{\rm r})_k = [2 + q_k - K_kq_k/(2 + K_k)]\,x_{{\rm r},k} = a_*^k\,x_{{\rm r},k}$, by \autoref{eq:astar_channel}. The real part of $\mathbf{x}_{\rm r}^*\tenW\mathbf{x}_{\rm r}$ is then $\Sigma\,|\mathbf{x}_{\rm r}|^2 = \sum_k a_*^k|x_{{\rm r},k}|^2 \geq a\,|\mathbf{x}_{\rm r}|^2$, which contradicts $\Sigma < a$ unless $\mathbf{x}_{\rm r} = 0$, and then $\mathbf{x}_{\rm g} = 0$. \qed
\end{pf}

The condition $a_d > 0$ only excludes a step without transport, $\Delta t = 0$, at which the exchange equilibria are neutral, $g = 1$.

By continuity, Lemma~\ref{lem:strict} implies that just above the boundary the checkerboard grows faster than every mode with $\theta$ bounded away from zero. The longest waves need a separate argument, since at $\theta = 0$ the exchange equilibria are neutral; an expansion of the eigenvalues in $\zeta$ shows that they are damped at a rate proportional to $\zeta$, so that the checkerboard grows fastest among them as well. We do not reproduce this expansion; the numerical tests below confirm that the checkerboard is the fastest-growing mode just above the boundary in every case.

For $\calM \neq 0$ the exchange couples the energy and momentum channels, and no constant rescaling makes it symmetric, so the proof does not apply and \autoref{eq:numax} is no longer exact. We test this case numerically.

\subsection{Numerical tests}
\label{app:numerics}

\subsubsection{Moving backgrounds}

We next consider whether Proposition~\ref{prop:checkerboard} continues to apply for moving backgrounds, keeping the assumption of an isotropic radiation field, by testing whether it remains the case that the checkerboard mode, $\theta = \pi$, is the first to become unstable as the local CFL number $\nu$ increases. We use the full amplification matrix $\tenG_\theta$ of \autoref{eq:ampl}, with the transport symbol of \autoref{eq:hll_symbol} and the exchange matrix
of \autoref{ssec:linear_waves} evaluated for the general case $\calM \neq 0$ (cf.~\autoref{eq:S_static}) is
\begin{equation}
\tenM = \calL \left(\begin{array}{ccccc}
0 & 0 & 0 & 0 & 0 \\
0 & -\frac{4\calC \calP}{3\calC^2-\calM^2} & -\frac{4\calP\calM}{\calC} & -\frac{\calP\calM}{3\calC} & \calP \\
0 & 0 & -\frac{4\left(\calC^2-\calM^2\right)\calP\gamma_1}{\calC} & \frac{\left(3\calC^2+\calM^2\right)\calP\gamma_1}{3\calC} & -2\calP\calM\gamma_1 \\
0 & \frac{4\phi \calC\calM}{3\calC^2-\calM^2} & 4\phi\calC & -\phi\calC & \phi\calM \\
0 & \frac{4\phi\calC^2}{3\calC^2-\calM^2} & 4\phi\calM & \frac{\phi\calM}{3} & -\phi\calC
\end{array}\right).
\end{equation}
We test at $\calC = 10^4$. The density perturbation is neither transported nor exchanged in the radiation step, so its eigenvalue is exactly unity; we exclude it and take the spectral radius $\rho(\tenG)$ over the remaining variables $(\delta\hv, \delta\hT, \delta\hE, \delta\hF)$.

For each parameter set we first find $\nu_\pi$, the largest $\nu \leq \nu_c$ (i.e.\ $\phi \leq 1$) for which $\rho(\tenG_\pi) \leq 1$, by bisection to a relative precision of $10^{-9}$. We then evaluate $\rho(\tenG_\theta)$ on 420 values of $\theta$: 20 logarithmically spaced in $[10^{-4}\pi, 4\times10^{-3}\pi]$ and 400 uniformly spaced in $[5\times 10^{-3}\pi, \pi]$. We make two tests.
\begin{enumerate}
    \item At 30 values of $\nu$, logarithmically spaced between $10^{-3}\nu_\pi$ and $(1 - 10^{-3})\nu_\pi$, $\rho(\tenG_\theta) \leq 1$ for every $\theta$. No mode is unstable below the limit of the checkerboard.
    \item Where $\nu_\pi < \nu_c$, so that the boundary is active, at $\nu = (1 + 10^{-3})\nu_\pi$ the largest $\rho(\tenG_\theta)$ occurs at $\theta = \pi$. The checkerboard is the first mode to become unstable.
\end{enumerate}
Together the two tests place the boundary of the complete scheme within $10^{-3}$ of $\nu_\pi$. We scan all combinations of
\begin{equation}
\begin{gathered}
    \sigma \in \{1/\sqrt{3}, 1\}, \qquad \gamma \in \{5/3, 7/5\}, \qquad \nu_c \in \{30, 300, 3000\}, \\
    \calP \in \{10^{-6}, 10^{-4}, 10^{-3}, 10^{-2}, 0.1, 1, 10\}, \\
    \tau_{\rm c} \in \{10^{-3}, 0.1, 0.5, 1, 2, 5, 20, 10^3, 10^6\}
\end{gathered}
\label{eq:theta_params}
\end{equation}
for a static background, $\calM = 0$, and the same set with $\sigma = 1/\sqrt{3}$ (Eddington closure) for $\calM = 0.5$ and 3. \autoref{tab:theta} lists the results for each Mach number $\calM$ and signal speed $\sigma$ of the background: the number of parameter sets (cases), the number of cases in which the boundary is active, $\nu_\pi < \nu_c$, the largest spectral radius minus one over all $\theta$ and all tested $\nu < \nu_\pi$ (test 1), the number of active cases in which the fastest-growing mode at $\nu = (1 + 10^{-3})\nu_\pi$ is $\theta_* = \pi$ (test 2), and the largest relative difference between $\nu_\pi$ and the closed form $\nu_{\max}$ of \autoref{eq:numax} (capped at $\nu_c$). Both tests pass in every case. The largest $\rho - 1$ below $\nu_\pi$ is at the level of floating-point round-off; it is not exactly zero because in the radiation step the matter modes are nearly neutral at every $\theta$. The last column compares $\nu_\pi$ with the closed form, which is derived from the checkerboard at $\calM = 0$: the two agree to the precision of the bisection for a static background, as Proposition~\ref{prop:checkerboard} requires, and, on this grid, to $10^{-3}$ for $\calM \leq 3$.

A search over continuous parameters finds larger differences for a moving background, up to $6\times10^{-3}$ at $\calM = 3$, in cells with $\tau_{\rm c}$ just below $2\sigma$ and $\gamma$ close to one. There \autoref{eq:numax} slightly overestimates the stable $\nu$, but the checkerboard remains the least stable mode.

\begin{figure}[t]
    \centering
    \includegraphics[width=0.6\textwidth]{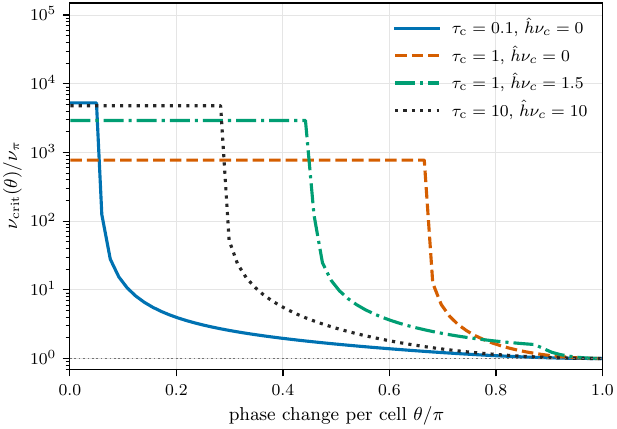}
    \caption{Stability limit $\nu_{\rm crit}(\theta)$ of each Fourier mode, divided by that of the checkerboard, $\nu_\pi$, for four cells with $\nu_c = 10^4$, $\sigma = 1/\sqrt{3}$, $\gamma = 5/3$ and $\calM = 0$, labeled by the cell optical depth $\tau_{\rm c}$ and $\hat{h}\nu_c$. A mode is stable for $\nu \leq \nu_{\rm crit}(\theta)$. The plateaus at small $\theta$ are the cap $\nu_{\rm crit} \leq \nu_c$, i.e.\ $\phi = 1$ is stable for those modes. Every curve lies above unity for $\theta < \pi$ and reaches it at $\theta = \pi$.}
    \label{fig:theta}
\end{figure}

\autoref{fig:theta} shows the same result mode by mode for four cells. For each $\theta$ we find the largest $\nu \leq \nu_c$ at which that single mode is stable, $\nu_{\rm crit}(\theta)$. In each cell $\nu_{\rm crit}(\theta)/\nu_\pi$ exceeds unity for every $\theta < \pi$ and reaches it at $\theta = \pi$. Long waves are stable at much larger $\nu$ than the checkerboard, often up to $\phi = 1$.

\subsubsection{Anisotropic M1 states}

For M1 about an anisotropic state the proof of \ref{app:proof} does not apply because the flux Jacobian is not that of a fixed Eddington tensor, and its largest characteristic speed exceeds $\sigma_d = T_{dd}^{1/2}$, by up to a factor of 1.7 for the closure of \citet{Levermore84a}. Anisotropic states are typical of optically thin cells, so we first scan every mode of an optically thin cell, with the Jacobian of the M1 flux and the HLL speeds $\mp\sigma_d\hat{c}$, floored at $0.1\,\hat{c}$ as in QUOKKA and Honghu.jl, for radiation states up to $|\vecF|/(cE) = 0.999$ in directions that include those where this factor is largest. All modes are stable, in one, two and three dimensions, both at the step of Proposition~\ref{prop:rsla}, $\nu_0 = 1/\sqrt{D}$, and at the bound itself, $\nu\sum_d\sigma_d = 1$, where only the checkerboard reaches $|g| = 1$.

An anisotropic state can also occur where the exchange is strong. A collimated beam absorbed in a layer thinner than a cell keeps $|\vecF| = cE$ at any optical depth, since absorption removes photons without redirecting the survivors; an example is starlight absorbed at the surface of a dusty disk, in the frequency groups of the starlight. We therefore also scan absorbing cells, with the same Jacobian and HLL speeds, absorption into a gas that holds the heat capacity ($K = 0$), cell optical depths from $10^{-3}$ to $10^3$, reduced fluxes up to $0.9994$ in every direction, and the step that the limiter \autoref{eq:phi_sb} sets with $\beta = 0.9$, or steps up to $\nu = 10^4$ where it sets no bound. On grids of one, two and three dimensions, with the flux in any direction that the grid resolves (along the axis in one dimension, since a one-dimensional simulation assumes symmetry about the axis and so carries no transverse flux), every mode is stable in all 300, 1197 and 1365 cases, and remains so at the bound itself, $\beta = 1$. For radiation that is diffusive rather than directed, \ref{app:m1} shows that the bound remains a stability condition for the complete M1 update.

\begin{table}
    \centering
    \small
    \begin{tabular}{lcccccc}
    \hline
    $\calM$ & $\sigma$ & cases & $\nu_\pi < \nu_c$ & $\displaystyle\max_{\nu < \nu_\pi,\,\theta}\rho - 1$ & $\theta_* = \pi$ & $\max|\nu_\pi/\nu_{\max} - 1|$ \\
    \hline
    0   & $1/\sqrt{3}$ & 378 & 290 & $2.7\times10^{-15}$ & 290/290 & $1.3\times10^{-8}$ \\
    0   & 1            & 378 & 303 & $2.2\times10^{-15}$ & 303/303 & $1.0\times10^{-7}$ \\
    0.5 & $1/\sqrt{3}$ & 378 & 290 & $2.7\times10^{-15}$ & 290/290 & $1.2\times10^{-4}$ \\
    3   & $1/\sqrt{3}$ & 378 & 290 & $2.2\times10^{-15}$ & 290/290 & $1.0\times10^{-3}$ \\
    \hline
    \end{tabular}
    \caption{Test that the checkerboard sets the stability boundary of the IMEX PD-ARS scheme with HLL transport.}
    \label{tab:theta}
\end{table}

\end{appendix}

\end{document}

%% file: tab_media.tex
\begin{table}
    \centering
    \small
    \setlength{\tabcolsep}{2.5pt}
    \begin{tabular}{lcccccccc}
    \hline
    System & $\rho$ & $T$ & $\kappa$ & $v$ & $\mathcal{L}$ & $\mathcal{D}$ & $\hat{h}$ & $\phi/\phi_0$ \\
    & [g cm$^{-3}$] & [K] & [cm$^2$ g$^{-1}$] & [km s$^{-1}$] \\
    \hline
    PPD midplane, 0.1 au & $10^{-7}$ & $1000$ & $1$ & 94 & $3\times10^{3}$ & $0.9$ & $3.8\times10^{-6}$ & $100$ \\
    PPD midplane, 1 au & $10^{-9}$ & $280$ & $1$ & 30 & $500$ & $0.05$ & $8.3\times10^{-6}$ & $100$ \\
    First hydrostatic core & $5\times10^{-12}$ & $100$ & $1$ & 1 & $380$ & $10^{-3}$ & $1.2\times10^{-4}$ & $100$ \\
    \makecell[l]{RGB star, middle\\ of envelope} & $8\times10^{-4}$ & $1.8\times10^{5}$ & $5\times10^{2}$ & 0.5 & $2.8\times10^{10}$ & $4\times10^{4}$ & $1.1\times10^{-3}$ & $100$ \\
    \makecell[l]{Sun, middle of\\ convection zone} & $0.054$ & $9.6\times10^{5}$ & $10^{2}$ & 0.1 & $1.9\times10^{10}$ & $8\times10^{3}$ & $2.5\times10^{-3}$ & $100$ \\
    \makecell[l]{Massive protostellar\\ envelope} & $10^{-14}$ & $100$ & $1$ & 3 & $15$ & $2\times10^{-4}$ & $0.061$ & $1.0$ \\
    \makecell[l]{AGB star, base\\ of envelope} & $3$ & $7\times10^{7}$ & $0.34$ & 6 & $4.5\times10^{9}$ & $9\times10^{4}$ & $17$ & $1.0$ \\
    Massive-star envelope & $10^{-9}$ & $2\times10^{5}$ & $1$ & 10 & $100$ & $3\times10^{-3}$ & $1.2\times10^{3}$ & $1.0$ \\
    ULIRG nucleus & $2.5\times10^{-19}$ & $80$ & $1$ & 100 & $32$ & $0.01$ & $1.2\times10^{3}$ & $1.0$ \\
    AGN dusty torus & $10^{-18}$ & $300$ & $5$ & 200 & $15$ & $0.01$ & $10^{4}$ & $1.0$ \\
    \hline
    \end{tabular}
    \caption{Representative parameters of optically thick astrophysical systems and the gain $\phi/\phi_0$ of the ARSA.}
    \label{tab:media}
\end{table}